\documentclass{svjour3} 
\smartqed 
\usepackage{graphicx}
\usepackage{longtable}
\def\lsim{\,\lower2truept\hbox{${<\atop\hbox{\raise4truept\hbox{$\sim$}}}$}\,}
\def\gsim{\,\lower2truept\hbox{${>\atop\hbox{\raise4truept\hbox{$\sim$}}}$}\,}

\usepackage{natbib} \bibpunct{(}{)}{;}{a}{}{,} 

\begin{document}

\title{Radio and Millimeter Continuum Surveys and their Astrophysical
Implications
}

\titlerunning{Radio and Millimeter Surveys}        

\author{Gianfranco De Zotti        \and Marcella Massardi \and Mattia Negrello
\and Jasper Wall
}


\institute{G. De Zotti and Marcella Massardi \at
              INAF - Osservatorio Astronomico di Padova,
              Vicolo dell'Osservatorio 5,
              I-35122 Padova, Italy \\
              and
              SISSA/ISAS,
              Via Beirut 2--4,
              I-34014 Trieste, Italy \\
              Tel.: +39-049-8293444,
              Fax: +39-049-8759840\\
              \email{gianfranco.dezotti,marcella.massardi@oapd.inaf.it}           
           \and
           Mattia Negrello \at
           Department of Physics and Astronomy,
           Open University,
           Walton Hall,
           Milton Keynes MK7 6AA, United Kingdom \\
           \email{m.negrello@open.ac.uk}
                      \and
           Jasper V. Wall \at
           Department of Physics and Astrophysics,
           University of British Columbia,
           Vancouver, Canada V6T 1Z1 \\
           \email{jvw@astro.ubc.ca}
}

\date{Received: date / Accepted: date}

\maketitle

\begin{abstract}
We review the statistical properties of the main populations of
radio sources, as emerging from radio and millimeter sky surveys.
Recent determinations of local luminosity functions are presented
and compared with earlier estimates still in widespread use. A
number of unresolved issues are discussed. These include: the
(possibly luminosity-dependent) decline of source space densities at
high redshifts; the possible dichotomies between evolutionary
properties of low- versus high-luminosity and of flat- versus
steep-spectrum AGN-powered radio sources; and the nature of sources
accounting for the upturn of source counts at sub-mJy levels. It is
shown that straightforward extrapolations of evolutionary models,
accounting for both the far-IR counts and redshift distributions of
star-forming galaxies, match the radio source counts at flux-density
levels of tens of $\mu$Jy remarkably well. We consider the
statistical properties of rare but physically very interesting
classes of sources, such as GHz Peak Spectrum and ADAF/ADIOS
sources, and radio afterglows of $\gamma$-ray bursts. We also
discuss the exploitation of large-area radio surveys to investigate
large scale structure through studies of clustering and the
Integrated Sachs-Wolfe effect. Finally we briefly describe the
potential of the new and forthcoming generations of radio
telescopes. A compendium of source counts at different frequencies
is given in an appendix.

\end{abstract}

\section{Introduction}
\label{intro}
For several decades, extragalactic radio surveys remained the most powerful tool to probe the distant universe. Even
`shallow' radio surveys, those of limited radio sensitivity, reach sources with redshifts predominantly above 0.5. Since
the 1960s, the most effective method for finding high-$z$ galaxies has been the optical identification of radio sources,
a situation persisting until the mid-1990's, when the arrival of the new generation of 8-10\ m class optical/infrared
telescopes, the refurbishment of the Hubble Space Telescope, the Lyman-break technique \citep{1996AJ....112..352S} and
the Sloan Digital Sky Survey \citep{2000AJ....120.1579Y} produced an explosion of data on high-redshift galaxies.

This is not a historical account \citep[see][]{Sullivan}; but listing the revolutions in astrophysics and cosmology
wrought by radio surveys serves to set out concepts and terminology. On the astrophysics side we note the following:

\begin{enumerate}
\item{\bf Active Galactic Nuclei (AGNs)} The discovery of radio galaxies \citep{1949Natur.164..101B, 1950MNRAS.110..508R}
whose apparently prodigious energy release \citep{1959ApJ...129..849B} suggested Compton catastrophe, calling the
cosmological interpretation of redshifts into question.

\item {\bf Synchrotron emission} The identification of synchrotron emission \citep{Gin51, Shk52}
as the dominant continuum process producing the apparent power-law spectra of radio sources.


\item {\bf Quasars} The discovery of quasars, starting with 3C\,273
\citep{Haz63, 1963Natur.197.1040S}, leading to the picture of the collapsed supermassive nucleus
\citep{1963MNRAS.125..169H}, and hence to the now-accepted view of the powerful Active Galactic Nucleus (AGN) --
massive black-hole - accretion disk systems \citep{Lyn69} powering double-lobed \citep{1953Natur.172..996J} radio
sources via `twin-exhaust' relativistic beams \citep{Bla74, Sch74}.

\item {\bf Relativistic beaming} The discovery of superluminal motions
of quasar radio components \citep{1971ApJ...170..207C}, this non-anisotropic emission \citep[anticipated
by][]{1967MNRAS.135..345R} resolving the Compton non-catastrophe \citep{1966ApJ...146..597W} and leading to the
development of unified models of radio sources: quasars and radio galaxies are one and the same, with orientation of
the axis to the viewer's line of sight determining classification via observational appearance \citep{Ant85,
1989SciAm.260...20B, 1995PASP..107..803U}.



\end{enumerate}

\noindent On the cosmology side we note the following:

\begin{enumerate}

\item {\bf Scale of the observable Universe} An irrefutable argument
by \citet{1955RSPSA.230..448R} placed the bulk of `radio stars' beyond 50 Mpc, and it was quickly realized when
arcmin positional accuracy became available \citep{Smi52} that the majority of the host galaxies were beyond the
reach of the optical telescopes of the epoch. \citet{1960PASP...72..354M} measured a redshift of 0.46 for 3C\,295,
the redshift record for a galaxy for 10 years. Astronomers had discovered a set of objects substantially `beyond' the
recognized Universe. By 1965 the redshift record was 2.0 for the quasar 3C\,9 \citep{Sch65}. Only after the turn of
the century did the redshift record become routinely set by objects discovered in surveys other than at radio
wavelengths \citep[e.g.][]{Ste00}.

\item {\bf History of the Universe} Early radio surveys generated a passionate and personal debate, the Steady-State
vs Big-Bang controversy. It was rooted in the simplest
statistics to be derived from any survey: the {\it integral} source counts, the number of objects per unit sky area
above given intensities or flux densities. As discussed by \citet{1955RSPSA.230..448R}, the source count from the 2C
radio survey \citep{1955MmRAS..67..106S} showed a cumulative (integral) slope of $\sim -3$, far steeper than that
expected from the Steady-State prediction, any reasonable Friedman model, or from a static Euclidean universe. For
each of these, the initial slope at the highest flux densities is $-3/2$. (Euclidean case: the number of sources,
$N$, is proportional to the volume, i.e. to $r^3$ for a sphere; the flux density is $\propto r^{-2}$, so that
$N\propto S^{-3/2}$.) \citet{1948MNRAS.108..252B} together with \citet{1948MNRAS.108..372H} were uncompromising
proponents of the new Steady-State theory. Ryle et al. interpreted the 2C apparent excess of faint sources in terms
of the radio sources having far greater space density at earlier epochs of the Universe. Confusion, the blending of
weak sources to produce a continuum of strong sources, was then shown to have disastrous effects on the early
Cambridge source counts. From an independent survey in the South, \citet{1958AuJPh..11..360M} found an initial slope
of -1.65 after corrections for instrumental effects, significantly lower than that found for 2C.
\citet{1957PCPS...53..764S} developed the P(D) technique, circumventing confusion and showing that the interferometer
results of 2C were consistent with the findings of Mills et al. But the damage had been done: cosmologists, led by
Hoyle, believed that radio astronomers did not know how to interpret their data.

\end{enumerate}

\noindent In 1965 the `source-count controversy' became irrelevant in one sense. \citet{Pen65} found what was immediately
interpreted \citep{1965ApJ...142..414D} as the relic radiation from a hot dense phase of the Universe. The Big Bang was
confirmed.

Ryle was right all the time. Integral source-count slopes of $-1.8$ or even as shallow as $-1.5$ were nowhere near what
the known redshifts plus Steady State cosmology -- or even any standard Friedman cosmology -- predicted. These all come
out at $-1.2$ or $-1.3$, shallower than the asymptotic $-1.5$ as sources of infinite flux density are not observed, and
nobody has ever claimed the initial source count slope at any frequency to be as flat as this.
The discovery of the fossil radiation \citep[see][]{peebles} may indeed have shown that a Big Bang took place; but the
source counts demonstrated further that {\it objects in the Universe evolve either individually or as a population} -- a
concept not fully accepted by the astronomy community until both galaxy sizes and star-formation rates were shown to
change with epoch.

Source counts from radio and mm surveys -- with errors and biases now understood -- are currently recognized as essential
data in delineating the different radio-source populations and in defining the cosmology of AGNs. These counts are
dominated down to milli-Jansky (mJy) levels by the canonical radio sources, believed to be powered by supermassive
black-holes \citep[e.g.][]{1984RvMP...56..255B} in AGNs. At fainter flux-density levels, a flattening of slope in the
Euclidean normalized {\it differential} counts (i.e. counts of sources with flux density $S$, within $dS$, multiplied by
$S^{2.5}$, see \S\,\ref{sec:counts}) was found \citep{1984A&AS...58....1W, 1984Sci...225...23F, 1984AJ.....89..610C},
interpreted at the time as the appearance of a new population whose radio emission is, to some still-debated extent,
associated with star-forming galaxies.

Radio-source spectra are usually described as power laws ($S_\nu \propto \nu^{-\alpha}$)\footnote{We note that this
negative sign convention for $\alpha$ is not universal; however the convention has been adopted for the K-corrections of
optical quasars and for the extrapolation from optical to X-rays (`$\alpha_{ox}$').}; the early low-frequency
meter-wavelength (e.g. 178 MHz) surveys found radio sources with spectra almost exclusively of steep power-law form, with
$\alpha \sim 0.8$. Later surveys at cm-wavelengths (higher frequencies, e.g. 5 GHz) found objects of diverse spectral
types, some with spectra rising to the high frequencies, some with steep low-frequency portions flattening and rising to
the high frequencies, and yet others with a hump in the radio regime, or indeed two or more humps. In general, anything
which was not `steep-spectrum' in form was called `flat-spectrum', an inaccurate nomenclature: very few truly
flat-spectrum sources have been found and even then the flatness persists over only a limited frequency range.
Nevertheless AGN-powered radio sources are traditionally classified in two main categories: steep- ($\alpha > 0.5$) and
flat-spectrum ($\alpha < 0.5$). Broadly speaking, to radio telescopes the steep-spectrum objects showed extended
double-lobed structures, while the flat-spectrum objects were point sources, unresolved until the Very-Long-Baseline
Interferometry (VLBI) technique provided sub-arc-second mapping. The compact nature of flat-spectrum sources led to the
conventional interpretation of  synchrotron self-absorption at frequencies below the bump(s), implying brightness
temperatures of $\sim10^{11}$\,K for the estimated magnetic field strengths.

From a physical point of view, it is appropriate to consider the integrated spectra as composites, built of the
combination of different {\it components} of radio sources. {\it Unified models} provide a framework for such a
discussion.

\begin{figure}
 \includegraphics[width=11.8cm]{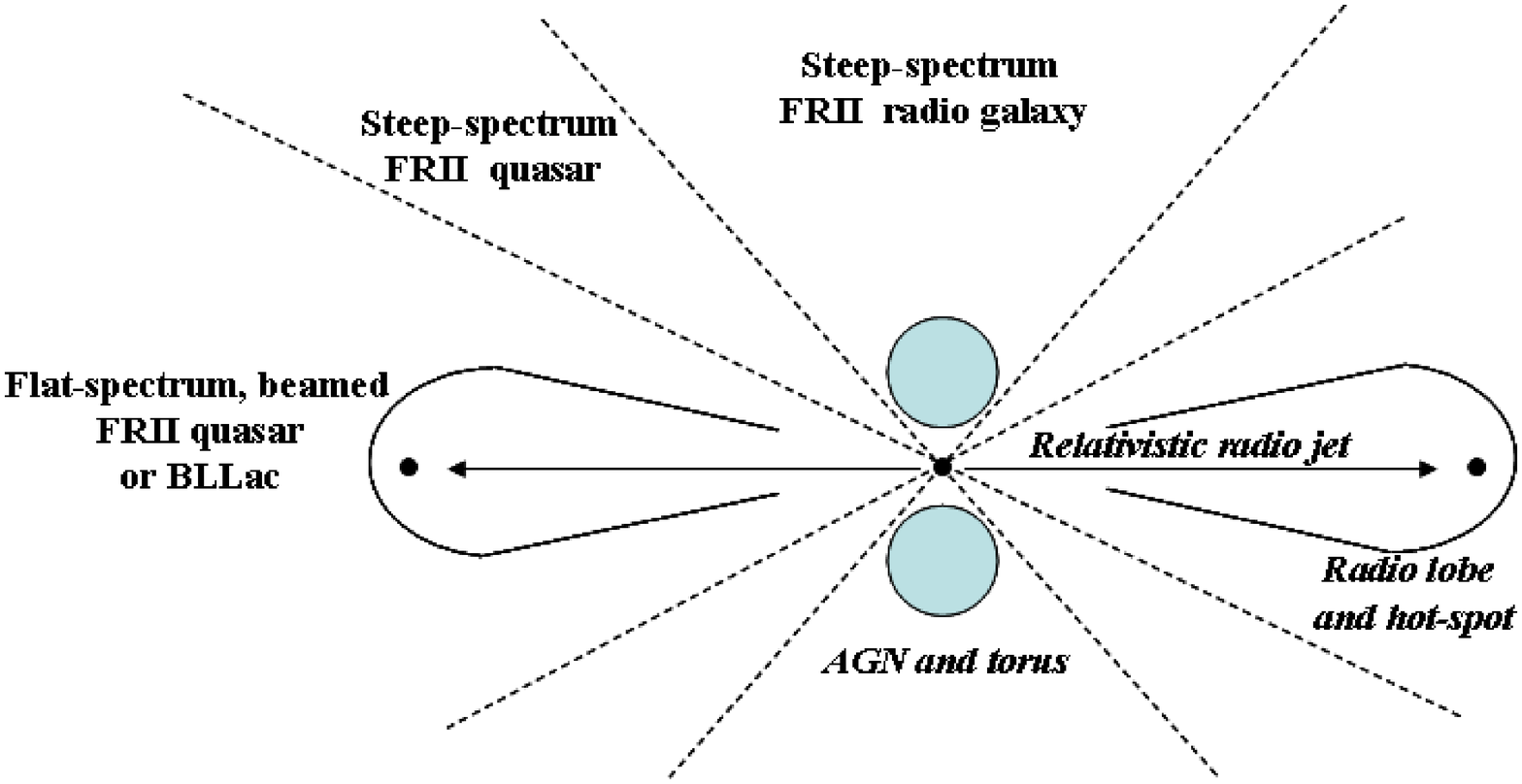}
\caption{Unified scheme for high radio-power Fanaroff-Riley (1974;
FRII) sources (following Jackson \& Wall 1999).}\label{fig:unified}
\end{figure}

In the widely accepted `unification' scheme
\citep{1979Natur.277..182S, 1982MNRAS.200.1067O,
1987slrs.work..104S, 1989SciAm.260...20B} the appearance of sources,
including this steep-spectrum/flat-spectrum dichotomy, depends
primarily on their their axis orientation relative to the observer.
This paradigm stems from the discovery of relativistic jets
\citep{1971ApJ...170..207C, 1972IAUS...44..228M} giving rise to
strongly anisotropic emission. In the radio regime
(Fig.~\ref{fig:unified}), a line-of-sight close to the source
jet-axis offers a view of the compact, Doppler-boosted,
flat-spectrum base of the approaching jet. Doppler-boosted
low-radio-power \citep[][type I (FRI;
edge-dimmed)]{1974MNRAS.167P..31F} sources are associated with BL
Lac objects, characterized by optically-featureless continua, while
the powerful type~II (FRII; edge-brightened) sources are seen as
flat-spectrum radio quasars (FSRQs). The view down the axis offers
unobstructed sight of the black-hole -- accretion disk nucleus at
wavelengths from soft X-rays to UV to IR, and this accretion-disk
radiation may outshine the starlight of the galaxy by 5 magnitudes.
The source appears stellar, either as a FSRQ or a BL\,Lac object.
FSRQs and BL Lacs are collectively referred to as {\it blazars.} In
the case of a side-on view, the observed low-frequency emission is
dominated by the extended, optically-thin, steep-spectrum
components, the radio lobes; and the optical counterpart generally
appears as an elliptical galaxy. A dusty torus \citep{Ant85} hides
the black-hole -- accretion-disk system from our sight
(Fig.~\ref{fig:unified}). At intermediate angles between the
line-of-sight and the jet axis, angles at which we can see into the
torus but the alignment is not good enough to see the
Doppler-boosted jet bases, the object appears as a `steep-spectrum
quasar'.

\begin{figure}
 \includegraphics[width=11.8cm]{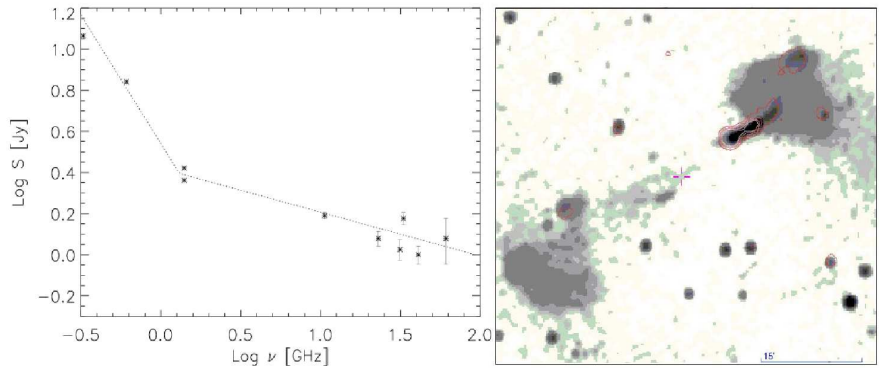}
\caption{Spectral behaviour in the millimeter band of the radio galaxy
NGC6251 (left panel) and (right panel) 11-GHz isophotes overlaid on
the 0.3 GHz map (Mack et al. 1997). The low-frequency spectrum is due
to the steep-spectrum outer lobes while at higher frequencies the
flatter-spectrum core-jet system dominates. }\label{fig:sp_morph}
\end{figure}

In general, then, each source has both a compact, flat-spectrum core and extended steep-spectrum lobes
(Fig.~\ref{fig:sp_morph}). This already implies that a simple power-law representation of the integrated radio spectrum
can only apply to a limited frequency range. The reality is even more complex \citep{1994AuJPh..47..625W}. External
absorption or, more frequently, self-absorption (synchrotron and free-free) can produce spectra rising with frequency at
the low-frequency optically-thick regime, while at high frequencies the synchrotron emission becomes optically thin,
power law, and energy losses of relativistic electrons \citep[``electron ageing'',][]{1966ApJ...146..621K} translate into
a spectral steepening.

Two classes of ultra-steep-spectrum ($\alpha > 1.3$) sources have
been discovered. One is associated with galaxy clusters; the objects
are of relatively low luminosity and generally are not associated
with any host galaxy. They are diffuse and of several types,
including cluster `radio halos', `radio relics' and `mini-halos',
and each type appears to involve reactivation of the hot
intra-cluster medium by shocks or cooling flows, the observed form
depending on the cluster evolutionary state
\citep{2008MmSAI..79..176F}. These `radio ghosts' will not be
discussed further here. The second class of ultra-steep-spectrum
source is very radio-luminous and these are mostly identified with
very-high-redshift radio galaxies. The high redshifts tempt the
suggestion that the steep spectral index is due to the effect of
redshift moving the steepest part of the spectrum (where electron
ageing effects are strong) into the observed frequency range.
However, \citet{2006MNRAS.371..852K} demonstrated that this is not
the dominant mechanism, and that high-redshift radio galaxies,
discovered by the steep-spectrum technique, have intrinsically
power-law spectra. The selection of ultra-steep-spectrum sources is
a very effective, but not the only \citep{2009arXiv0907.1447J}, way
to find {\it high redshift radio galaxies} \citep[see][for a
comprehensive review]{2008A&ARv..15...67M}, including the one
holding the current record, TN J0924-2201 at $z=5.19$
\citep[$\alpha_{0.365}^{1.4} \simeq 1.6$;][]{1999ApJ...518L..61V}.
The highest-redshift radio-loud {\it quasar} known to date, the
$z=6.12$ QSO J1427+3312 \citep{2006ApJ...652..157M}, also has a
steep radio spectrum ($\alpha^{8.4}_{1.4}=1.1$) although it was not
discovered through this characteristic.

In very compact regions, synchrotron self-absorption can occur up to very high radio frequencies, giving rise to sources
with spectral peaks in the GHz range. At high radio luminosities this category comprises the GHz Peaked Spectrum (GPS)
sources \citep{1998PASP..110..493O} some of which peak at tens of GHz \citep[High Frequency
Peakers;][]{1998ASPC..144..187E, 2000A&A...363..887D, 2002NewAR..46..299D, 2005A&A...432...31T}. At low luminosities,
high-frequency spectral peaks, again due to strong synchrotron self-absorption, may be indicative of radiatively
inefficient accretion, thought to correspond to late phases of the AGN evolution, with luminosities below a few percent
of the Eddington limit (advection-dominated accretion flows \citep[ADAF,][]{1999ApJ...520..298Q} or adiabatic
inflow-outflow scenarios \citep[ADIOS,][]{1999MNRAS.303L...1B, 2004MNRAS.349...68B}).

As the `flat' spectra are actually the superposition of emitting regions peaking over a broad frequency range
\citep{1969ApJ...155L..71K, 1980ApJ...238L.123C}, whose emission is strongly amplified and blue-shifted by relativistic
beaming effects, a power-law description is a particularly bad approximation. The spectral shapes are found to be
complicated, and generally show single or multiple humps. Many of these show flux-density variations, attributed to the
birth and expansion of new components and shocks forming in relativistic flows in parsec-scale regions. The variations
may be on times scales from hours to months or even years, and substantial resources have been devoted to monitoring
these variable sources, led by groups at Michigan (USA) and Metsah\"{o}vi (Finland) \citep[e.g.][]{2003ASPC..300..159A,
2008Natur.452..851V}. The latter reference shows how global (multi-wavelength and multi-telescope) these monitoring
programmes have become; moreover the quasi-periodicity for the object in question, OJ\,287, indicates that it is probably
a binary black-hole system. With regard to flux variations, we also note the `Intra-Day Variables' (IDVs), blazars whose
flux densities vary wildly on time scales from minutes to days: these are flat-spectrum objects with extremely small
components that show inter-stellar scintillation (ISS) via the turbulent, ionized inter-stellar medium  (ISM) of our
Galaxy \citep[e.g.][]{2007ASPC..365..279L}. Detailed discussion of all these variable objects is beyond the scope of this
review.
\begin{figure}
\includegraphics[height=11cm,angle=90]{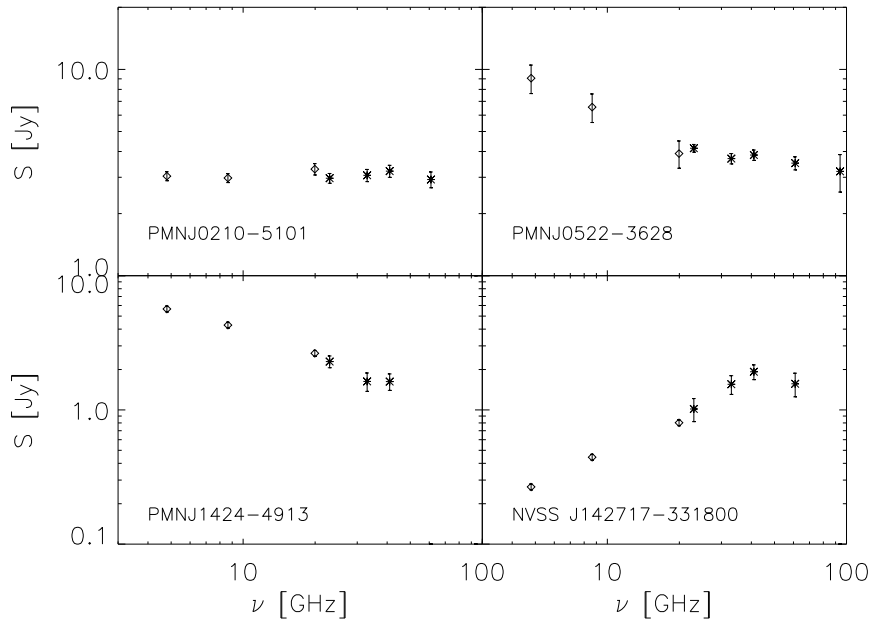}
\vspace{0.5cm} 
\caption{Examples of radio-source spectra at mm wavelengths: a
flat-spectrum source (\emph{top left panel}); a steep-spectrum source
(\emph{bottom left panel}); a source whose spectrum flattens at
$\nu\sim 10$~GHz (\emph{top right panel}); a High Frequency Peaker
(HFP) source (\emph{bottom right panel}). Data from the NEWPS
Catalogue \citep[][\emph{asterisks}]{2007ApJS..170..108L} and from
the AT20G Survey
\citep[\emph{diamonds},][]{2008MNRAS.384..775M}.}\label{fig:sour_spectra}
\end{figure}


The discovery of Compact Steep Spectrum sources \citep[CSS;][]{1981A&AS...43..381K, 1982MNRAS.198..843P,
1998PASP..110..493O} originally appeared to be an exception to the conventional wisdom that steep and flat spectra are
associated with extended and compact sources respectively. CSS sources are unresolved or barely resolved by standard
interferometric observations (arcsec resolution), and the integrated spectra show peaks at $< 0.5\,$GHz, above which the
spectral indices (on average, $\alpha \sim 0.75$) are typical of extended radio sources. There is compelling evidence
that these objects, as well as GPS and associated types of object (HFPs and CSOs -- Compact Symmetric Objects) are {\it
young radio galaxies}, as summarized concisely by \citet{2008arXiv0802.1976S}.

It follows from the above that the conventional two-population approach (flat- and steep-spectrum) assuming power-law
spectra is particularly defective at high radio frequencies, where several different factors (emergence of compact cores
of powerful extended sources, steepening by electron energy losses, transition from the optically-thick to the
optically-thin synchrotron regime of very compact emitting regions, etc.) combine to produce complex spectra (see
Fig.~\ref{fig:sour_spectra}). Nevertheless, for many practical applications the conventional approach remains useful in
describing the bulk population properties of AGN-powered radio sources.

The radio emission of star-forming galaxies is mostly optically-thin synchrotron from relativistic electrons interacting
with the galactic magnetic field, but with significant free-free contributions from the ionized interstellar medium
\citep{1992ARA&A..30..575C, 2002A&A...392..377B, 2008A&A...477...95C}. At mm wavelengths, however, the radio emission is
swamped by (thermal) dust emission, whose spectrum rises steeply with increasing frequency. The well-known tight
correlation between radio and far-IR emission of star-forming galaxies \citep{1985ApJ...298L...7H, 1986ApJ...305L..15G,
1991ApJ...376...95C} vastly increases the body of data relevant to characterize, or at least constrain the evolutionary
properties of this population. However, to date few attempts have been made to build comprehensive models encompassing
both radio and far-IR/sub-mm data \citep[but see][]{2003MNRAS.341L...1G}.

In this paper we first review the observed radio to mm-wave source counts (\S\,\ref{sec:counts}), the data on the local
luminosity function of different radio source populations (\S\,\ref{sec:AGNLLF}), and the source spectral properties
(\S\,\ref{sec:spectra}). Next (\S\,\ref{sec:AGN}) we look at evolutionary models for the classical radio sources as well
as for individual populations, such as GPS sources, ADAF/ADIOS sources, and (\S\,\ref{starforming}) star-forming galaxies
and $\gamma$-ray afterglows at radio wavelengths. We deal briefly with the Radio Background (\S\,\ref{sec:bck}) and the
Sunyaev-Zeldovich effect on cluster and galaxy scales(\S\,\ref{SZ}).
Section~\ref{clust} contains a summary of the information on large scale structure stemming from large-area radio
surveys. Finally, in \S\,\ref{sec:future} we summarize perspectives for the future, and \S\,\ref{concl} contains some
conclusions.

Unless otherwise noted, we adopt a flat $\Lambda$CDM cosmology with $\Omega_{\Lambda}=0.7$ and
$H_0=70\,\hbox{km}\,\hbox{s}^{-1}\,\hbox{Mpc}^{-1}$.

\begin{figure}
\includegraphics[height=11.8cm, width=10cm, angle=90]{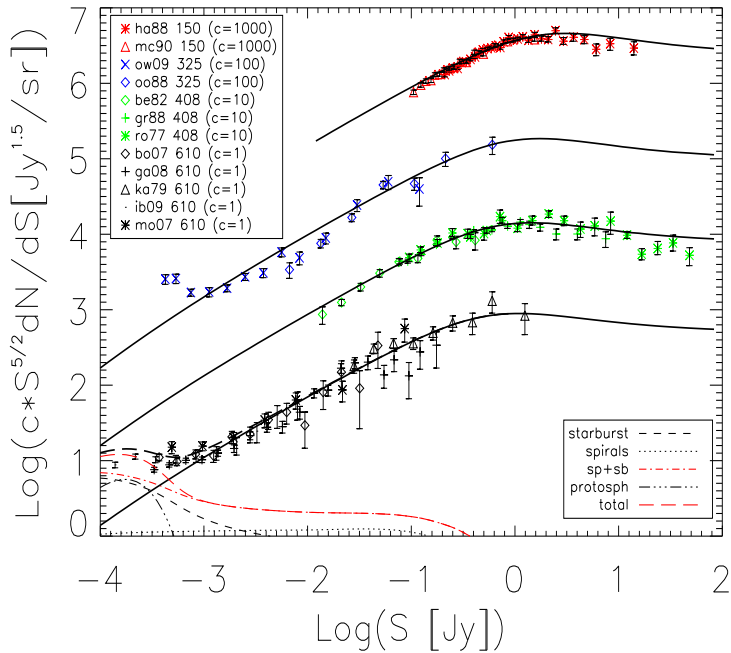}
\caption{Differential source counts at 150, 325, 408, 610 MHz normalized to
$c S^{-2.5}_\nu$, with $c=1000$, 100, 10, 1 respectively. Reference codes
are spelt out in the notes to Tables~\protect\ref{tab:150MHz},
\protect\ref{tab:408MHz}, and \protect\ref{tab:610MHz}. The lines are
fits yielded by an updated evolution model (Massardi et al., in
preparation). }\label{fig:count_low}
\end{figure}

\begin{figure}
\includegraphics[height=11.8cm, width=10cm, angle=90]{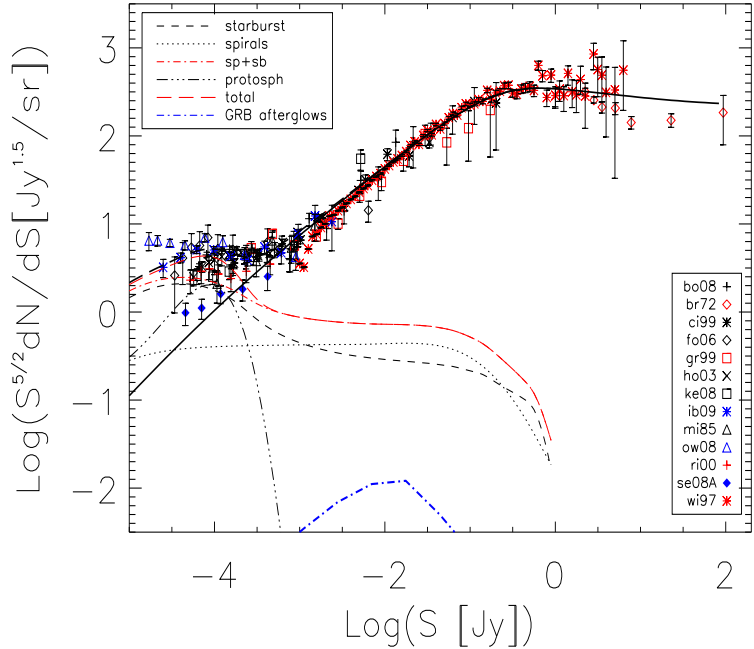}
\caption{Normalized differential source counts at 1.4 GHz. Note that
the filled diamonds show the counts of AGNs only, while all the other
symbols refer to total counts. Reference codes are spelt out in the
note to Table~\protect\ref{tab:1d4GHz}. A straightforward
extrapolation of evolutionary models fitting the far-IR to mm counts
of populations of star-forming (normal late type (spirals or sp), starburst (sb), and
proto-spheroidal) galaxies, exploiting the well established
far-IR/radio correlation, naturally accounts for the observed counts
below $\sim 30\,\mu$Jy (see \S\,\protect\ref{starforming}). At higher flux densities the counts are
dominated by radio-loud AGNs: the thick solid line shows the fit of
the same model as in Fig.~\protect\ref{fig:count_low}. The dot-dashed
line shows the counts of $\gamma$-ray burst (GRB) afterglows predicted
by the \citet{2000ApJ...540..687C} model.}\label{fig:1d4GHz}
\end{figure}

\begin{figure}
\includegraphics[height=11.8cm, width=10cm, angle=90]{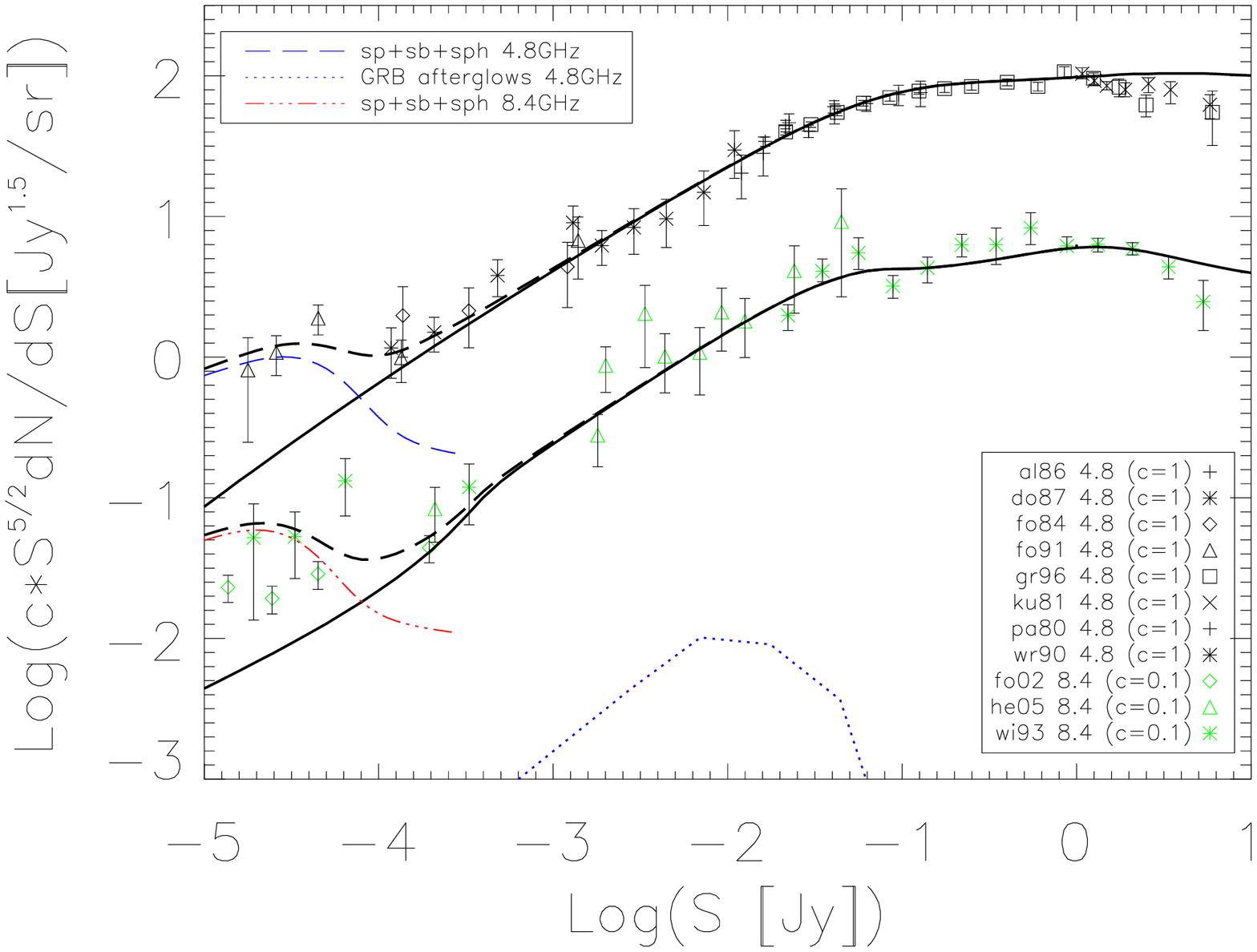}
\caption{Differential source counts at 4.8 and 8.4 GHz normalized to
$c\times S^{-2.5}_\nu$, with $c= 1$ and 0.1, respectively. Reference
codes are spelt out in the notes to Tables~\protect\ref{tab:4d8GHz}
and \protect\ref{tab:8d4GHz}. In the upper inset sp, sb, and sph stand for
spiral, starburst and proto-spheroidal galaxies, respectively. The fit to the 4.8 GHz counts is from
the same model as in Figs.~\protect\ref{fig:count_low} and
\protect\ref{fig:1d4GHz}, while at 8.4 GHz we show the fit yielded by
the \citet{2005A&A...431..893D} model, tailored for data above 5
GHz. The dotted line shows the counts of GRB afterglows at 3~GHz
predicted by the \citet{2000ApJ...540..687C} model.
}\label{fig:4d8GHz}
\end{figure}

\section{Observed source counts}\label{sec:counts}

\subsection{From surveys to counts}

The counts of sources demand cosmic evolution, but by themselves
provide limited information on this evolution. Since even bright
radio sources are frequently optically faint to invisible, the
traditional way to characterize the evolutionary properties relies
heavily on source counts from blind surveys, with limited and
incomplete cross-waveband identification and redshift information.
However, as discovered in the 2C survey, getting from a sky survey
to a source count is difficult, and modern instrumentation, while
generally avoiding the confusion issue which bedevilled 2C,  does
not remove the difficulties.

It is surface brightness, or rather differential surface brightness
above a background (CMB, Galactic radiation, ground radiation),
which is measured in radio/mm surveys. Discrete sources stand out
from this background by virtue of apparent high differential {\it
surface brightness}, $\Delta T_b$. The simple relations linking
$\Delta T_b$ to point-source flux density (via the Rayleigh-Jeans
approximation  and the radiometer equation incorporating telescope
and receiver parameters) appear in basic radio astronomy texts, e.g
\citet{1997ira..book.....B}.

Surveys are complete only to a given limit in $\Delta T_b$,
translating to {\it Jy per beam area}\footnote{1 Jy (Jansky) =
$10^{-26}\,\hbox{W}\,\hbox{Hz}^{-1}\,\hbox{m}^{-2}$ or
$10^{-23}\,\hbox{erg}\,\hbox{cm}^{-2}\,\hbox{s}^{-1}\,\hbox{Hz}^{-1}$}.
For point sources, this limit is clearly defined. For extended
sources, the total flux density
\begin{equation}
S_{\nu} = \int_{\Omega} B_{\nu}(\theta,\phi)d\Omega,
\label{s_vs_b}
\end{equation}
i.e. the incremental brightness $B$ must be integrated over the
extent of the source to find the total flux density. If a source is
extended and its brightness temperature is constant across the beam
response, then given the Rayleigh-Jeans approximation $B =
2k_BT_b/\lambda^2$ ($k_B$ is the Boltzmann constant, $\lambda$ is
wavelength), for a survey sensitivity limit of $S_{\rm lim}$ per
beam, we have from eq.~(\ref{s_vs_b})
\begin{equation}
T_{b,{\rm min}} = \frac{\lambda^2 S_{\rm lim}}{2 k_B \int d\Omega}.
\label{tmin}
\end{equation}
The integral is the beam solid angle; for a circular Gaussian beam
with full width at half maximum of FWHM arcsec, this may be
approximated as $2.66\cdot10^{-11}\, \hbox{FWHM}^2$ sterad. Two
iconic sky surveys at 1.4 GHz with the NRAO Very Large Array
illustrate the brightness limit issue. For the FIRST survey
\citep{1995ApJ...450..559B} with $\hbox{FWHM}= 5$~arcsec and $S_{\rm
min} =1$~mJy, eq.~(\ref{tmin}) gives $T_{\rm min} \approx 24\,$K,
while for the NVSS survey \citep{1998AJ....115.1693C} with a 45
arcsec beam and $S_{\rm min} = 3$~mJy, $T_{\rm min} \approx 0.9\,$K.
There are significant selection effects which arise as a
consequence, most notably the lack of sensitivity in FIRST to the
majority of spiral galaxies, near and far, as well as to low surface
brightness features such as ghost or relic radiation. The redeeming
features of its higher resolution are emphasized below.

It is a major undertaking to proceed from a list of deflections in
Jy per beam, either apparently unresolved, or resolved as regions of
emission, to a complete catalogue of radio/mm sources. In the first
place, there is the surface brightness limitation described above;
in the compromises of survey design it is critical to decide just
what population(s) of sources will be incompletely represented.
There is the issue of overlap: for instance Centaurus A, NGC\,5128,
the nearest canonical radio galaxy, extends over 9 degrees of the
southern sky; there are many discrete distant sources catalogued
within the area covered by Cen~A. There is also the double nature of
radio-galaxy emission: this requires that components found as
individual detections be `matched up' or assembled to find the true
flux density of single sources, cores as well as double lobes.
Moreover many sources show extended regions of lower surface
brightness which are poorly aligned. If source scale is large
enough, pencil-beam or filled aperture telescopes are better at
finding and mapping these than are aperture-synthesis
interferometers. The issue of `missing flux' is notorious for
interferometers, because of their limited response to the longer
wavelengths of the spatial Fourier transform of the brightness
distribution.

The difficulties have been brought to sharp focus by the excellent
decision to carry out the two major VLA surveys, FIRST
\citep{1995ApJ...450..559B} and NVSS \citep{1998AJ....115.1693C},
both at 1.4 GHz but differing in resolution by a factor of 9. From
these highly complementary surveys, the reality of how different
resolutions affect raw source lists may be seen immediately
\citep{2002MNRAS.337..993B}. FIRST and NVSS are far more than the
sum of the parts. The low resolution of NVSS gains the spiral
galaxies and much other low-surface-brightness detail not seen in
FIRST. The relatively high resolution of FIRST can be used to sort
out the blends and overlaps in NVSS, and it enables direct
cross-waveband identifications, a shortcoming of the lower NVSS
resolution. Used together they can provide samples complete on many
criteria; but significant effort in examining many individual
emission features is still required.

With regard to surveys using interferometers, the noise level in an
interferometric image is given by:
\begin{equation}\label{eq:sensitivity}
 \sigma_{\rm image}=\frac{\sqrt{2} k_B T_{\rm sys}}{A \eta_e \eta_q}\sqrt{\frac{1}{t\  N_{\rm base}\Delta\nu}}.
\end{equation}
where $T_{\rm sys}$ is the system temperature, $A$ is the antenna
surface area, $\eta_e$ is the aperture efficiency, the ratio of
effective collecting area to surface area, $\eta_q$ is the sampling
efficiency depending on digitization levels and sampling rate, $t$
is the integration time, $N_{\rm base} = N(N-1)/2$ is the number of
baselines, $N$ is the number of antennas, and $\Delta\nu$ is the
bandwidth. ($\eta_e$ is generally 0.3 to 0.8, and $\eta_q$ 0.7 to
0.9.) The integration time per pointing needed to reach a detection
limit of say $S_{\rm lim} = 5\sigma_{\rm image}$ can be
straightforwardly obtained from eq.~(\ref{eq:sensitivity}).  The
number of pointings necessary to cover a sky solid angle $\Omega_s$
with a telescope field of view FOV is\footnote{This assumes uniform
response over the field-of-view. The inevitable non-uniformity
across the FOV implies an additional factor of $\sim 2$ for uniform
sky coverage. The data from separate pointings are combined by
squaring the relative response to weight the data by the square of
the signal-to-noise ratio (SNR). If the beam is approximated by a
Gaussian, then this process effectively halves the beam size; see
\citet{1998AJ....115.1693C}.}
\begin{equation}\label{eq:num_point}
n_p=\Omega_s/\hbox{FOV}.
\end{equation}
If the integral counts of sources scale as $S^{-\beta}$, the number
of sources detected in a given area scales as $t^{\beta/2}$. For a
given flux density, the number of detections is proportional to the
surveyed area, i.e. to $t$.  Thus, to maximize the number of
detections in a given observing time it is necessary to go deeper if
$\beta>2$ and to survey a larger area if $\beta<2$. The `narrow and
deep' vs. `wide and shallow' argument for maximizing source yield
always resolves, at radio frequencies, in favour of the latter,
because $\beta > 2$, implying a differential count slope of less
than $-3$, has never been observed at any flux-density level. On the
other hand, very steep counts are observed at millimeter and
sub-millimeter wavelengths \citep{2009arXiv0907.1093A,
2006MNRAS.372.1621C}.

Compilation of complete and reliable catalogues, complete samples,
almost invariably involves data at other frequencies.
Source-component assembly for example is an iterative process which
may require cross-waveband identification of the host object,
galaxy, quasar, etc. The identification process leads on to the
construction of complete samples, complete at both the survey
frequency and at some other wavelength, i.e. in optical/IR
identifications. Such samples are rare and require great
observational effort. One of the best known of these, the '3CRR'
sample \citep{1983MNRAS.204..151L} is a revised version of the
revised 3C catalogue \citep{1962MmRAS..68..163B} from the original
3C survey of \cite{1959MmRAS..68...37E}. (The sample is also the
most extreme sample of high-power radio AGN, and its contents are
far from typical of the radio-mm survey population.) The process
will become easier with large-area optical surveys such as SDSS
\citep{2000AJ....120.1579Y} and with the advent of synoptic
telescopes such as LSST.

\begin{figure}
\includegraphics[height=11.8cm, width=9cm, angle=90]{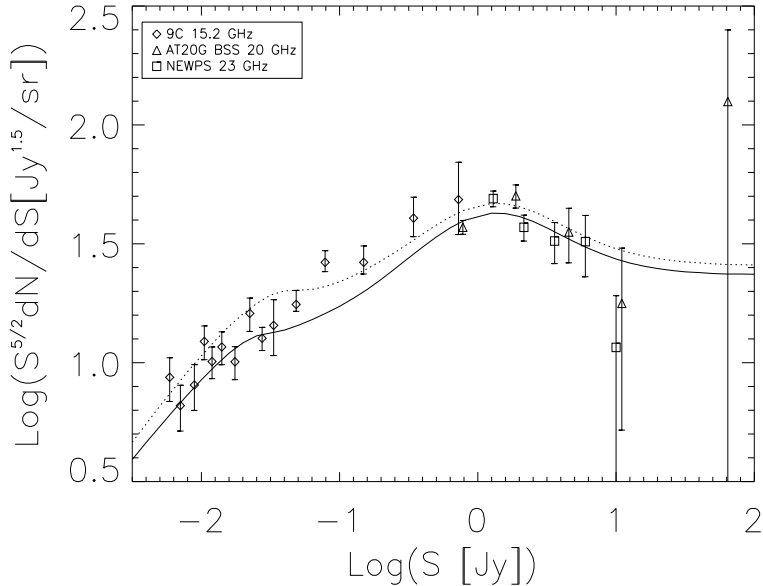}
\caption{Differential source counts normalized to $S^{-2.5}_\nu$ for
the 15\,GHz 9C survey \citep{2003MNRAS.342..915W,2009arXiv0908.0066W}, for the 20\,GHz
ATCA Bright Source Sample \citep{2008MNRAS.384..775M}, and for the
WMAP 23~GHz survey \citep{2009MNRAS.392..733M}. The model by
\citet{2005A&A...431..893D} is also shown for comparison (solid line:
20 GHz, dotted: 15 GHz). }\label{fig:count_high}
\end{figure}

Given complete samples, then, we can compile source counts. (It
should be noted that these are frequently constructed by
approximations from raw deflection lists, to circumvent the labour
discussed above. {\it Caveat emptor.}\footnote{The buyer should also
beware of confusing as complete samples (a) lists of sources in
which large volumes of data are assembled from different surveys and
different completeness algorithms \citep[e.g.
PKSCat90,][]{1990PKS...C......0W}, and (b) spectral samples, in
which flux-density measurements at different frequencies are
assembled to obtain the integrated spectra of samples of sources not
necessarily selected by survey completeness
\citep[e.g.][]{1966ApJS...13...65P, 1969ApJ...157....1K}.}) Today
the task of checking for systematic effects from approximations or
statistical procedures is made easier because the counts from
different survey samples -- except for the very deepest ones --
overlap at various flux-density levels. The counts are usually
presented in `relative differential' form, the differential counts
$dN/dS$ giving the number of sources per unit area with flux density
$S$ within $dS$, subsequently and conveniently normalized to the
'Euclidean' form, i.e. multiplied by $cS^{2.5}$, with $c$ being a
suitably chosen constant. (A uniform source distribution in a static
Euclidean universe yields $dN/dS \propto S^{-2.5}$ as described
earlier). A summary of the available source counts at different
frequencies is given in Tables~\ref{tab:150MHz}--\ref{tab:WMAP} (see
also Figs.~\ref{fig:1d4GHz}--\ref{fig:count_high}).

In the case of surveys covering small areas, the field-to-field
variations arising from the source clustering (sampling variance)
further adds to the uncertainties. The fractional variance of the
counts is \citep{1980lssu.book.....P}:
\begin{equation}
\left\langle {n-\langle n \rangle \over \langle n\rangle}
\right\rangle^2 = {1\over \langle n \rangle} + \sigma_v^2
\end{equation}
with
\begin{equation}\label{eq:sigmav}
\sigma_v^2={1\over \Omega^2}\int \int w(\theta)\,d\Omega_1\,d\Omega_2
\end{equation}
where $\theta$ is the angle between the solid angle elements
$d\Omega_1$ and $d\Omega_2$, and the integrals are over the solid
angle covered by the survey.

The angular correlation function of NVSS and FIRST sources (see
\S\,\ref{clust}) is consistent with a power-law shape
\citep{2002MNRAS.329L..37B, 2002MNRAS.337..993B,
2003A&A...405...53O}:
\begin{equation}\label{eq:w}
w(\theta) \simeq 10^{-3} (\theta/\hbox{deg})^{-0.8},
\end{equation}
for angular separations up to at least $4^\circ$. Inserting
eq.~(\ref{eq:w}) in eq.~(\ref{eq:sigmav}) we get
\begin{equation}
\sigma^2=2.36\times 10^{-3} (\Omega/\hbox{deg}^2)^{-0.4}.
\end{equation}
The errors given in Tables~\ref{tab:150MHz}--\ref{tab:8d4GHz}
include this contribution for surveys over areas $\le
25\,\hbox{deg}^2$.

Differences between source counts for independent fields are in
general far larger than these errors imply
\citep{2007ASPC..380..189C}. There is little doubt that different
calibrations, beam corrections and resolution corrections are the
dominant if not exclusive culprits. Further advances in calibration
procedures and characterization of the structures of faint sources
will be required before sampling variance comes to dominate the
errors in faint counts of radio sources.

\subsection{Low frequency surveys} \label{sec:lowfreq}

Low-frequency surveys have a long and illustrious (but initially
chequered) history, as we have mentioned. The most extensive ones,
both in terms of area (see also \S\,\ref{clust}) and of depth, are
those at $\sim 1\,$GHz and at $\sim 5\,$GHz. The NRAO VLA Sky Survey
\citep[NVSS;][]{1998AJ....115.1693C} covers the sky north of $\delta
=-40^\circ$ (82\% of the celestial sphere) at 1.4 GHz, down to $\sim
2.5\,$mJy. It has resolution of 45 arcsec FWHM and the raw catalogue
contains $1.8 \times 10^6$ entries. It is complemented by the Sydney
University Molonglo Sky Survey \citep[SUMSS;][]{2003MNRAS.342.1117M}
at 0.843~GHz. The survey was completed in 2007 with the Molonglo
Galactic Plane Survey \citep[MGPS;][]{2007MNRAS.382..382M}, and now
covers the whole sky south of declination $-30^\circ$.

The VLA 1.4-GHz FIRST survey \citep[for Faint Images of the Radio
Sky at Twenty-cm;][]{1995ApJ...450..559B} is the high-resolution (5
arcsec FWHM) counterpart of NVSS, and has yielded accurate ($<1$
arcsec rms) radio positions of faint compact sources. The new
catalog, released in July 2008 (format errors corrected in October
2008), covers $\sim 8444\,\hbox{deg}^2$ in the North Galactic cap
and $611\,\hbox{deg}^2$ in the south Galactic cap, for a total of
$9055\,\hbox{deg}^2$ yielding a list of $\sim 816,000$ objects.
Northern and Southern areas were both chosen to coincide
approximately with the area covered by the SDSS. The typical flux
density detection threshold of point sources is of about 1 mJy/beam,
decreasing to 0.75 mJy/beam in the southern Galactic cap equatorial
stripe.

Almost full-sky coverage was also achieved at $\sim 5\,$GHz --
albeit to a much higher flux-density level -- by the combination of
the Northern Green Bank GB6 survey with the Southern Parkes-MIT-NRAO
(PMN) survey. The GB6 catalog \citep{1996ApJS..103..427G} covers the
range $0^\circ \le \delta \le 75^\circ$ down to $\sim 18\,$mJy/beam,
the FWHM major and minor diameters are of $3'.6$ and $3'.4$,
respectively. The flux-density limit of the PMN catalog
\citep{1993AJ....105.1666G} is typically $\sim 30$~mJy/beam but
varies with declination, which spans the range from $-87.5^\circ$ to
$+10^\circ$; the FWHM is of $\simeq 4'.2$.

Other large-area, low-frequency surveys:

\begin{itemize}

\item the VLA Low-Frequency Sky Survey
\citep[VLSS;][]{2007AJ....134.1245C} is a 74-MHz
continuum survey covering the entire sky North of $\delta =
-30^\circ$ to a typical point-source detection limit of 0.7 Jy;

\item the Cambridge 6C survey at 151 MHz \citep[][and references
therein]{1993MNRAS.263...25H} covers most of the extragalactic
sky above $\delta = 30^\circ$, but generally away from the
Galactic plane, with $4.2'\times 4.2' \csc\delta$ resolution.
The 7C survey \citep{2007MNRAS.382.1639H}, at the same
frequency, covers a similar region of the sky with higher
resolution ($70''\times 70''\csc(\delta)$). A somewhat
lower-resolution survey has been carried out in the
low-declination strip $9h < RA < 16h$, $20^\circ < \delta <
35^\circ$ \citep{1996MNRAS.282..779W}.

\item The 8C survey \citep{1990MNRAS.244..233R, 1995MNRAS.274..447H}
covers the polar cap above $\delta = 60^\circ$ at 38 MHz with a
typical limiting flux density of about 1~Jy/beam.

\item The Westerbork Northern Sky Survey
\citep[WENSS;][]{1997A&AS..124..259R, 2000yCat.8062....0D}
covers the 3.14 sr north of $\delta=+30^\circ$ at 326 MHz with
$54''\times 54''\csc(\delta)$ resolution in total intensity and
linear polarization, to a flux-density limit of approximately 18
mJy/beam.

\end{itemize}

\noindent For more complete references to low-frequency radio
surveys, see Tables~\ref{tab:150MHz}--\ref{tab:8d4GHz}.

\subsection{Deep surveys and sub-mJy counts} \label{sec:submJy}

The deepest surveys cover small areas of sky on the scales of the
primary beams of synthesis telescopes; they are carried out with
such telescopes in single long exposures, or in nested overlapping
sets of such exposures. Because source counts are steep, only small
survey areas are required to obtain large enough samples of faint
sources to be statistically significant.

From such surveys, the deepest counts at 1.4 to 8.4 GHz show an
inflection point at $\lsim 1\,$mJy \citep{1985AJ.....90.1957M,
1985ApJ...289..494W, 1998MNRAS.296..839H, 2000ApJ...533..611R,
2003A&A...403..857B, 2003A&A...398..901C, 2003AJ....125..465H,
2004MNRAS.352..131S, 2005AJ....130.1373H, 2006A&A...457..517P,
2006ApJS..167..103F, 2006MNRAS.372..741S, 2007A&A...463..519B,
2007ApJ...660L..77I, 2008ApJ...681.1129B, 2008AJ....136.1889O}. The
point of inflection was originally interpreted as signalling the
emergence of a new source population
\citep[e.g.][]{1984ApJ...287..461C,1989ApJ...338...13C}.
\citet{1985ApJ...289..494W} suggested that the majority of sub-mJy
radio sources are faint blue galaxies, presumably undergoing
significant star formation (SF), and \citet{1987ApJ...318L..15D}
successfully modeled the sub-mJy excess counts with evolving
starburst galaxies, a model that also described the IRAS $60\,\mu$m
counts.

More recent data and analyses have confirmed that starburst galaxies
are indeed a major component of the sub-mJy 1.4 GHz source counts,
perhaps dominating below 0.3--0.1\,mJy \citep{1993MNRAS.263...98B,
1993MNRAS.263..123R, 1998MNRAS.296..839H, 2000ExA....10..419H,
2004MNRAS.352..131S, 2008MNRAS.386.1695S, 2005MNRAS.358.1159M,
2007MNRAS.378..995M,2009ApJ...694..235P}. However, spectroscopic results by
\citet{1999MNRAS.304..199G} suggested that early-type galaxies were
the dominant population at sub-mJy levels. Further, it was recently
suggested (and modeled) that the flattening of the source counts may
be caused by 'radio-quiet' AGN (radio-quiet quasars and type 2 AGN),
rather than star forming galaxies \citep{2006MNRAS.372..741S}.
Distinct counts for high and low-luminosity radio galaxies show that
low-luminosity FRI-type galaxies probably make a substantial
contribution to the counts at 1 mJy and somewhat below
\citep{2008MNRAS.390..819G}. Based on a combination of optical and
radio morphology as an identifier for AGN and SF galaxies,
\citet{2006ApJS..167..103F} suggested that at most 40\% of the
sub-mJy radio sources are AGNs, while \citet{2007ASPC..380..205P}
indicated that this fraction may be 60--80\%.
\citet{2008AJ....135.2470H} found that the host galaxy colors and
radio-to-optical ratios indicate that low-luminosity (or
``radio-quiet'') AGN make up a significant proportion of the sub-mJy
radio population. \citet{2008ApJS..177...14S}, using a newly
developed rest-frame-colour based classification in conjunction with
the VLA-COSMOS 1.4 GHz observations, concluded that the radio
population in the flux-density range of $\sim 50\,\mu$Jy to 0.7 mJy
is a mixture of 30--40\% of star forming galaxies and 50--60\% of
AGN galaxies, with a minor contribution ($\sim 10\%$) of QSOs.

The origin of these discrepancies can be traced to three main
reasons (see also \S\,\ref{sec:AGNLLF}). First, the identification
fraction of radio sources with optical counterparts, which is
generally taken to be representative of the full radio population,
spans a wide range (20\% to 90\%) in literature depending on the
depth of both the available radio and optical data. Second, it is
important to make a distinction between the presence of an AGN in
the optical counterpart of a radio source, and its contribution to
the radio emission \citep{2008MNRAS.386.1695S}. Non-radio AGN
indicators like optical/IR colours, emission lines, mid-IR SEDs,
X-ray emission, etc. are not well correlated with the radio emission
of the AGNs and therefore are not necessarily valid diagnostics of
radio emission powered by accretion onto a supermassive black hole
\citep{2005MNRAS.358.1159M}. Third, there are uncertainties in
specifying survey level: deep surveys normally cover but one primary
beam area, heavily non-uniform in sensitivity. A survey claimed
complete at some specified flux density in the central region alone
is in fact heavily biased to sources of 5 to 10 times this flux
density; the survey as a result is biased to the higher-flux-density
population, namely AGNs.

\cite{2008MNRAS.386.1695S} used four diagnostics (radio morphology,
radio spectral index, radio/near-IR and mid-IR/radio flux-density
ratios) to single out, in a statistical sense, radio emission
powered by AGN activity. They were able to calculate the source
counts separately for AGNs and star-forming galaxies. The latter
were found to dominate below $\simeq 0.1\,$mJy at 1.4 GHz, while
AGNs still make up around one quarter of the counts at $\sim
50\,\mu$Jy.

\citet{2008ApJ...681.1129B} pointed to evidence of a decline of the
1.4 GHz counts below $\sim 0.1\,$mJy. It is possible that a new
upturn may be seen at $\lsim 1\,\mu$Jy, due to the emergence of
normal star-forming galaxies \citep{1999ASPC..193...55W,
2000ExA....10..419H}.

Essentially all surveys and catalogues are carried out and compiled
without reference to polarization (the NVSS being an important
exception): linear polarization is generally less than a few
percent, and certainly at mJy levels, below the uncertainties in
flux densities due to calibration, noise and confusion. An average
of circular polarizations is generally used. (Subsequent to surveys,
thousands of measurements of polarization on individual sources have
been carried out at different frequencies, with the rotation
measures thus derived used to map the details of the Galactic
magnetic field -- see e.g. \citet{2007ApJ...663..258B}). The DRAO
1.4-GHz survey of the ELAIS N1 field \citep{2007ApJ...666..201T} was
carried out expressly to examine polarization statistics. The data
at the faintest flux densities, 0.5 to 1.0 mJy, show a trend of
increasing polarization fraction with decreasing flux density,
previously noted by \citet{2002A&A...396..463M} and
\citet{2004MNRAS.349.1267T}, at variance with current models of
population mix and evolution.

\subsection{High frequency surveys and counts} \label{sec:highfreq}

High-frequency surveys up into the mm-wavelength regime vitally
complement their low-frequency counterparts. The early cm-wavelength
surveys (Parkes 2.7 GHz, NRAO 5 GHz) in the late 1960s and 1970s
found that flat-spectrum sources -- or at least sources whose
integrated spectra were dominated by components showing synchrotron
self-absorption -- constitute 50\% or more of all sources in high
flux-density samples. Modelling space density to examine evolution
demands determination of the extent and nature of this emergent
population, most members of which are blazars.

High-frequency surveys are very time-consuming. For telescopes with
diffraction-limited fields of view the number of pointings necessary
to cover a given area scales as $\nu^2$. For a given receiver noise
and bandwidth, the time per pointing to reach the flux level $S$
scales as $S^{-2}$ so that, for a typical optically-thin synchrotron
spectrum ($S\propto \nu^{-0.7}$), the survey time scales as
$\nu^{3.4}$. However usable bandwidth is roughly proportional to
frequency, so that the scaling becomes $\sim \nu^{2.4}$; but a
20~GHz survey still takes more than $\sim 25$ times longer than a
5~GHz survey to cover the same area to the same flux-density limit.

High-frequency surveys have an additional aspect of uncertainty:
variability. The self-absorbed components are frequently unstable,
young and rapidly evolving. Variability by itself would not be an
issue except for the fact that it leads to serious biases. This is
primarily because a survey will always select objects in a high
state at the expense of those in a low state, and the steep source
count at high flux densities exacerbates this situation. A second
issue concerns the spectra. Sources are predominantly detected
`high'; to return after the survey for flux-density measurements at
other frequencies guarantees (statistically) that these new
measurements will relate to a lower state. Non-contemporaneous
spectral measurements -- if above the survey frequency -- will be
biased in the sense of yielding spectra apparently steeper than at
the survey epoch. The bias can have serious consequences for e.g.
K-corrections in space-density studies, as described below.

Cosmic Microwave Background (CMB) studies, boosted by the on-going
NASA WMAP mission and by the forthcoming ESA Planck mission, require
an accurate characterization of the high-frequency properties of
foreground radio sources both in total intensity and in
polarization. Radio sources are the dominant contaminant of
small-scale CMB anisotropies at mm wavelengths. This can be seen by
recalling that the mean contribution of unresolved sources with flux
$S_i$ to the antenna temperature $T_A$ measured within a solid angle
$\Omega$ is:
\begin{equation}
T_A=\frac{\sum_i S_i \lambda^{2}}{2k_{\rm B}\Omega}=\frac{\sum_i
S_i\ell^{2}\lambda^{2}}{8k_{\rm B}\pi^{2}},
\end{equation}
where $k_{\rm B}$ is the Boltzmann constant, $\lambda$ is the
observing wavelength, and we have taken into account that for high
multipoles ($\ell \gg 1$), $\Omega \simeq (2\pi/\ell)^2$. If sources
are randomly distributed on the sky, the variance of $T_A$ is equal
to the mean, and their contribution to the power spectrum of
temperature fluctuations grows as $\ell^2$ while the power spectra
of the CMB and of Galactic diffuse emissions decline at large
$\ell$'s (small angular scales). Therefore, Poisson fluctuations due
to extragalactic sources are the dominant contaminant of CMB maps on
scales $\lsim 30'$, i.e. $\ell \gsim 400$
\citep{1999AIPC..476..204D, 1999ASPC..181..153T}.

The diversity and complexity of radio-source spectra, particularly
for sources detected at the higher frequencies, make extrapolations
from low frequencies, where extensive surveys exist, unreliable for
the purpose of establishing CMB contamination. Removing this
uncertainty was the primary motivation of the Ryle-Telescope 9C
surveys at 15.2 GHz \citep{2001MNRAS.327L...1T,
2003MNRAS.342..915W}. These were specifically designed for source
subtraction from CMB maps produced by the Very Small Array (VSA) at
34 GHz. The surveys have covered an area of $\simeq
520\,\hbox{deg}^2$ to a $\simeq 25\,$mJy completeness limit.
\citet{2009arXiv0908.0066W} reported on a series of deeper regions,
amounting to an area of $115\,\hbox{deg}^2$ complete to
approximately 10 mJy, and of $29\,\hbox{deg}^2$ complete to
approximately 5.5 mJy. The counts over the full range 5.5 mJy -- 1
Jy are well described by a simple power-law:
\begin{equation}\label{eq:wal}
 {dN\over dS}\simeq 51 \left({S\over \hbox{ Jy}}\right)^{-2.15}\
\hbox{Jy}^{-1}\,\hbox{sr}^{-1}.
\end{equation}
A 20-GHz survey of the full Southern sky to a limit of $\simeq
50\,$mJy has been carried out by exploiting the Australia Telescope
Compact Array (ATCA) fast-scanning capabilities (15$^\circ$
min$^{-1}$ in declination along the meridian) and the 8-GHz
bandwidth of an analogue correlator. The correlator was originally
developed for the Taiwanese CMB experiment AMiBA
\citep{2001AIPC..586..172L} but has been applied to three of the 6
22~m dishes of the ATCA. A pilot survey \citep{2004MNRAS.354..305R,
2006MNRAS.371..898S} at 18.5~GHz was carried out in 2002 and 2003.
It detected 173 sources in the declination range $-60^\circ$ to
$-70^\circ$ down to 100~mJy. The full survey was begun in 2004 and
was completed in 2008. More than 5800 sources brighter that 45 mJy
were detected below declination $\delta = 0^\circ$. An analysis of a
complete
 flux-limited sub-sample ($S_{20\rm GHz} > 0.50\,$Jy)
comprising 320 extragalactic radio sources was presented by
\citet{2008MNRAS.384..775M}.

Shallow (completeness levels $\gsim 1\,$Jy) all-sky surveys at 23,
33, 41, 61, and 94 GHz have been carried out by the Wilkinson
Microwave Anisotropy Probe (WMAP). Analyses of WMAP 5-year data have
yielded from 388 \citep{2009ApJS..180..283W} to 516
\citep{2009MNRAS.392..733M} detections. Of the latter, 457 are
identified with previously-catalogued extragalactic sources, 27 with
Galactic sources; 32 do not have counterparts in lower frequency all
sky surveys and may therefore be just high peaks of the highly
non-Gaussian fluctuation field.

Counts at $\sim 30$ GHz have been estimated from DASI data over the
range 0.1 to 10 Jy \citep{2002Natur.420..772K}, from CBI maps in the
range 5--50 mJy \citep{2003ApJ...591..540M}, and down to 1 mJy from
the SZA blind cluster survey (Muchovej et al. 2009, in prep.).

\citet{2005MNRAS.360..340C} used 33-GHz observations of sources
detected at 15 GHz to extrapolate the 9C counts in the range
$20\,\hbox{mJy} \le S_{33}\le 114\,\hbox{mJy}$.
\citet{2009arXiv0901.4330M} carried out Green Bank Telescope (GBT)
and Owens Valley Radio Observatory (OVRO) 31-GHz observations of
3165 NVSS sources; 15\% of them were detected. Under the assumption
that the $S_{31\rm GHz}/S_{1.4\rm GHz}$ flux ratio distribution is
independent of the 1.4 GHz flux density over the range of interest,
they derived the maximum likelihood 1.4 to 31 GHz spectral index
distribution, taking into account 31-GHz upper limits, and exploited
it to estimate the 31-GHz source counts at mJy levels: $N(>S) =
(16.7 \pm 0.4)\,\hbox{deg}^{-2}\,(S/1\,\hbox{mJy})^{-0.80 \pm 0.01}$
($0.5\,\hbox{mJy} < S < 10\,\hbox{mJy}$). The derived counts were
found to be 15\% lower than predicted by the
\citet{2005A&A...431..893D} model.

Preliminary indications of a spectral steepening of flat-spectrum
sources above $\sim 20\,$GHz, beyond the expectations of the blazar
sequence model \citep{1998MNRAS.299..433F, 1998MNRAS.301..451G} have
been reported. \citet{2007MNRAS.379.1442W} used the spectral-index
distributions over the range 1.4--43 GHz based on `simultaneous'
multifrequency follow-up observations \citep{2004MNRAS.354..485B} of
a sample of extragalactic sources from the 9C survey at 15 GHz to
make empirical estimates of the source counts at 22, 30, 43, 70, and
90 GHz by extrapolating the power-law representation of the 15-GHz
counts [eq.~(\ref{eq:wal})]. \citet{2008MNRAS.385.1656S} carried out
simultaneous 20- and 95-GHz flux densities measurements for a sample
of AT20G sources. The inferred spectral-index distribution was used
to extrapolate the AT20G counts to 95 GHz.  The extrapolated counts
are lower than those predicted by the \citet{2005A&A...431..893D}
model, and (except at the brightest flux densities) also lower than
the extrapolation by \citet{1994AAS...185.1212H} of the 5-GHz
counts. On the other hand, they are within the range of the
\citet{2007MNRAS.379.1442W} estimates in the limited flux density
range where both data sets are valid, although the slopes are
significantly different. Both \citet{2007MNRAS.379.1442W} and
\citet{2008MNRAS.385.1656S} assume that the spectral index
distribution is independent of flux density. This can only be true
for a limited flux density interval, since the mixture of steep-,
flat-, and inverted-spectrum sources varies with flux density. In
fact, the median 20--95 GHz spectrum ($\alpha=0.39$) found by
\citet{2008MNRAS.385.1656S} is much flatter than that
($\alpha=0.89$) measured at 15--43 GHz by
\citet{2007MNRAS.379.1442W} for a fainter sample.

Of course, extrapolations from low frequencies can hardly deal with
the full complexity of source spectral and variability properties,
and may miss sources with anomalously inverted spectra falling below
the threshold of the low-frequency selection. They are therefore no
substitute for direct blind high-frequency surveys. On the other
hand, the recent high frequency surveys (9C, AT20G, WMAP) did not
produce ``surprises'', such as a population of sources not present
in samples selected at lower frequencies. The analysis of WMAP 5-yr
data by \citet{2009MNRAS.392..733M} has shown that the counts at
bright flux densities are consistent with a constant spectral index
up to 61 GHz, although at that frequency there is a marginal
indication of a spectral steepening. The WMAP counts at 94 GHz are
highly uncertain because of the limited number of detections and the
lack of a reliable flux calibration. However, taken at face value,
the WMAP 94-GHz counts are below the predictions by the
\citet{2005A&A...431..893D} model by $\simeq 30\%$. This indication
is confirmed by recent measurements of the QUaD collaboration
\citep{2009AAS...21334006F} who suggest that the model counts should
be rescaled by a factor of 0.7 and of 0.6 at 100 and 150 GHz,
respectively.

An indication in the opposite direction, albeit with very poor
statistics, comes from the MAMBO 1.2-mm (250 GHz) blank-field
imaging survey of $\sim 0.75\,\hbox{deg}^2$ by
\citet{2006A&A...448..823V}. This survey has uncovered 3
flat-spectrum radio sources brighter than 10 mJy, corresponding to
an areal density several times higher than expected from
extrapolations of low-frequency counts without spectral steepening.

A 43-GHz survey of $\sim 0.5\,\hbox{deg}^2$, carried out with $\sim
1600$ snapshot observations with the VLA in D-configuration, found
only one certain source down to 10 mJy (Wall et al., in
preparation). A statistical analysis of the survey data yielded a
source-count law in good agreement with predictions of
\citet{2007MNRAS.379.1442W} and \citet{2008MNRAS.385.1656S}. There
is no strong indication of a previously unrecognized population
intruding at this level.

\begin{figure}
\includegraphics[height=9.cm, width=11.8cm]{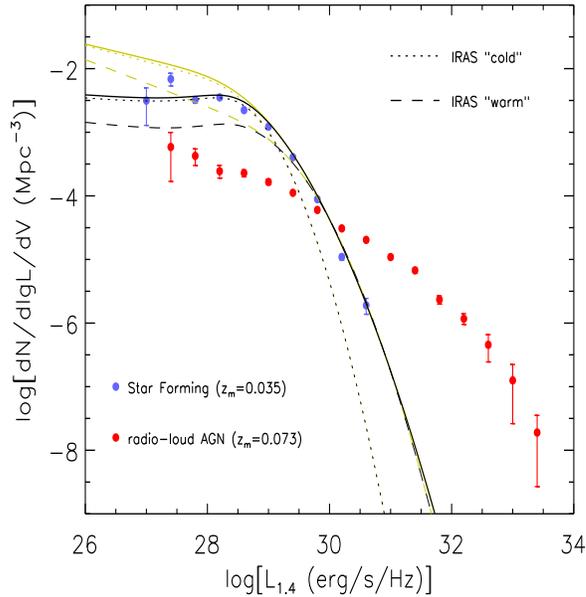}
\caption{Local luminosity functions at 1.4~GHz of radio AGNs (red
dots) and of star-forming galaxies (blue dots), as estimated by
\citet{2007MNRAS.375..931M}. The lines show extrapolations to
1.4 GHz of the $60\mu{\rm m}$ local luminosity functions of ``warm''
(usually interpreted as starburst; dashed) and ``cold'' (normal late type; dotted)
IRAS galaxies by \citet{2003ApJ...587L..89T}; the solid lines are the
sum of the two contributions. The yellow lines refer
to the linear radio/far-IR relationship of eq.~(\ref{eq:corr}), while
the black lines are based on that of eq.~(\ref{eq:corr1}), which
deviates from linearity at low luminosities. }\label{fig:rlf1d4GHz}
\end{figure}

\begin{figure}
\includegraphics[height=10cm, angle=90]{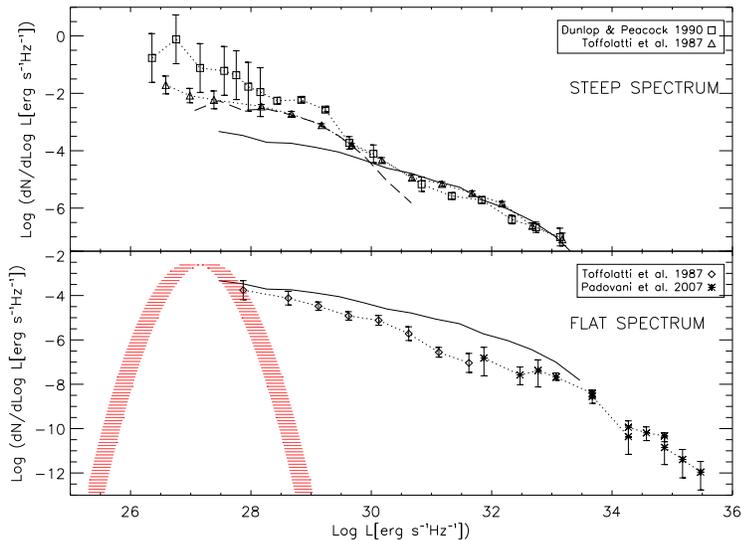}
\caption{ Contributions     of     steep-     (upper     panel)     and
flat/inverted-spectrum (lower  panel) sources to  the local luminosity
functions at  1.4~GHz. In  both panels the  solid line shows  Mauch \&
Sadler's (2007) estimate of the {\it total} (i.e., flat/inverted- plus
steep-spectrum) local radio luminosity  function of AGNs. In the upper
panel, the dashed line shows  Mauch \& Sadler's (2007) estimate of the
local    luminosity   function    of   starburst    galaxies   (mostly
steep-spectrum),             that             dominate             for
$\log\,L(\hbox{erg}\,\hbox{s}^{-1}\,\hbox{Hz}^{-1})    <    30$;   the
estimates by Toffolatti et al.  (1987) and Dunlop \& Peacock (1990) of
the local luminosity functions  of steep-spectrum sources include both
starburst  galaxies  and   AGNs.  The  Toffolatti's  local  luminosity
function of  flat-spectrum sources  (lower panel) joins  smoothly with
the estimated local luminosity function  of BL Lac objects obtained by
Padovani et al. (2007). The cross-hatched area shows the range spanned
by the estimates by Pierpaoli  \& Perna (2004) of the local luminosity
functions of ADAF sources.  }\label{fig:fl_flat_steep}
\end{figure}

\section{Local luminosity functions} \label{sec:AGNLLF} The local luminosity function (LLF)
describes the local space density of sources as a function of luminosity: it constitutes an important boundary condition
for evolutionary models. Its determination is complicated by several factors discussed e.g. by
\citet{1987A&A...184....7T}. Ideally we would like to have a large, complete radio-selected sample of sources, all with
redshift measurements, all at low enough redshifts for evolutionary effects to be insignificant, yet distant enough for
the redshifts to be accurate distant estimators. The sources should be distributed over large enough volumes for
clustering effects to average away.

In practice, however, the well-known fact that the redshift
distribution of complete samples of radio sources peaks at $z \sim
1$ for all flux-density levels down to $\sim 10\,$mJy, implies that
local sources are swamped by the much more numerous distant ones.
Singling them out by means of complete redshift measurements is
therefore impractical, and we must confine ourselves to those
brighter than some optical magnitude limit, i.e. we must deal with
both radio and optical selection. If the magnitude limit for
redshift measurements is too shallow, we lose the contribution of
optically-faint galaxies. To some extent this bias may be corrected
using bivariate (optical/radio) luminosity functions
\citep{1976ApJ...207..700F, 1977A&A....57...41A}. Alternatively,
radio surveys of optically-selected samples can be used.

Spectacular progress has been recently achieved combining large-area
spectroscopic surveys (Las Campanas, SDSS, 2dF and 6dF) with the
NVSS and FIRST surveys \citep{2000A&A...360..463M,
2002MNRAS.333..100M, 2002MNRAS.329..227S, 2005MNRAS.362....9B,
2007MNRAS.375..931M}. A more local sample has been used by
\citet{2002AJ....124..675C}.

If full redshift information is available for a flux-limited
radio-selected sample containing too few local sources for a
meaningful LLF to be directly derived, the LLF can be estimated from
the luminosity distribution of the sample, for any chosen evolution
function \citep{1980MNRAS.193..683W} . The results are, by definition,
model dependent, although the evolution function may be tightly
constrained by source counts and other data.

The key to this process is disentangling the star-forming galaxies
from the AGNs. Radio AGNs dominate above $S_{1.4\rm GHz} \simeq
10\,$mJy; at lower flux densities an increasing fraction of nearby
galaxies whose radio emission is fuelled by active star formation
appears. Optical spectra can be used to identify the dominant process
responsible for the radio emission of each source. Star-forming
galaxies have spectra typical of HII regions with strong narrow
emission lines of H$\alpha$ and H$\beta$, while AGNs may have a
variety of spectra, including pure absorption lines (like spectra of
giant elliptical galaxies), LINER or conventional type 1 or type 2 AGN
spectra. Optical AGN spectra, however, do not necessarily imply that
the radio emission is of nuclear origin. In fact, there is a body of
evidence that the star-formation and nuclear activities are tightly
connected, but the radio and optical emissions of AGNs are largely
uncorrelated -- about 90\% of AGNs are radio-quiet. An important
diagnostic tool to distinguish between galaxies whose radio emission
is due to star formation and those harbouring a radio-loud AGN is the
very-well-established, remarkably tight and nearly linear correlation
between FIR and radio continuum emission from star-forming galaxies
\citep{1985ApJ...298L...7H, 1986ApJ...305L..15G,
1991ApJ...376...95C}. A frequently-used criterion
\citep{2002AJ....124..675C} is that galaxies with radio to far-IR flux
ratio more than three times higher than the mean for star-forming
galaxies are classified as AGN-powered. \citet{2007MNRAS.375..931M}
found disagreement between spectroscopic classification and the
radio/FIR diagnostic at the $\sim 10\%$ level; a similar reliability
was estimated for their classification based on optical
spectroscopy. Objects with composite AGN $+$ starburst radio emission
are probably a primary source of classification ambiguity. This
suggests that the classification uncertainties may contribute
significantly to the overall errors on the local luminosity function
of each population. Nevertheless, rather accurate estimates of the
separate 1.4-GHz local luminosity functions of AGNs and starburst
galaxies are now available (see Fig.~\ref{fig:rlf1d4GHz}).

With the star-formers disentangled from the radio AGNs, a further
dichotomy in the local luminosity function evaluation is required.
Evolutionary models for radio AGNs generally split the total radio
AGN local luminosity function into the contributions of the steep-
and flat-spectrum sources. As discussed in \S\,\ref{sec:spectra},
this is a rather crude, but frequently unavoidable, simplification.
(Obviously, also the radio spectra of star-forming galaxies must be
known for evolution models, but the problem is simpler because in
most cases the spectra are ``steep'', with mean $\alpha \sim 0.7$
and a relatively narrow dispersion.)

The 1.4-GHz selection emphasizes steep-spectrum sources, but the flat-spectrum sources may be important in some
luminosity ranges. The estimates of separated local luminosity functions for the two populations go back to
\cite{1981MNRAS.196..597W}, \cite{1985MNRAS.217..601P}, \cite{1987A&A...184....7T}, \cite{1987MNRAS.225..297S}, with
little progress thereafter. \citet{2008MNRAS.385..310R} estimated that the density of steep-spectrum sources with $\log
L(1.4\,{\rm GHz})/\hbox{erg}\,\hbox{s}^{-1}\,\hbox{Hz}^{-1} > 32$ is $\simeq (3.0 \pm 1.2)\times
10^{-7}\,\hbox{Mpc}^{-3}$.

Moderate to low-luminosity flat- or inverted-spectrum sources are mostly classified as BL Lac objects. Very weak,
inverted-spectrum radio sources in the centers of otherwise quiescent ellipticals may correspond to late phases of AGN
evolution (ADAF or ADIOS sources, see \S\,\ref{ADAF}). The observational information on this latter population is very
limited. \citet{2004MNRAS.354.1005P} assumed that their space density equals that of elliptical galaxies brighter than
$L_\star$, and adopted a log-normal luminosity function with mean $\log L(2.7\,{\rm
GHz})/\hbox{erg}\,\hbox{s}^{-1}\,\hbox{Hz}^{-1}$ in the range 27--28, and dispersion $\sigma=0.25$. As illustrated by
Fig.~\ref{fig:fl_flat_steep}, the data on the local luminosity function of flat/inverted-spectrum sources already
constrain the space density of these sources.

As for BL Lacs, a serious hindrance in the determination of the
luminosity function is their essentially featureless spectrum,
complicating (or defeating) redshift determination. However, several
lines of evidence suggest that their luminosity function evolves
weakly if at all \citep{2007ApJ...662..182P}, so that the useful
volume for computing the local luminosity function extends up to
substantial redshifts. The estimate by \citet{2007ApJ...662..182P}
compares well with the LLF of flat-spectrum sources obtained by
\citet{1987A&A...184....7T}. On the contrary, high luminosity
flat-spectrum sources are very rare locally and evolve strongly, so
that a model independent estimate of the local luminosity function is
essentially impossible.

\section{Source spectra and evolution}
\label{sec:spectra}

This discussion of radio spectra is far from exhaustive: it sets out to serve two purposes. One is related to the
K-corrections, the correction for spectral form which must be used to derive luminosities at rest-frame frequencies.
Getting these corrections right is essential in determining space density. The second issue concerns relating source
counts at different frequencies, and in particular modelling the poorly-determined high-frequency counts from the
well-defined low-frequency counts. This limited discussion thus ignores some aspects of radio spectral measurements which
are critical to the astrophysics of radio AGNs, such as variability and monitoring, mentioned briefly in the
Introduction.

We have noted that spectra of radio sources are frequently represented
as simple power-laws, $S\propto \nu^{-\alpha}$, with the spectral
index, $\alpha \sim 0.8$ for steep-spectrum sources and $\sim 0$ for
the flat-spectrum ones. However, all radio galaxies deviate from this
simple behaviour. Various physical mechanisms contribute to shaping
the emission spectrum. At low rest-frame frequencies spectra generally
show a sharp decline with decreasing frequency, attributed to
synchrotron self-absorption; a low energy cut-off to the spectrum of
relativistic electrons may also have a role
\citep{1989MNRAS.239..401L}. The decline is mostly observed at rest
frequencies of tens of MHz, but the absorption turnover frequency can
be orders of magnitude higher, as in GPS and ADAF/ADIOS sources.

In the optically-thin regime, the spectral index of synchrotron emission, the dominant radiation mechanism encountered in
classical radio astronomy, reflects the index of the energy distribution of relativistic electrons. This distribution is
steepened at high energies by synchrotron losses as the source radiates, and by inverse Compton losses on either the
synchrotron photons themselves or on photons of the external environment \citep{1991AJ....102.1659K}. Inverse Compton
losses off the cosmic microwave background (CMB) increase dramatically with redshift since the radiation energy density
grows as $(1+z)^4$. As a consequence, a decline with increasing redshift of the frequency at which the spectral
steepening occurs can be expected.

While inverse Compton losses are most important to sources with weak magnetic fields, powerful sources may possess more
intense magnetic fields enhancing the synchrotron emission. The faster electron energy losses yield a more pronounced
steepening, correlated with radio power ($P$). Disentangling the effects of radio power and redshift is difficult because
in flux-limited samples the more powerful sources are preferentially found at higher redshifts. A further complication
arises because a convex spectral shape means that redshifting produces an apparent systematic steepening of the spectrum
between two fixed observed frequencies as redshift increases. Since the redshift information is frequently missing,
K-corrections cannot be applied and a $P$--$\alpha$ correlation may arise from any combination of these three causes.

This is the situation for the correlation reported by
\citet{1980MNRAS.190..903L}. Employing (a large proportion of)
redshift estimates for a sample drawn from the 38 MHz 8C survey,
\citet{1993MNRAS.263..707L} showed that the high-frequency (2~GHz)
spectral index correlated more closely with redshift than with
luminosity. While at first sight this may suggest the dominant
importance of inverse Compton losses on high-frequency spectra,
\citet{1993MNRAS.263..707L} pointed out that the correlation between
spectral index and redshift weakens when the radio K-correction is
applied. This means that such correlation may be induced, at least
in part, by the spectral curvature due to self absorption at the
very low selection frequency. In fact, magnetic fields in the very
luminous Lacy et al. sources should be strong enough for synchrotron
losses to dominate inverse Compton losses.
\citet{1999AJ....117..677B}, studying a number of complete samples
of radio sources selected at frequencies close to 151 MHz, with a
coverage of the $P$--$z$ plane (see \S~\ref{sec:AGN}) substantially
improved over previous studies, concluded that:

\begin{enumerate}

\item The rest-frame spectral index at low frequency depends on the
source luminosity ($P$--$\alpha$ correlation), but also on physical
size ($D$--$\alpha$ correlation) in the sense that sources with larger
physical sizes $D$ have steeper spectra.

\item The rest-frame spectral index at high frequency (GHz) depends on
the source redshift.

\end{enumerate}

\noindent Simple expressions for the average rest-frame spectra of FRI
and FRII radio galaxies as a function of radio power and
Fanaroff--Riley type \citep{1974MNRAS.167P..31F} were derived by
\citet{2001ASPC..227..242J}.

With regard to the second issue, relating source counts at different frequencies, relevant aspects are to what extent
a power-law approximation of the source spectra may be viable, i.e. to what extent low-frequency self-absorption,
electron ageing effects at high
frequencies etc. can be neglected; up to what frequencies do blazars have ``flat'' spectra; are source spectra correlated
with other parameters (luminosity, redshift) and, if so, how can these correlations be described?

In this regard we have noted that surveys at frequencies of 5~GHz and
higher are dominated (at least at the higher flux densities) by
`flat-spectrum' sources. The spectra of these sources are generally not
power-laws, but have complex and individual behaviour, showing
spectral bumps, flattenings or inversions (i.e. flux increasing with
increasing frequency), frequently bending to steeper power-laws at
higher frequencies. Examples are shown in
Fig.~\ref{fig:sour_spectra}. The complex behaviour results from the
superposition of the peaked (self-absorption) spectra of up to several
components. These components are generally beamed relativistically
with the object-axis close to the line of sight; they are the
parts of jet-base components racing towards the
observer at highly relativistic speeds.

The dominant populations of flat-spectrum sources are BL Lac objects
and flat-spectrum radio quasars (FSRQs), collectively referred to as
``blazars''. Their spectral energy distributions (i.e. the distributions of $\nu L_\nu$) show two broad
peaks. The low-energy one, extending from the radio to the UV and
sometimes also to X-rays, is attributed to synchrotron emission from a
relativistic jet, while the high-energy one, in the $\gamma$-ray
range, is interpreted as an inverse-Compton component arising from
upscattering of either the synchrotron photons themselves
\citep[synchrotron self-Compton process, SSC,
e.g.][]{1992ApJ...397L...5M, 1993AIPC..280..578B} or photons produced
by the accretion disc near the central black-hole and/or scattered/reprocessed in the broad-line region
\citep{1993NYASA.688..311B, 1994ApJ...421..153S}.

\citet{1994MNRAS.267..167G} investigated the radio to sub-millimeter
spectra of a random sample of very luminous BL Lacs and radio-loud
violently variable quasars. They found generally flat or slowly
rising 5--37 GHz spectra (median $\alpha_5^{37}\simeq 0$ for both
populations), and declining 150--375 GHz spectra, with a
statistically significant difference between BL Lacs and quasars,
the former having flatter spectra (median $\alpha_{150}^{375}\simeq
0.43$) than the latter (median $\alpha_{150}^{375}\simeq 0.73$).

Indication of strong spectral curvature was reported by
\citet{2000MNRAS.319..121J} for a quasi-complete sample drawn from
the 2.7 GHz Parkes half-Jansky flat-spectrum
sample \citep{1997MNRAS.284...85D}. 
It should be noted
however that the data used to construct the radio spectra are
heterogeneous, and the bias which this introduces is very serious
\citep{2005A&A...434..133W}. An extrapolation of this spectral
behaviour to frequencies above 20 GHz would be in conflict with WMAP
finding of a median spectral index $\alpha \simeq 0$
\citep{2003ApJS..148....1B}.

\citet{1998MNRAS.299..433F} found evidence for an anticorrelation
between the frequency of the synchrotron peak $\nu_p$ and the blazar
luminosity, and proposed a scenario, dubbed ``the blazar sequence'',
for a unified physical understanding of the vast range of blazar
properties. The scenario was further extended by
\citet{1998MNRAS.301..451G}. The idea is that blazars indeed
constitute a spectral sequence, the source power being the fundamental
parameter. The more luminous sources are ``redder'', i.e. have both
the synchrotron and the inverse Compton components peaking at lower
frequencies than the lower-luminosity ``blue'' blazars. If so, the
sub-mm steepening could be a property of only the brightest
sources. The validity of the scenario has been questioned, however
\citep[e.g.,][]{1999A&A...351...59G, 2003ApJ...588..128P,
2004MNRAS.348..937C, 2005MNRAS.356..225A, 2006A&A...445..441N,
2006ApJ...637..183L, 2007Ap&SS.309...63P, 2008A&A...488..867N}. 

On the whole, the spectral curvature question is still subject to
dispute. Substantial progress is expected from the forthcoming
surveys by the Planck mission \citep{2006astro.ph..4069T}, covering
the range 30--857 GHz, that will provide the first complete samples
allowing unbiased studies of the high frequency behaviour. 

\section{Evolutionary models: radio AGNs}
\label{sec:AGN}

By 1966 the counts produced from the Cambridge 3C and 4C surveys at
178 MHz \citep{1966MNRAS.133..151G} had already defined the
characteristic shape confirmed and refined by later surveys: in the
Euclidean normalized presentation, a `Euclidean' portion at the
brightest flux densities, followed at lower flux densities by a
`bump' \citep[dubbed `bulge' by][]{1994AuJPh..47..625W}, and then a
roll-off toward the faint intensities. The `bump' is the signature
of strong cosmic evolution, implying that the universe must have
evolved from a state of `violent activity' in the past to a more
quiescent phase at the present epoch, in contradiction to the steady
state model as pointed out by \citet{1961MNRAS.122..349R}. It is
also in complete contradiction to uniformly-populated Friedman
models \citep[e.g.][]{1980MNRAS.193..683W}. A further implication of
strong evolution, also noted by \citet{1961MNRAS.122..349R}, is that
it swamps the effect on the source counts of different cosmological
models, frustrating the original hopes that the counts could inform
on the geometry of the universe. The evolution is indeed
spectacular, yielding co-moving source densities at $z \sim 1$--2 a
factor of $\gsim 10^3$ higher than those of local sources of
similarly high luminosities \citep{1966MNRAS.133..421L}. As stressed
by \citet{1989ApJ...338...13C}, this means that the fraction of
nearby radio sources is low even at large flux densities; the median
redshifts of radio sources in complete samples with limits ranging
from mJy to Jy are close to 0.8.

Two reference evolutionary scenarios for radio sources (not to be
interpreted literally) were used in the 1960s to interpret the counts;
they remain popular today. In the \emph{luminosity evolution}
scenario, the comoving density of sources is constant but luminosities
vary with cosmic epoch, while in the \emph{density evolution} scenario
the comoving density of sources of any luminosity varies.
A density {\it and} luminosity evolution of the luminosity function at
the frequency $\nu$ can be described as:
\begin{equation}\label{eq:evol}
\rho(L,z;\nu) = f_d(z)\rho(L/f_l(z),z=0;\nu),
\end{equation}
where $f_d(z)$ and $f_l(z)$ are the functions describing the density
and the luminosity evolution, respectively.

In his pioneering investigation, \citet{1966MNRAS.133..421L} showed
that either $f_{d}(z) = (1+z)^m$ or $f_{l}(z) = (1+z)^{m'}$ would
fit his data, provided that only sources with high radio luminosity
($\log(L_{178\rm MHz})\gsim 33.5$ in the case of luminosity
evolution or $\log(L_{178\rm MHz})\gsim 34.9$ in the case of density
evolution; luminosities in
$\hbox{erg}\,\hbox{s}^{-1}\,\hbox{Hz}^{-1}$) evolve. This is called
power-law evolution in which the function is a power of the
expansion factor of the Universe. (To be precise, Longair quoted
evolution functions proportional to $t^n$ in an Einstein--de Sitter
universe, where the cosmic time $t$ is proportional to
$(1+z)^{-3/2}$). The power-law evolution diverges at high redshifts,
and must be truncated. The problem is avoided in the exponential
evolution model proposed both by \cite{1970MNRAS.147..139D} and by
\cite{1970MNRAS.149..365R}, $f_{d}(z)= \exp(k\tau(z))$ or $f_{l}(z)=
\exp(k'\tau(z))$, where $\tau$ is the lookback time in units of the
Hubble time $H_0^{-1}$, and $k^{-1}$  or $k'^{-1}$ are the
evolutionary timescales in the same units.

A further step in exploring radio-AGN evolution was taken by
\citet{1968ApJ...151..393S} and \citet{1968MNRAS.138..445R}: the
development of the $V_{\rm max}$ or `luminosity -- volume' test. The
test avoids mentioning source counts, at the time still lingering
under a cloud of suspicion. It requires a complete sample, and it
requires that if two or more limits to this completeness are in
play, they all be considered. For example
\citet{1968ApJ...151..393S} applied the test to the quasars of the
3CR catalogue, a sample defined by the radio flux-density survey
limit and by the optical magnitude limit of the original Palomar
Observatory Sky Survey. The test is simple: take each object and
``push'' it out in increasing redshift until it becomes faint enough
to encounter one of the two limits, radio or optical. (The process
requires that the optical and radio spectra be known so as to define
K-corrections in each band.) Stop ``pushing'' at the {\it first}
limit encountered; this defines $z_{\rm max}$ and hence $V_{\rm
max}$, the maximum co-moving volume in which the object could be
found. Calculate the co-moving $V$ for the object, the volume
defined by the object's redshift, and then form the ratio $V/V_{\rm
max}$. Do this for all objects in the sample. It is easy to show
that if they are uniformly distributed in space, the values should
be uniformly distributed between 0 and 1.0, i.e. $\langle V/V_{\rm
max}\rangle = 0.5 \pm 1/\sqrt{12N}$ where $N$ is the number of
objects in the sample. The statistic is a maximum-likelihood
estimator and is unbiased. It remains in widespread use and has
undergone many refinements, e.g the $C^-$ method of
\citet{1971MNRAS.155...95L}, the application to combined samples
\citep{1980ApJ...235..694A}, $V_{\rm max}$ methods to extract an
evolution function \citep[][and references
therein]{1987MNRAS.226..273C}, methods to estimate the luminosity
function directly \citep{1976ApJ...207..700F, 1993ApJ...404...51E},
and the banded $V/V_{\rm max}$ method \citep{1990MNRAS.247...19D}.
There is a vast literature; \citet{1997AJ....114..898W} is a good
starting point.

\citet{1968ApJ...151..393S} found strong evolution for the 3CR
quasars, in accord with the evolution derived by
e.g. \citet{1966MNRAS.133..421L}. \citet{1970MNRAS.151...45L} showed
how very closely the test was related to the source-count test and
emphasized that the two methods were far from independent. However,
$V/V_{\rm max}$ is versatile, simple in concept, model-free, and comes
with a statistical pedigree, in contrast to source counts, the trial
and error process of guessing suitable evolution functions, and the
frequently less-than-transparent statistical methods used to compare
these with source-count and redshift data.

However, the evolution functions mentioned above may carry physical
significance. For example a physical interpretation was offered by
\citet{1985ApJ...296..402C} who pointed out that models for
gravitational energy release near collapsed massive objects yield
$dL/dt=-A(t) L^{1+p}$. If $A=\hbox{constant}$ and $p=0$ we get the
exponential luminosity evolution function with timescale
$k^{-1}=A^{-1}$. If $A(t) \propto t^{-1}$ we get a power-law
evolution. It must be stressed that these evolutionary laws apply to
{\it source populations}, not to {\it individual sources}: the
evolutionary timescales, $k^{-1}$, are found to be of order of a few
to several Gyrs, while the lifetimes of bright radio AGN phases are
at least one and probably two orders of magnitude shorter
\citep{2008ApJ...676..147B}. \citet{1977A&A....54..475G} made an
early attempt to build a model explicitly constrained by
astrophysical factors. They linked the radio-source formation to
that of parent galaxies and assumed radio-emitting lifetimes
inversely proportional to radio luminosities.

Most evolution models have ignored the distributions of spectral
indices around the mean values (which may be luminosity dependent) for
both steep- and flat-spectrum sources \citep[but
see][]{1984ApJ...287..461C}. If such distributions can be approximated
as Gaussians with dispersion $\sigma$, and the differential counts
have a power-law shape, $n(S)=kS^{-\beta}$, the mean spectral index
$\bar{\alpha}$ of sources with given value of $S$ shifts with
frequency \citep{1964ApJ...140..969K, 1984ApJ...287..461C,
1984A&A...131L...1D}:
\begin{equation}\label{eq:spind}
\bar{\alpha}=\bar{\alpha}_0+\sigma^2(1-\beta)\ln(\nu/\nu_0),
\end{equation}
and the amplitude of the counts scales with frequency as
$k=k_0(\nu/\nu_0)^q$, with:
\begin{equation}\label{eq:spind}
q=\bar{\alpha}_0(1-\beta)+0.5\sigma^2(1-\beta)^2\ln(\nu/\nu_0).
\end{equation}
The dispersion of the mean spectral indices thus counteracts the
effect on counts of the high-frequency steepening
(\S\,\ref{sec:spectra}). The effect is amplified by the increase
with frequency of the variability amplitude
\citep{1988AJ.....95..307I, 2004A&A...419..485C}. We note however
that (a) power-law approximations for source counts hold over very
limited flux-density ranges only; and (b) a Gaussian approximation
for spectral-index distributions holds only for low-frequency
surveys. By 1.4 GHz, spectral-index distributions have pronounced
tails towards flat spectra and by 5 GHz, the spectral-index
distribution is almost the sum of two Gaussians with peaks at $\sim
0.8$ (steep-spectrum sources) and $\sim 0.0$ (flat-spectrum
sources). Clearly, the above equations can be used to calculate
$\bar{\alpha}$ and $q$ for each population.

\subsection{Low versus high luminosities}

Extensive discussions can be found in the literature on whether or not
the cosmological evolution is a property of powerful radio sources
only. The origin of the discussion is \citet{1966MNRAS.133..421L}'s
classic study. He adopted a luminosity function extending over about 8
decades in luminosity and with a shape not far from a power-law, so
that density and luminosity evolution were essentially equivalent over
the flux density range covered by the counts available at the
time. Under these assumptions, the relative narrowness of the
evolution bump in normalized counts (width less than 2 orders of
magnitude) compared to the breadth of the luminosity function (8
orders of magnitude) could be accounted for if {\it only the brightest
sources evolve strongly}. It was a fundamental discovery of
\citet{1966MNRAS.133..421L}'s investigation that `differential
evolution' was therefore essential: if all sources evolved equally,
there are far too many faint sources, i.e. the model bump is far too
broad.

As more data accumulated, the evolutionary models were refined.
\citet{1978MNRAS.182..617R, 1980MNRAS.190..143R} and
\citet{1980MNRAS.193..683W} factorized the evolving luminosity
function as $\rho(L,z) = F(L,z)\rho_0(L)$, where $F(L,z)$ is the
evolution function, able to represent density or luminosity
evolution, or a combination of both. Models assuming a
luminosity-independent evolution function were found to produce
unsatisfactory results, while (following the qualitative description
of the previous paragraph) good fits of the data were obtained only
by assuming a far stronger $F(L,z)$ for more luminous sources.
\citet{1987ApJ...318L..15D} proposed a luminosity-dependent
luminosity evolution model in which the luminosity evolution
timescale increases with decreasing luminosity and exceeds
$H_0^{-1}$, so that sources evolve weakly (if at all), below some
critical luminosity $L_s \sim 10^{31}\,\hbox{erg} \,\hbox{s}^{-1}
\,\hbox{Hz}^{-1}$. This is in keeping with the indications that a
variety of physical processes can sustain a steady low-level fueling
of the central engine for times longer than the Hubble time
\citep{1985ApJ...296..402C}. The \citet{1987ApJ...318L..15D} model
was exploited by \citet{1998MNRAS.297..117T} to carry out remarkably
successful predictions of radio source contributions to small scale
anisotropies measured by Cosmic Microwave Background experiments.

The differential evolution is suggestive of two populations. Pushing the high/low luminosity dichotomy to the extreme,
some investigators explicitly considered two populations, one of non-evolving low-luminosity sources, and the other of
high-luminosity, strongly evolving sources. \citet{1980RSPTA.296..367W} identified the two populations with FR\,I and
FR\,II radio galaxies. The border between FR\,I and FR\,II is, approximately, at $L_{1.4\rm GHz}\sim 10^{32 }\,\hbox{erg}
\,\hbox{s}^{-1} \,\hbox{Hz}^{-1}$, although the division appears to be dependent on both radio power and optical
luminosity of the host galaxy \citep[cf.][]{1996AJ....112....9L}. \citet{1999MNRAS.304..160J} extended the scheme to
flat-spectrum sources, assuming that BL Lac objects and flat-spectrum radio quasars are the beamed counterparts of FR\,I
and FR\,II objects respectively, as discussed in \S~\ref{unified}.

Before the advent of 2dF and SDSS sky surveys, the space
distribution of low-luminosity radio AGN was a matter of
speculation. For example, \citet{1983MNRAS.204..151L} showed that
the low-luminosity radio galaxies of 3CRR [$\log(L_{\rm 178 MHz} /
\hbox{erg} \,\hbox{s}^{-1} \,\hbox{Hz}^{-1}$)$ < 34$] gave values of
$\langle V/V_{\rm max}\rangle \approx 0.50$, suggesting little or no
evolution. This view was supported by \citet{2004MNRAS.352..909C}
who found that the comoving space density of sources fainter than
$L_{1.4\rm GHz}\sim 4 \cdot 10^{32}\,\hbox{erg} \, \hbox{s}^{-1}
\,\hbox{Hz}^{-1}$ remains approximately constant with increasing
redshift up to $z\simeq 0.5$. However, numbers were small and
uncertainties undoubtedly permitted some mild evolution. On the
basis of discovering two distant FR\,I galaxies in the Hubble Deep
Field, \citet{2001MNRAS.328..897S} proposed that FR\,I galaxies
showed significant evolution. With recourse to the 2dF sample of
galaxy redshifts, \citet{2007MNRAS.381..211S} quantified this: they
found substantial cosmological evolution over the redshift range $0
< z < 0.7$ of radio galaxies with $10^{31} < L_{1.4\rm GHz} < 10^{32
}\,\hbox{erg} \,\hbox{s}^{-1} \,\hbox{Hz}^{-1}$, i.e. in the
luminosity range of FR\,I sources. They also found indications that
the most powerful radio galaxies in their sample (with $L_{1.4\rm
GHz} > 10^{33}\,\hbox{erg} \,\hbox{s}^{-1} \,\hbox{Hz}^{-1}$)
undergo more rapid evolution over the same redshift range. The
latter findings are consistent with those by
\citet{2001MNRAS.322..536W}, who used emission-line strength rather
than morphology (FR\,I/FR\,II) to discriminate between radio-source
populations. The critical radio luminosity dividing sources with
weak/absent emission lines (the less radio-luminous population of
FR\,I and FR\,II), and the more radio-luminous population of
strong-line FR\,II radio galaxies and quasars, is approximately
$L_{1.4\rm GHz}\sim 6 \cdot 10^{32 }\,\hbox{erg} \,\hbox{s}^{-1}
\,\hbox{Hz}^{-1}$. Both populations show evidence for evolution, but
the comoving density of the more powerful sources rises far more
dramatically than that of the low-luminosity population.

It must be noted, however, that a luminosity-dependent evolution
function, $F(L,z)$, does not necessarily imply a
luminosity-dependent luminosity evolution. For example, in the case
of a luminosity function $\rho\propto L^{-\gamma}$ a uniform
power-law evolution, $L(z)=L(0)(1+z)^a$, yields $\rho[L(z)] =
\rho[L(0)](1+z)^{a\gamma}$; the evolution function depends on the
slope of the luminosity function. If the luminosity function levels
off below some `bending' luminosity $L_b$, a luminosity-independent
luminosity evolution translates into a constant comoving space
density at low luminosities, with strong variations with epoch
confined to the steep portion of the luminosity function. As shown
by \citet{1984ApJ...284...44C}, it is then possible to fit a wide
range of radio data with a model assuming that all sources evolve
equally.

Finally, we note that the debate on the evolution of low luminosity
radio AGNs has been somewhat muddled for some time by the poor
knowledge of their local luminosity function. As discussed in
\S\,\ref{sec:AGNLLF}, it is now clear that the faint portion of the
radio luminosity function is dominated by starburst galaxies, while
the luminosity function of radio AGNs flattens below $\log
L(1.4\,{\rm GHz})/\hbox{erg}\,\hbox{s}^{-1}\,\hbox{Hz}^{-1} \simeq
31$.

\subsection{Steep- versus flat-spectrum sources}
\label{steepvsflat}

The width of the bump in the Euclidean normalized counts increases
with increasing frequency. The bump of the steep-spectrum population
dominating the low-frequency counts shifts to fainter flux densities
and gradually fades as survey frequency increases. But as survey
frequency increases the evolution bump of flat-spectrum sources
becomes steadily more prominent, combining with the steep-spectrum
bump to produce an increasingly broad overall maximum, a broader range
of flux densities over which the count slope is close to Euclidean
\citep{1987IAUS..124..545K}. This initially misled investigators to
consider different space distributions and evolutionary laws for
steep- and flat-spectrum sources, with less evolution for the latter.
Demonstrating the lack of independence of $V/V_{\rm max}$ and
source-count results, similar indications of little evolution for
flat-spectrum objects came from this
direction. \citet{1968ApJ...151..393S} found a slight indication in
the 3CR sample of quasars that $\langle V/V_{\rm max} \rangle$
appeared to be less for the flatter-spectrum objects. When substantial
flat-spectrum quasar samples became available from cm-wavelength
surveys, it was found that $\langle V/V_{\rm max}\rangle \simeq 0.5$,
consistent with a uniform distribution \citep{1976ApJ...209L..55S,
1977MNRAS.180..193M}. This is in contrast to $\langle V/V_{\rm
max}\rangle \simeq 0.7$ for steep-spectrum quasars, indicative of a
strong increase of their space density with $z$.

\citet{1981MNRAS.196..611P} stressed that the data available at the
time poorly constrained the luminosity function over most areas of the
luminosity-redshift plane \citep[see also][]{1981MNRAS.196..597W}, and
pointed out that within the regions where the luminosity function was
reasonably well defined, steep- and flat-spectrum sources behave
similarly: both spectral types undergo differential evolution, the
strength of evolution increasing with luminosity. The analysis by
\citet{1984ApJ...287..461C} confirmed that the two spectral classes
may indeed evolve similarly within the constraints of the data. This
position received further support from comparative analysis of
evolution of AGNs in the radio, optical and X-ray bands by
\citet{1985A&A...143..277D}, who noted that the apparently weaker
evolution of flat-spectrum sources might be related to relativistic
beaming effects that boost the apparent radio luminosity. In
particular, BL Lac objects are probably associated with low-luminosity
steep-spectrum radio sources which also show weak evolution. More
extensive data and further analyses \citep{1987ApJ...318L..15D,
1990MNRAS.247...19D} reconciled the epoch-dependent spatial
distributions of the two populations, showing that they were
essentially identical. This finding produced a necessary consistency
for the success of the unified scheme for radio-loud AGNs; it is
clearly not possible for side-on and end-on versions of the same
populations to have different space distributions. `Unified
evolutionary schemes', in which flat-spectrum quasars and BL-Lac
objects are `beamed' versions of FR I and FR II sources, were
presented by \citet{1995PASP..107..803U} and
\citet{1999MNRAS.304..160J} and are discussed in \S~\ref{unified}.

Some of the apparent discrepancy between these analyses can perhaps
be resolved noting that, while $\langle V/V_{\rm max}\rangle > 0.5$
means an increase with redshift of the source density, $\langle
V/V_{\rm max}\rangle \simeq 0.5$ does not necessarily mean no
evolution. Evolution increasing the source density up to $z\sim 2$
and decreasing it afterwards may yield $\langle V/V_{\rm max}\rangle
\simeq 0.5$ for quasars visible up to high $z$; at high-$z$ the
visibility of flat/inverted-spectrum quasars is enhanced compared to
the steep-spectrum ones by the more favourable K-correction. This
may be part of the explanation; but probably the majority of the
explanation for disparate $\langle V/V_{\rm max}\rangle$ results
arises from selecting the steep-spectrum samples from their initial
steep portion of source count, while selecting the flat-spectrum
sources from an effectively fainter (and flatter) portion of the
source count.

\subsection{High-$z$ evolution} \label{sec:highz}

At some epoch, radio galaxies and quasars have to be born; some epoch
after recombination has to have assembled galaxies suitable to harbour
both massive black hole cores and fuel systems for these to produce
collimated twin-beam radio AGNs. We have described how easily we can
trace the very strong increase of the co-moving number density of
powerful radio sources between redshift $\simeq 0$ and 2. How hard is
it to find the more distant epoch at which this relatively high
co-moving density falls to signify the epoch of AGN birth?

This is an important issue. The radio activity is associated with
processes driving the growth of supermassive black holes (SMBHs) in
galactic nuclei, which in turn is tied to galaxy formation and
evolution. The radio activity may also drive feedback processes that
have a key role in the evolution of black holes and their host
galaxies \citep{2004ApJ...600..580G, 2006MNRAS.368L..67B,
2006MNRAS.370..645B, 2006MNRAS.365...11C}. Powerful radio galaxies and
quasars trace the most massive SMBHs \citep{2003MNRAS.340.1095D,
2004MNRAS.351..347M} which are hosted by the most massive galaxies
\citep{2005SSRv..116..523F}. Radio survey data are unaffected by dust
extinction and may thus provide crucial tests of the optical/IR
results on galaxy formation.

One of the achievements of the outstanding \citet{1990MNRAS.247...19D}
free-form evolution analysis was to provide evidence for a decline (or
high redshift ``cut-off'') in the number density of sources with both
steep and flat spectra at redshifts beyond 2.5 to 3.0. However their
samples were incomplete in redshift information, and the results were
dependent on the accuracy of photometric redshifts ascribed to sources
of the optically-faintest host galaxies.

Tracing the redshift cutoff is still under
investigation. \citet{1996Natur.384..439S} reported evidence of a
strong decrease of the space density of flat-spectrum radio-loud
quasars (FSRQs) for $z > 3$. \citet{2000MNRAS.319..121J} disputed
this, pointing out that the apparently curved nature of many of the
radio spectra involved, and in particular a spectral steepening to the
high frequencies, might reduce or remove the significance of an
apparent redshift cutoff. Ignoring such steepening leads to the
prediction of more high-$z$ sources than are actually seen, and this
could be misinterpreted as evidence for a decline of the space
density. They concluded that the comoving volume covered by the
available samples is too small to make definitive statements about any
redshift cut-off for the most luminous flat-spectrum sources, although
both a constant comoving density and a decline as abrupt as those
envisaged by \citet{1996Natur.384..439S} were found to be marginally
inconsistent with the data. Using a larger sample with a more rigorous
analysis, \citet{2005A&A...434..133W} found evidence, significant at
$>4\sigma$, of a diminution in space density of flat-spectrum quasars
at $z > 3$, consistent with the redshift-cutoff forms of both
optically-selected \citep{1995AJ....110...68S, 2004AJ....128..515F}
and X-ray selected \citep{2005A&A...441..417H, 2005ApJ...624..630S}
quasars. Wall et al. drew attention to a major source of bias in the
Jarvis et al. sample, the non-contemporaneous nature of the radio
flux-density measurements. It is certain that this introduces too much
apparent high-frequency curvature and decline into the radio
spectra. The effect of the bias was demonstrated by Wall et al., using
both contemporaneous and non-contemporaneous spectral data.

\begin{figure}
\includegraphics[width=0.7\textwidth,angle=-90]{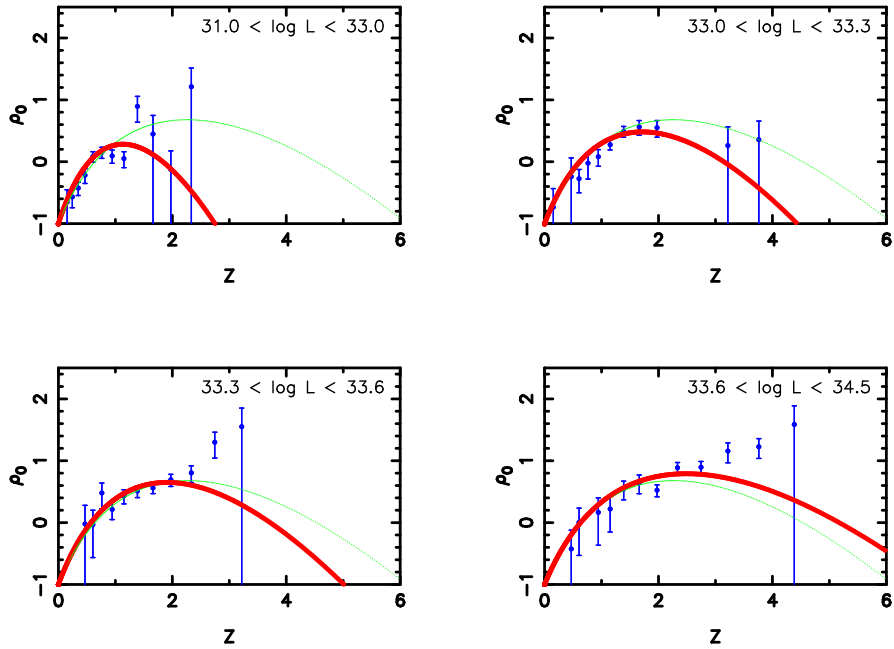}
\caption{Relative space density of flat-spectrum radio quasars from
the Parkes Quarter-Jansky sample, as derived from a Maximum-Likelihood
analysis similar to that of \citet{2008MNRAS.383..435W}. The sample
and Single-Object Survey (SOS) technique used were described by
\citet{2005A&A...434..133W}. The peak of FSRQ activity is a monotonic
function of 2.7-GHz monochromatic radio luminosity ($L$
in units of $\hbox{ergs}\,\hbox{s}^{-1}\,\hbox{Hz}^{-1}$) in the
pseudo-downsizing sense, i.e. lower luminosities have the peak of their activity at
lower redshifts. The green curve in each diagram represents the
global solution for the entire sample. This `down-sizing' is similar
to that found for quasars selected at X-ray wavelengths (Hasinger et
al. 2005) and for submm galaxies (SMG) from JCMT surveys (Wall, Pope
\& Scott 2008). } \label{walletal}
\end{figure}

\begin{figure}
\includegraphics[width=0.75\textwidth]{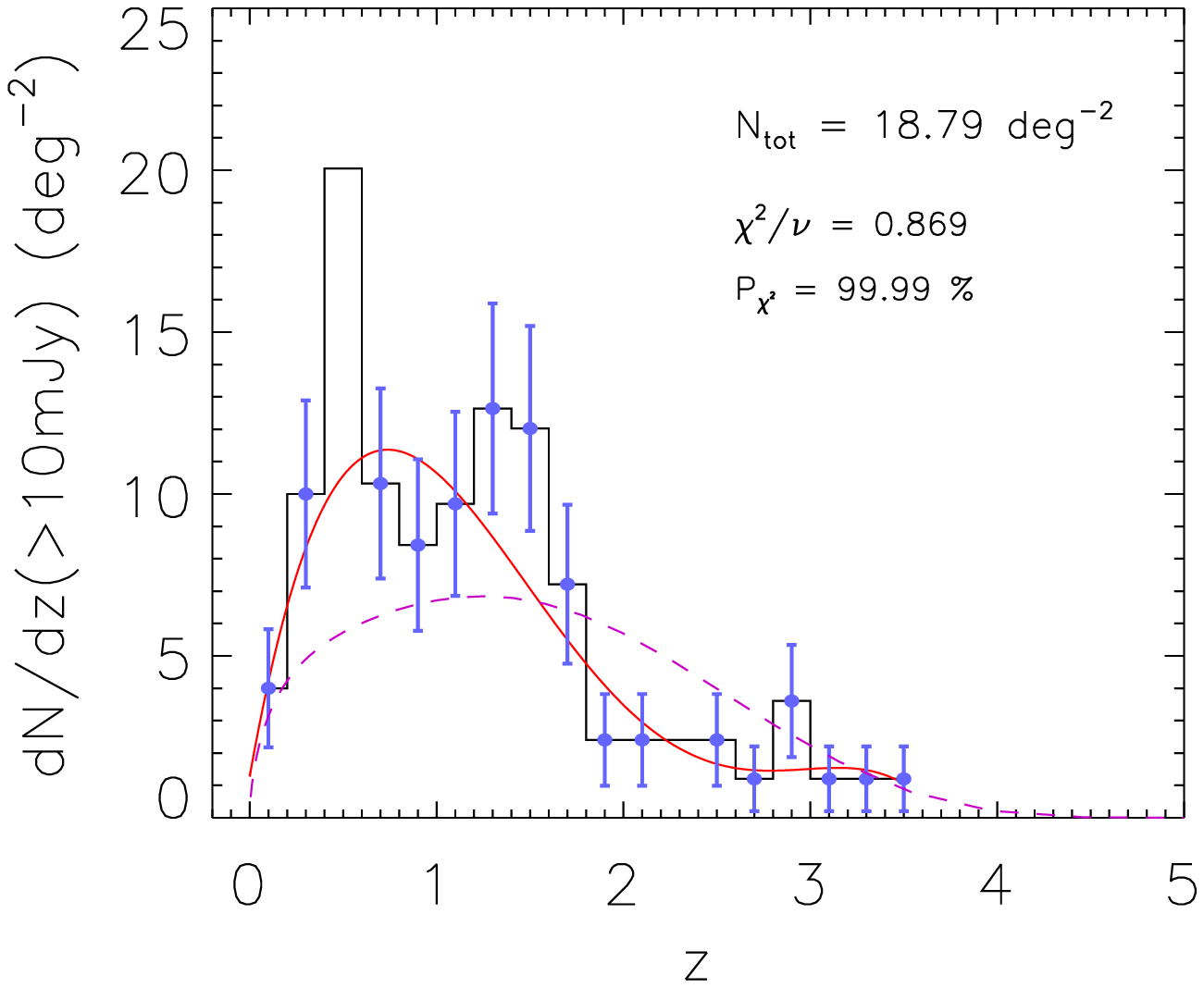}
\caption{Redshift distribution of sources brighter than 10 mJy at 1.4
 GHz in the CENSORS sample \citep{2008MNRAS.385.1297B}, compared with
that predicted by the PLE model by \citet[][dashed
 line]{1990MNRAS.247...19D} and with the fit of
 eq.~(\protect\ref{eq:nzB}). As the model applies to the AGN
 population, the two sources classified as starburst galaxies (nos. 95
 and 124) were excluded from the histogram.The Kaplan-Meier estimator
 was used to take into account the lower limits into
account.}\label{fig:nz}
\end{figure}

Using three samples selected at low frequencies,
\citet{2001MNRAS.327..907J} found that the space density of the most
radio-luminous {\it steep-spectrum} radio sources is consistent with
being constant between $z\simeq 2.5$ and $z \simeq 4.5$ and excluded a
decline as steep as suggested by \citet{1996Natur.384..439S,
1999ASPC..156..163S}. This conclusion was confirmed by
\citet{2007MNRAS.375.1349C}. However, the samples remain small and
incomplete in redshift information and as for
\citet{1990MNRAS.247...19D}, the faintest host galaxies require
redshift estimation from a $K - z$ plot. It is much harder to track
any redshift diminution for steep-spectrum radio galaxies then for
FSRQ and the statistical uncertainties are inevitably greater; showing
that the steep-spectrum samples are consistent with a uniform
distribution does not disprove the redshift cutoff found by
\citet{2005A&A...434..133W} for FSRQ.

These apparently contradictory results may again be accounted for by
luminosity-dependent evolution. \citet{1998ASPC..146...17H} reported
indications that the high-$z$ decline of the space density of
flat-spectrum quasars is more pronounced and starts at lower
redshifts for less powerful sources. These indications were
confirmed by subsequent studies that did not distinguish between
flat- and steep-spectrum sources \citep{2001MNRAS.328..882W,
2003ApJ...591...43V, 2005MNRAS.357.1267C, 2006MNRAS.371..695C}. The
luminosity dependence of the high-$z$ decline is qualitatively
similar to the {\it downsizing} observed for galaxies and optically
and X-ray selected quasars \citep{1996AJ....112..839C,
2005AJ....129..578B, 2008ApJ...675..234P}. A new analysis in
progress by Wall et al. (2009 in preparation) uses a Bayesian
modelling process similar to that described by
\citet{2008MNRAS.383..435W}, and this pseudo-downsizing effect is
very clear for FSRQ, as shown in Fig.~\ref{walletal}.

There remains further need for complete redshift information on
faint samples of steep-spectrum radio sources to clarify their
high-redshift evolution. An important step in this direction has
been the Combined EIS-NVSS Survey Of Radio Sources
\citep[CENSORS,][]{2008MNRAS.385.1297B} that included spectroscopic
observations of 143 out of a total of 150 sources with $S_{1.4\rm
GHz}
> 3.8\,$mJy. Of these, 137 form a complete sample to a flux-density
limit of 7.2\,mJy. The resulting redshift distribution agrees well
with the distribution in Fig.~28 of \citet{1984ApJ...287..461C} but
is not well reproduced by any of the \citet{1990MNRAS.247...19D}
models (see, e.g., Fig.~\ref{fig:nz}). These data promise
substantial improvement in this field.

\subsection{Unified evolutionary schemes}
\label{unified}

\citet{1995PASP..107..803U} carried out a thorough examination of
the most widely accepted version of the unification scheme
encompassing steep- and flat-spectrum radio sources. This scheme is
based on the premise that relativistic beaming of lobe-dominated,
steep-spectrum, moderate radio power FR\,I and high-radio-power
FR\,II radio galaxies gives rise to core dominated, flat-spectrum BL
Lac objects and radio loud quasars, respectively, when the line of
sight is close to the jet axis (Fig.~\ref{fig:unified}). Urry \&
Padovani based their analysis on a comparison of luminosity
functions for the different objects, and they showed that with
reasonable assumptions for the beaming parameters, these were in
agreement. (We remind the reader that in the case of beamed
emission, the true luminosities are lower than those inferred from
the observed fluxes assuming isotropic emission by a factor
$\omega/4\pi$, where $\omega$ is the solid angle of the beaming
cone(s). For example, in the case of twin beams each with
semi-aperture angle of $7^\circ$, $\omega/4\pi \simeq 0.094$.)

More recent analyses \citep{2007ApJ...667..724L,
2008ApJ...674..111C}, using observations of the jet kinematics and
the apparent superluminal speeds for a complete sample
\citep{2008ASPC..386..240L}, have confirmed the general validity of
the scheme and have improved the accuracy in parameter
determination. The assumption that the observed luminosity function
of FR\,II galaxies has the same power-law shape as the intrinsic
luminosity function of radio-loud quasars was successfully tested
using a maximum likelihood method. \citet{1992ApJ...387..449P} had
originally developed their method of comparing the calculated beamed
luminosity functions of flat-spectrum quasars and BL Lac objects
with the observed ones before their comprehensive review paper. Now
the observed distribution of Lorentz factors of relativistic jets
\citep{2008A&A...488..897H} is found to be in good agreement with
their estimate. The observed distribution of viewing angles also
turned out to be consistent with predictions of their unified model.
\citet{2008A&A...488..897H} found that the transition from blazars
to ordinary radio loud quasars occurs at a viewing angle of
$15^\circ$ to $20^\circ$, to be compared with
\citet{1995PASP..107..803U}'s estimate of $14^\circ$.

\citet{1997MNRAS.290L..17W} and \citet{1999MNRAS.304..160J} derived
an evolutionary model aimed at explaining the behaviour of both the
flat- and steep-spectrum populations within a unified scheme.
Following \citet{1995PASP..107..803U}, they assumed BL\,Lacs to be
the beamed versions of FR\,I galaxies, and FSRQs to be the beamed
versions of FR\,II galaxies. However the analysis differed from that
of Urry \& Padovani in deriving model parameters directly from the
data rather than by comparing luminosity functions. As a first step,
Jackson \& Wall derived evolution models separately for FR\,I and
FR\,II radio galaxies, using data (counts and redshifts) {\it solely
from low-frequency surveys} in order to avoid beamed objects and to
establish space-density behaviour for these parent populations. The
evolution models they derived happened to have redshift cutoffs, and
for the FRI galaxies the approximation of space density constant
with epoch was adopted. These both gave good fits to the
low-frequency data, but neither feature is essential to the outcome
of the experiment. Together with observed ratios of beamed (core)
flux to unbeamed flux, Jackson \& Wall then used a Monte Carlo
process to orient statistically-large samples of FR\,I and FR\,II
sources randomly to the line of sight. Using a grid of beaming
parameters (range of Lorentz factors, torus opening angles), they
then calculated the number of beamed objects (BL\,Lacs and FSRQs)
produced at different flux densities. Identifying these objects with
the cm-excess sources found in cm-wavelength surveys, they then
closed the loop by using the higher-frequency source counts
(primarily at 5 GHz) to find the permissible range of beaming
parameters. The process was able to reproduce the exact form of the
higher-frequency counts with their broader bump
(\S~\ref{steepvsflat}) as well as torus opening angles and Lorentz
factors in reasonable agreement with observations.
\citet{2007ApJ...667..724L} found that the Lorentz factor
distribution is much steeper for low-redshift ($z < 0.1$)
low-luminosity sources than for the more powerful, high-$z$ ones,
although the uncertainties are large. Therefore the most extreme
relativistic jets are rarer in the low-$z$ population. This
indicates that the low- and high-redshift groups are likely to be
from different parent populations, consistent with the
dual-population scheme of \citet{1999MNRAS.304..160J}.

The \citet{1999MNRAS.304..160J} unification scheme was adopted by
\citet{2008MNRAS.388.1335W} to reproduce the observed variety of
radio-loud AGNs, including radio galaxies, steep- and flat-spectrum
quasars, and strongly beamed sources such as BL\,Lacs and FSRQs. These
authors produced a simulation of a sky area of $20\times
20\,\hbox{deg}^2$ out to $z=20$, and down to a flux density of
$10\,$nJy at several frequencies from 151 MHz to 18 GHz. The model
comprises 5 source populations with different evolutionary properties:
FR\,I and FR\,II AGNs, ``radio-quiet'' AGNs (defined as all X-ray -
selected AGNs), and star-forming galaxies in quiescent and
star-bursting phases. The simulation includes redshift-dependent size
distributions both for radio-loud AGNs and for star-forming
galaxies. Clustering is modeled by attributing to each population an
effective halo mass and computing the corresponding bias factors,
$b(z)$. The simulations have been built with the SKA in mind and will
serve to inform design of the SKA, design of analysis software and
design of observing programmes.

\subsection{Special classes of sources}

\subsubsection{GHz peaked spectrum (GPS) sources}

GPS sources are powerful ($\log L_{\rm 1.4\, GHz} \gsim
32\,\hbox{erg}\,\hbox{s}^{-1}\,\hbox{Hz}^{-1}$), compact ($\!\lsim
1\,$kpc) radio sources with convex spectra peaking at GHz
frequencies \citep[see][for a comprehensive
review]{1998PASP..110..493O}. They are identified with both galaxies
and quasars. It is now widely agreed that GPS sources correspond to
the early stages of the evolution of powerful radio sources, when
the radio-emitting region grows and expands within the interstellar
medium of the host galaxy, before plunging out into the
intergalactic medium and becoming an extended radio source
\citep{1995A&A...302..317F, 1996ApJ...460..634R,
1996cyga.book..209B, 2000MNRAS.319..445S}. Conclusive evidence that
these sources are young came from VLBI measurements of propagation
velocities. Speeds of up to $\simeq 0.4c$ were measured, implying
dynamical ages $\sim 10^3$ years \citep{1999NewAR..43..657P,
2000ApJ...541..112T, 2000A&A...360..887T}. The identification and
investigation of these sources is therefore a key element in the
study of the early evolution of radio-loud AGNs.

The model by \citet{2000A&A...354..467D} implies that extreme GPS
quasars, peaking at $\nu > 20\,$GHz, should comprise a substantial
fraction of bright radio sources in the WMAP survey at $\nu \simeq
20\,$GHz, while GPS galaxies with similar $\nu_{\rm peak}$ should be
about 10 times less numerous. For a maximum rest-frame peak
frequency $\nu_{p,i} =200\,$GHz, the model predicts about 10~GPS
quasars with $S_{30{\rm GHz}} > 2\,$Jy peaking at $\geq 30\,$GHz
over the 10.4 sr at $|b| >10^\circ$. Although the number of quasars
with spectral peaks at $\geq 30\,$GHz in the WMAP survey is
consistent with this, when additional data
\citep{2003BSAO...55...90T} are taken into account such sources look
more like blazars caught during a flare that is optically thick up
to high frequencies. Furthermore, \citet{2005A&A...432...31T} showed
that most (perhaps two thirds) of the quasars in the sample of High
Frequency Peaker (HFP, GPS sources peaking above a few GHz)
candidates selected by \citet{2000A&A...363..887D} are likely to be
blazars, while all the 10 candidates classified as galaxies are
consistent with being truly young sources. This conclusion was
strengthened by the VLBA variability and polarization studies of
\citet{2005A&A...435..839T}, \citet{2006A&A...450..959O,
2007A&A...475..813O}, and \citet{2008A&A...479..409O}.

An implication of these results is that the samples of confirmed GPS
{\it quasars} are too small to allow meaningful study of their
evolutionary properties. The situation is somewhat better for GPS {\it
galaxies}. \citet{2006A&A...445..889T} found that the observed
redshift and peak frequency distributions of these sources can be
satisfactorily accounted for in the framework of the self-similar
expansion model proposed by \citet{1996cyga.book..209B,
1999mdrg.conf..173B}. According to this model, the properties of the
sources are determined by the interaction of a compact, jet-driven,
over-pressured, non-thermal radio lobe with a dense interstellar
medium. Fits of the redshift and peak frequency distributions require
a decrease of the emitted power and of the peak luminosity with source
age or with decreasing peak frequency, consistent with expectations
from Begelman's model, but at variance with the
\citet{2000MNRAS.319..445S} model.

\subsubsection{Late stages of AGN evolution}\label{ADAF}

Late stages of the AGN evolution, characterized by low
radiation/accretion efficiency, were brought into sharper focus by
the discovery of ubiquitous, moderate-luminosity hard X-ray emission
from nearby ellipticals. VLA studies at high radio frequencies (up
to 43~GHz) have shown, albeit for a limited sample of objects, that
all the observed compact cores of elliptical and S0 galaxies have
spectra rising up to $\simeq 20$--30 GHz
\citep{1999MNRAS.305..492D}.

There is growing evidence that essentially all massive ellipticals
host super-massive black holes \citep[e.g.][]{2005SSRv..116..523F}.
Yet nuclear activity is not observed at the level expected from
\citet{1952MNRAS.112..195B} spherical accretion theory, in the
presence of extensive hot gaseous halos, and for the usually-assumed
radiative efficiency $\simeq 10\%$ \citep{1999MNRAS.305..492D}.
However as proposed by \citet{1982Natur.295...17R}, the final stages
of accretion in elliptical galaxies may occur via
Advection-Dominated Accretion Flows (ADAFs), characterized by a very
low radiative efficiency \citep{1995MNRAS.277L..55F}. The ADAF
scenario implies strongly self-absorbed thermal cyclo-synchrotron
emission due to a near-equipartition magnetic field in the inner
parts of the accretion flows, most easily detected at cm to mm
wavelengths. However the ADAF scenario is not the only possible
explanation of the data, and is not problem-free. {\it Chandra}
X-ray observations of Sgr~A at the Galactic Center are suggestive of
a considerably lower accretion rate compared to Bondi's prediction
\citep{2003ApJ...591..891B}, so that the very low ADAF radiative
efficiency may not be required.

A stronger argument against a pure ADAF in the Galactic Center is
that the emission is strongly polarized at mm/sub-mm wavelengths
\citep{2000ApJ...534L.173A, 2000ApJ...538L.121A}. Moreover
\citet{1999MNRAS.305..492D} and \citet{2001ApJ...547..731D} found
that the high-frequency nuclear radio emission of a number of nearby
early-type galaxies is substantially below the prediction of
standard ADAF models. The observations are more consistent with the
adiabatic inflow-outflow solutions (ADIOS) developed by
\citet{1999MNRAS.303L...1B}, whereby the energy liberated by the
accretion drives an outflow at the polar region. This outflow
carries a considerable fraction of the mass, energy and angular
momentum available in the accretion flow, thus suppressing the radio
emission from the inner regions. Both the intensity and the peak of
the radio emission depend on the mass loss rate.

Tentative estimates of the counts due to these sources and of the
associated small-scale fluctuations were presented by
\citet{2004MNRAS.354.1005P}. As shown in
Fig.~\ref{fig:fl_flat_steep} their model A assumes a local
luminosity function (upper boundary of the cross-hatched area)
higher than current estimates of the local luminosity function of
all flat/inverted spectrum sources. Consistency is obtained only for
their minimal model B (lower boundary of the cross-hatched area).

\section{Evolutionary models: star-forming galaxies}
\label{starforming}

\subsection{Star-forming, normal and sub-mm galaxies}

The radio emission of star-forming galaxies correlates with their
star for\-ma\-tion rate, as demonstrated by the well-established
tight correlation with far-IR emission \citep{1985ApJ...298L...7H,
1986ApJ...305L..15G, 1992ARA&A..30..575C, 2002A&A...384L..19G}.
\citet{2001ApJ...554..803Y} found that the overall trend in the
range $L_\nu(60\mu{\rm m}) \simeq
10^{30}$--$10^{32.5}\,\hbox{erg}\,\hbox{s}^{-1}\,\hbox{Hz}^{-1}$ is
indistinguishable from a linear relation:
\begin{equation}
L_{1.4\rm GHz}=1.16\times 10^{-2}L_\nu(60\mu{\rm m}). \label{eq:corr}
\end{equation}
Galaxies with $L_\nu(60\mu{\rm m})<
10^{30}\,\hbox{erg}\,\hbox{s}^{-1}\,\hbox{Hz}^{-1}$ are found to
have radio to far-IR luminosity ratios systematically lower than
those given by eq.~(\ref{eq:corr}). The apparent deviation from
linearity in the radio/far-IR correlation at low luminosities is
supported by a comparison of $60\,\mu$m and 1.4 GHz local luminosity
functions \citep{2001ApJ...554..803Y, 2005MNRAS.362....9B}. Simply
shifting the $60\,\mu$m luminosity function
\citep{1990MNRAS.242..318S, 2003ApJ...587L..89T} along the
luminosity axis according to eq.~(\ref{eq:corr}) yields a good match
to the radio luminosity function \citep{2005MNRAS.362....9B,
2007MNRAS.375..931M} for $L_{1.4\rm GHz}\gsim
10^{28}\,\hbox{erg}\,\hbox{s}^{-1}\,\hbox{Hz}^{-1}$. At yet lower
luminosities, however, the extrapolated luminosity function lies
increasingly above the observed one. Full agreement is recovered
(Fig.~\ref{fig:rlf1d4GHz}) by replacing eq.~(\ref{eq:corr}) with
\begin{equation}
L_{1.4\rm GHz}=1.16\times 10^{-2}L_b \left[\left({L_\nu(60\mu{\rm
m})\over L_b}\right)^{-3.1}+ \left({L_\nu(60\mu{\rm m})\over
L_b}\right)^{-1}\right]^{-1}, \label{eq:corr1}
\end{equation}
in which $L_b = 8.8\times
10^{29}\,\hbox{erg}\,\hbox{s}^{-1}\,\hbox{Hz}^{-1}$.

While a radio/far-IR correlation is expected since young stars are
responsible both for dust heating and for the generation, via
supernova explosions, of synchrotron emitting relativistic
electrons, a clear explanation of its tightness and of its linearity
over a large luminosity range is still missing. A decrease of the
$L_{1.4\rm GHz}/L_\nu(60\mu{\rm m})$ ratio with {\it increasing}
far-IR luminosity is expected from the increase of the effective
dust temperature, $T_d$, with luminosity
\citep{1996MNRAS.279..847B}. For a galaxy like the Milky Way, the
far-IR SED peaks at $170\,\mu$m, whereas for an Ultra Luminous
Infrared Galaxy (ULIRG) it peaks at about $60\,\mu$m
\citep{2005ARA&A..43..727L}. This factor of 3 increase in
temperature for a factor $\sim 10^3$ increase in luminosity
corresponds to $T_d \propto L_{\rm FIR}^{1/6}$. If the radio
luminosity is proportional to the global far-IR luminosity, this
increase in dust temperature results in a decrease of the $L_{1.4\rm
GHz}/L_\nu(60\mu{\rm m})$ ratio by a factor of 2.5--3.

On the other hand, there are different contributions to the global
far-IR luminosity. In Luminous and Ultra Luminous Infrared galaxies,
the emission is dominated by warmer dust, associated with
star-formation, while infrared ``cirrus'' emission, heated by older
stars, becomes increasingly important in galaxies with lower and
lower star-formation rates. The latter component may be weakly
correlated with radio emission, if at all. Moreover, in very low
luminosity galaxies interstellar magnetic fields may be so weak as
to let synchrotron emitting electrons escape into intergalactic
space or to lose energy primarily via inverse Compton scattering of
CMB photons. These processes may over-compensate the effect of
decreasing dust temperature\footnote{We are grateful to J. Condon
for enlightening comments on this issue.}.

Anyway, the tight empirical relationship between radio and far-IR
luminosities for star-forming galaxies allows us to take advantage of
the wealth of data at far-IR/sub-mm wavelengths to derive the radio
evolution properties. We expect a different evolution for starburst
and normal late-type galaxies as the starburst activity is likely
triggered by interactions and mergers that were more frequent in the
past, while in normal galaxies the star-formation rate has probably
not changed much over their lifetimes. The bulk of the sub-mm counts
measured by SCUBA surveys \citep{2006MNRAS.370.1057S,
2006MNRAS.372.1621C} is due to yet another population, the sub-mm
galaxies (SMGs), proto-spheroidal galaxies in the process of forming
most of their stars \citep{2004ApJ...600..580G}.

There have been a number of attempts to model the evolution of
star-forming galaxies and in particular to account for the apparent
intrusion of this population into the source counts at $S_{\rm 1.4
GHz} \leq 1$~mJy; see e.g. \citet{2004MNRAS.349.1353K}. A
straightforward extrapolation to radio frequencies of the
evolutionary models by \citet{2007MNRAS.377.1557N} for the three
populations (normal, starburst and sub-mm galaxies), exploiting
eq.~(\ref{eq:corr1}) and the SEDs of NGC$\,6946$ for normal
late-type galaxies and of Arp220 for starburst and proto-spheroidal
galaxies, yields the curves shown in
Figs.~\ref{fig:count_low}--\ref{fig:4d8GHz}, nicely reproducing the
counts at tens of $\mu$Jy levels. We note however that new
observational data, some of which is described in
\S~\ref{sec:submJy}, may permit substantial refinement of these
models.

The cross-over between synchrotron plus free-free emission prevailing
at cm wavelengths, and thermal dust emission, generally occurs at
$\lambda \simeq 2$--3 mm (in the rest frame), so that at frequencies
of tens of GHz there are contributions from both components
\citep[see][]{2005A&A...431..893D}.

\subsection{Radio afterglows of $\gamma$-ray bursts (GRBs)}

The afterglow emission of GRBs can be modelled as synchrotron
emission from a decelerating blast wave in an ambient medium,
plausibly the interstellar medium of the host galaxy
\citep{1997ApJ...485L...5W, 1999ApJ...523..177W,
1999A&AS..138..533M}. The radio flux above the self-absorption break
at $\lsim 5\,$GHz, is proportional to $\nu^{1/3}$ up to a peak
frequency that decreases with time. This implies that surveys at
different frequencies probe different phases of the expansion of the
blast wave. Owing to their high brightnesses, GRB afterglows may be
detected out to exceedingly high redshifts and are therefore
important tracers of (a) the early star formation in the Universe,
and of (b) the absorption properties of the intergalactic medium
across the reionization phase. Estimates of the counts of GRB
afterglows have been made by \citet{2000ApJ...540..687C}, who found
that at a fixed time-lag after the GRB in the observer's frame,
there is only a mild change in the observed flux density at radio
wavelengths with increasing redshift. This stems in part from the
fact that afterglows are brighter at earlier times and that a given
observed time refers to an intrinsic time in the source frame that
is earlier as the source redshift increases. According to
\citet{2000ApJ...540..687C} estimates, a large area survey at
$\simeq 1\,$cm to a flux limit $\simeq 1\,$mJy should discover some
GRBs \citep[see also][]{2001PASP..113....6S, 2005A&A...431..893D}.
Predictions of Ciardi \& Loeb's models at 1.4 and 3~GHz are shown in
Figs.~\ref{fig:1d4GHz} and \ref{fig:4d8GHz}.


\begin{figure}
  \includegraphics[height=11.8cm, width=0.75\textwidth,angle=90]{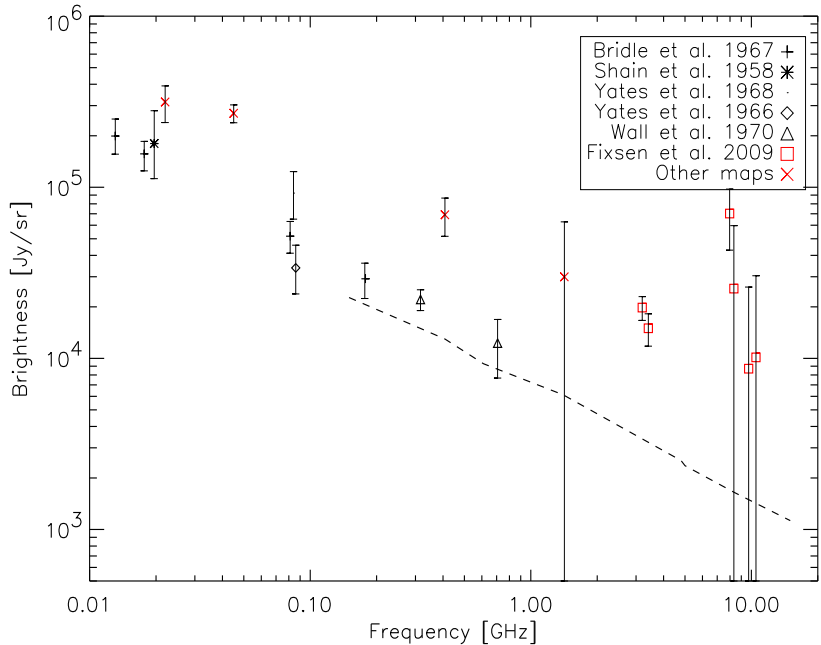}
\caption{Estimates of the extragalactic radio background at different frequencies.
The black symbols refer to estimates exploiting the methods mentioned in the first
two paragraphs of \S\,\protect\ref{sec:bck}, while the red symbols refer
to estimates by Fixsen et al. (2009) either using ARCADE 2 data (squares) or re-analyzing
published data from large area surveys ($\times$). The dashed line shows the extragalactic
background spectrum yielded by the models fitting the counts in
Figs.~\protect\ref{fig:count_low}--\protect\ref{fig:count_high}.  } \label{fig:bck}
\end{figure}

\section{The radio background}\label{sec:bck}

The radio background provides a key constraint on the counts of
sources too faint to be individually detected. At frequencies below
1~GHz, the extragalactic radio background is swamped by the much
more intense Galactic emission, primarily the synchrotron emission
from diffused and integrated supernova remnants. Estimates at meter
wavelengths made over 40 years ago \citep{1967MNRAS.136..219B} using
the $T$--$T$ plot method \citep{1962MNRAS.124..297T} yielded an
antenna temperature of the extragalactic background at 178 MHz of
$T_{\rm bkg}=30\pm 7\,$K, about one third of the minimum total sky
brightness at that frequency. Subtraction of the Galactic emission
assuming that it scales as $\csc|b|$, ($b =$ Galactic latitude) is
very inaccurate, as (i) Galactic emission towards the Galactic poles
is not removed, and (ii) there are major features such as the North
Galactic Spur which follow no such law. In fact the morphology is
complex; see for example the superb map at 408~MHz by
\citet{1982A&AS...47....1H}.

An independent estimate of the background intensity was obtained by
\citet{1970Natur.228..847C}, exploiting the low-frequency
measurements obtained with the Radio Astronomy Explorer (RAE-1)
satellite. While the $T$--$T$ plot method exploits the isotropy of
the background and its different spectral index to separate it from
Galactic emission, low-frequency measurements exploit the strong
attenuation of the extragalactic radiation below 1~MHz due to
free-absorption by electrons in the interstellar medium.
Measurements at these frequencies can therefore be used to single
out the Galactic-disk  component. Extrapolating to higher
frequencies and subtracting from measurements of the total flux
Clark et al. were able to obtain an estimate of the extragalactic
background intensity. Obviously the method works best in regions of
low Galactic emission, such as the `north halo minimum' region
($l\sim 150^\circ$, $b\sim 50^\circ$). Once again the isotropic
component was identified to be about one-third the minimum total
brightness observed at 100 MHz. The spectral index appeared to be
similar to the average spectral index observed for extra-galactic
sources, suggesting that the isotropic component does represent the
extragalactic background rather than an isotropic halo of the
Galaxy.

\citet{2008ApJ...682..223G} calculated the brightness of the
isotropic background anticipated from unresolved extragalactic
source, using modern compilations of source counts and fitting
smooth functions to these counts. Their results range from $T_b =
38600$~K at 151~MHz to $0.41$~K at 8.44~GHz; over this range the
$T_b$ -- frequency law is close to a power law.

Interest in the radio background was recently revived by the results
of the second-generation balloon-borne experiment ARCADE-2 (Absolute
Radiometer for Cosmology, Astrophysics, and Diffuse Emission). After
subtracting a model for the Galactic emission and the CMB,
\citet{2009arXiv0901.0555F} found excess radiation at 3~GHz about 5
times brighter than the estimated contribution from extragalactic
radio sources, as calculated by Gervasi et al. and in the present
work (see Fig.~\ref{fig:bck}). From a re-analysis of several
large-area surveys at lower frequencies to separate the Galactic and
extragalactic components, \citet{2009arXiv0901.0555F} obtained an
extragalactic background power-law spectrum of $T = 1.26\pm
0.09\,\hbox{K}\, (\nu/\nu_0)^{-2.60\pm 0.04}$, with $\nu_0=1\,$GHz
from 22 MHz to 10 GHz, in addition to a CMB temperature of $2.725\pm
0.001\,$K. These results are compared to earlier estimates in
Fig.~\ref{fig:bck}, where we show the background brightness,
$I_\nu$, in Jy/sr, as a function of frequency. $I_\nu$ is related to
the antenna temperature, $T_a$, by
\begin{equation}
T_a={c^2 I_\nu \over 2 k_B \nu^2} \simeq 3.25\times 10^{-5}\left({\nu\over
{\rm GHz}}\right)^{-2}{I_\nu\over \rm Jy/sr}.
\end{equation}
where the numerical coefficient holds for $T_a$ in K, and
$1\,\hbox{Jy}=10^{-23}\,\hbox{erg}\,\hbox{cm}^{-2}$
$\hbox{s}^{-1}\,\hbox{Hz}^{-1}$. As shown by the figure, the ARCADE
results are inconsistent with earlier measurements. We note that the
antenna temperature at 81.5 MHz implied by the power-law fit of
\citet{2009arXiv0901.0555F}, 854~K, exceeds the minimum {\it total}
sky brightness temperature of 680~K measured by
\citet{1967MNRAS.136..219B}.

\section{Sunyaev-Zeldovich effects}\label{SZ}

The Sunyaev \& Zeldovich (SZ) effect \citep{1972CoASP...4..173S}
arises from the inverse Compton scatter of CMB photons against hot
electrons. For a comprehensive background review see
\citet{1999PhR...310...97B}. The CMB intensity change is given by
\begin{equation}
\Delta I_\nu = 2{(k T_{\rm CMB})^3 \over (hc)^2 }y g(x)
\end{equation}
where $T_{\rm CMB}= 2.725\pm 0.002\,$K \citep{1999ApJ...512..511M}
is the CMB temperature and $x=h\nu/kT_{\rm CMB}$.

The spectral form of this ``thermal effect'' is described by the
function
\begin{equation}
g(x) = x^4\hbox{e}^x [x\cdot \coth(x/2) -4]/(\hbox{e}^x -1)^2,
\end{equation}
which is negative (positive) at values of $x$ smaller (larger) than
$x_0=3.83$, corresponding to a critical frequency $\nu_0=217$ GHz.

The Comptonization parameter is
\begin{equation}
y = \int {kT_e \over m_ec^2} n_e \sigma _T dl,
\end{equation}
where $m_e$, $n_e$ and $T_e $ are the electron mass, density and temperature
respectively, $\sigma _T$ is the Thomson cross section, and the
integral is over a line of sight through the plasma.

With respect to the incident radiation field, the change of the CMB
intensity across a galaxy or a cluster can be viewed as a net flux
emanating from the plasma cloud, given by the integral of intensity
change over the solid angle subtended by the cloud
\begin{equation} {\Delta F}_\nu =\int \Delta I_{\nu}d\Omega \propto
Y\equiv \int y d\Omega. \label{eq:ly}
\end{equation}
In the case of hot gas trapped in the potential well due to an object
of total mass $M$, the parameter $Y$ in eq.~(\ref{eq:ly}), called
integrated Y-flux, is proportional to the gas-mass-weighted electron
temperature $\langle T_e \rangle$ and to the gas mass $M_g=f_gM$:
\begin{equation}
Y\propto f_g \langle T_e \rangle M\ .
\end{equation}
At frequencies below 217 GHz, the Y-flux is negative and can therefore
be distinguished from the positive signals due to the other source
populations.

\subsection{Sunyaev-Zeldovich (SZ) effects in galaxy clusters}\label{sect:sz}

The SZ effect from the hot gas responsible for the X-ray emission of
rich clusters of galaxies has been detected with high
signal-to-noise and even imaged in many tens of objects
\citep{2002ARA&A..40..643C, 2004ApJ...617..829B,
2005MNRAS.357..518J, 2006ApJ...647...25B, 2009ApJ...701...42H,
2009ApJ...701...32S}. Detailed predictions of the counts of SZ
effects require several ingredients, generally not well known: the
cluster mass function, the gas fraction, the gas temperature and
density profiles. All these quantities are evolving with cosmic time
in a poorly-understood manner. Therefore, predictions are endowed
with substantial uncertainties. Current models generally assume a
self-similar evolution of the relationships between the main cluster
parameters \citep[mass, gas temperature, gas fraction, X-ray
luminosity;][]{2008ApJ...675..106B}. Several predictions for the SZ
counts are available \citep[e.g.][]{1995A&A...300..335D,
1997ApJ...479....1C, 2005A&A...431..893D, 2008arXiv0805.4361C}. SZ
maps have been constructed, mostly using the output of
hydrodynamical cosmological simulations, by
\citet{2005MNRAS.360...41G}, \citet{2008A&A...483..389P},
\citet{2009A&A...493..859W}, amongst others.

Our understanding of the physics of the intra-cluster plasma is
expected to improve drastically thanks to ongoing and forthcoming SZ
surveys such as those with the South Pole Telescope
\citep{2009arXiv0907.4445C}, the Atacama Cosmology Telescope
\citep{2006NewAR..50..969K}, APEX-SZ \citep{2006NewAR..50..960D}, AMI \citep{2008MNRAS.391.1545Z}, SZA \citep{2007ApJ...663..708M}, OCRA\_p \citep{2007MNRAS.378..673L}, and the Planck mission
\citep{2006astro.ph..4069T}.

\subsection{Galaxy-scale Sunyaev-Zeldovich effects}

The formation and early evolution of massive galaxies is thought to
involve the release of large amounts of energy that may be stored in
a high-pressure proto-galactic plasma, producing a detectable SZ
effect. (Note that the amplitude of the effect is a measure of the
pressure of the plasma.) According to the standard scenario
\citep{1977MNRAS.179..541R, 1978MNRAS.183..341W}, the collapsing
proto-galactic gas is shock-heated to the virial temperature, at
least in the case of large halo masses \citep[$\gsim
10^{12}\,M_{\odot}$,][]{2006MNRAS.368....2D}. Further important
contributions to the gas thermal energy may be produced by supernova
explosions, winds from massive young stars, and mechanical energy
released by central super-massive black-holes. The corresponding SZ
signals are potentially a direct probe of the processes governing
the early phases of galaxy evolution and on the history of the
baryon content of galaxies. They have been investigated under a
variety of assumptions by many authors \citep{1999ApJ...527...16O,
1999MNRAS.302..288N, 1999ApJ...522...66Y, 2000A&A...357....1A,
2001MNRAS.324..537M, 2002MNRAS.337..242P, 2003MNRAS.342L..20O,
2003ApJ...597L..93L, 2004MNRAS.348..669R, 2004AIPC..703..375D,
2007ApJ...661L.113C, 2008MNRAS.384..701M, 2008MNRAS.390..535C}.

Widely different formation modes for present day giant spheroidal
galaxies are being discussed in the literature, in the general
framework of the standard hierarchical clustering scenario. One mode
\citep{2004ApJ...600..580G, 2006ApJ...650...42L,
2009MNRAS.397..534C} has it that these galaxies generated most of
their stars during an early, fast collapse featuring a few violent,
gas rich, major mergers; only a minor mass fraction may have been
added later by minor mergers. Alternatively, spheroidal galaxies may
have acquired most of their stars through a sequence of, mostly dry,
mergers \citep{2007MNRAS.375....2D, 2008MNRAS.384....2G}.

The second scenario obviously predicts far less conspicuous
galaxy-scale SZ signals that the first one. In the framework of the
first scenario \citet{2008MNRAS.384..701M} find that the detection
of substantial numbers of galaxy-scale thermal SZ signals is
achievable by blind surveys with next generation radio telescope
arrays such as EVLA, ALMA and SKA. This population is detectable
even with a 10\% SKA, and wide-field-of-view options at high
frequencies on any of these arrays would greatly increase survey
speed. An analysis of confusion effects and contamination by radio
and dust emissions shows that the optimal frequency range is 10--35
GHz. Note that the baryon to dark matter mass ratio at virialization
is expected to have the cosmic value, i.e. to be about an order of
magnitude higher than in present day galaxies. Measurements of the
SZ effect will provide a direct test of this as yet unproven
assumption, and will constrain the epoch when most of the initial
baryons are swept out of the galaxies.

\section{Wide area surveys and large scale structure}
\label{clust}

Extragalactic radio sources are well suited to probe the large-scale
structure of the Universe: detectable over large cosmological
distances, they are unaffected by dust extinction, and can thus
provide an unbiased sampling of volumes larger than those usually
probed by optical surveys. On the other hand, their 3D
space-distribution can be recovered only in the very local Universe
\citep[$z\lsim 0.1$; see][]{1991MNRAS.253..307P, 2004MNRAS.350.1485M}
because the majority of radio galaxies detected in the available
large-area, yet relatively deep, surveys, carried out at frequencies
$\leq 1.4$~GHz, have very faint optical counterparts, so that redshift
measurements are difficult. As a result, only the {\it angular} (2D)
clustering can be measured for the entire radio AGN
population. High-frequency surveys have much higher identification
rates \citep{2006MNRAS.371..898S}, suggesting that this difficulty may
be overcome when such surveys cover sufficient sky and are linked to
wide-area redshift surveys.

\subsection{The angular correlation function and its implications}

Just the basic detection of clustering in the 2D distribution of radio
sources proved to be extremely difficult
\citep[e.g.][]{1976MNRAS.175...61W, 1981MNRAS.194..251S,
1989A&A...220...35S} because at any flux-density limit, the broad
luminosity function translates into a broad redshift distribution,
strongly diluting the spatial correlations when projected onto the
sky. Only with the advent of deep radio surveys covering large areas
of the sky, FIRST \citep{1995ApJ...450..559B}, WENSS
\citep{1997A&AS..124..259R}, NVSS \citep{1998AJ....115.1693C}, and
SUMSS \citep{2003MNRAS.342.1117M}, did it become possible to detect
the angular clustering of these objects with high statistical
significance: see \citet{1998MNRAS.297..486C} and
\citet{1998MNRAS.300..257M, 1999MNRAS.306..943M, 2002MNRAS.329L..37B}
for FIRST; \citet{2002MNRAS.337..993B, 2002MNRAS.329L..37B} and
\citet{2003A&A...405...53O} for NVSS; \citet{1999mdrg.conf..399R} for
WENSS; and \citet{2004MNRAS.347..787B}. Even then there remained
difficulties of interpretation due to spurious correlation at small
angular scales caused by the multiple-component nature of extended
radio sources \citep{2002MNRAS.337..993B}; the raw catalogues
constructed from these large surveys list {\it components} of sources
rather than single `assembled' sources. Amongst the cited surveys,
NVSS is characterized by the most extensive sky coverage and can thus
provide the best clustering statistics, despite its somewhat higher
completeness limit ($\sim 3\,$mJy vs $\sim 1\,$mJy of FIRST). The
two-point angular correlation function $w(\theta)$, measured for NVSS
sources brighter than 10 mJy, is well described by a power-law of
slope $-0.8$ extending from $\sim 0.1$ degrees up to scales of almost
10 degrees \citep{2002MNRAS.329L..37B}. A signal of comparable
amplitude and shape was detected in the FIRST survey at the same flux
density limit, on scales of up to 2-3 degrees \citep[see
e.g.][]{1998MNRAS.300..257M, 1999MNRAS.306..943M}, while at larger
angular separations any positive clustering signal - if present - is
hidden by the Poisson noise.

Most of the analyses performed so far with the aim of reproducing the
clustering of radio galaxies \citep[see e.g.][]{2002MNRAS.337..993B,
2002MNRAS.329L..37B, 2003A&A...405...53O} assumed a two-point spatial
correlation function of the form $\xi_{\rm
rg}(r)=(r/r_{0})^{-\gamma}$. The power-law shape is in fact preserved
when projected onto the sky \citep{1953ApJ...117..134L}, so that the
observed behaviour of the angular correlation is well recovered. The
correlation length $r_{0}$ was found to lie in the range 5--15 Mpc,
the large range reflecting the uncertainties in both the redshift
distribution of the sources and the time-evolution of
clustering. Despite the wide range in measurement of $r_0$, the above
results suggest that radio galaxies are more strongly clustered than
optically-selected galaxies.

A deeper examination of the power-law behaviour of the angular
two-point correlation function up to scales of the order of $\sim
10^\circ$ highlights interesting issues. Within the Cold Dark Matter
paradigm of structure formation, the spatial correlation function of
matter displays a sharp cut-off near a comoving radius of $r \sim
100\,\hbox{Mpc}$, which at the average redshift for radio sources
$<z>\sim 1$, corresponds to angular separations of only a few ($\sim
1^\circ-2^\circ$) degrees. This is in clear contrast to the
observations of the angular two-point correlation function. The
question is how to reconcile the clustering properties of these
sources with the standard scenario of structure formation. Some
authors have tried to explain the large-scale positive tail of the
angular correlation function $w(\theta)$ as due to a high-density
local population of star-forming galaxies
\citep{2004MNRAS.351..923B}. Others \citep{1999MNRAS.306..943M}
suggested that the results can be reproduced by a suitable choice of
the time-evolution of the bias parameter, i.e. the way radio
galaxies trace the underlying mass distribution. The first
hypothesis can be discarded on the basis of more recent
determinations of the space density of local star-forming galaxies
with a 1.4-GHz radio counterpart \citep[e.g.][]{2002MNRAS.333..100M,
2002MNRAS.329..227S, 2007MNRAS.375..931M}. Even the second approach,
although promising, suffers a number of limitations due to both
theoretical modelling and quality of data then available.

Theoretical predictions for the angular two-point correlation function of a given class of objects using
\citet{1953ApJ...117..134L} equation
%
%
%
\begin{equation}
w(\theta)=\int dz{\mathcal
N}^{2}(z)\int d(\delta z) \xi[r(\delta
z,\theta),z] \left/\left[\int dz{\mathcal
N}(z)\right]^{2}\right.
\label{eq:wth_obj}
\end{equation}
require two basic ingredients: their redshift distribution, ${\mathcal
N}(z)$, i.e. the number of objects brighter than the flux limit of the survey as a
function of redshift, and the value of the bias factor as a function
of redshift, $b(z)$. In equation~(\ref{eq:wth_obj}), $r(\delta
z,\theta)$ represents the {\it comoving} spatial distance between two
objects located at redshifts $z$ and $z+\delta z$ and separated by an
angle $\theta$ on the sky. For a flat universe and in the small-angle
approximation (still reasonably accurate for scales of interest here,
$0.3^{\circ}\lsim\theta\lsim 10^{\circ}$)
\begin{equation}
r^{2}=\left(\frac{c}{H_0}\right)^{2}\left[\left({\delta z\over
E(z)}\right)^2+\left(d_{C}(z)\theta\right)^2\right],
\label{eq:r}
\end{equation}
with
\begin{equation}
E(z)=\left[\Omega_M(1+z)^3 + \Omega_\Lambda\right]^{1/2},\
d_C(z)=\int_0^z{dz'\over E(z')}.
\label{eq:r}
\end{equation}
On sufficiently large scales, where the clustering signal is produced by galaxies residing in distinct dark-matter halos
and under the assumption of a one-to-one correspondence between sources and their host halos, the spatial two-point
correlation function can be written as the product of the correlation function of dark matter, $\xi_{\rm DM}$, times the
square of the bias parameter, $b$ \citep{1997MNRAS.286..115M,1998MNRAS.299...95M}:
\begin{equation}
\xi(r,z)=b^{2}(M_{\rm eff},z)\xi_{\rm DM}(r,z).
\label{eq:xi_obj}
\end{equation}
Here, $M_{\rm eff}$ is the effective mass of the dark matter haloes in
which the sources reside and $b$ is derived in the extended
\citet{1974ApJ...187..425P} formalism according to the prescriptions
of \citet{1999MNRAS.308..119S}. 

\citet{2006MNRAS.368..935N} adopted the ${\mathcal N}(z)$ from
\citet{1990MNRAS.247...19D}'s pure luminosity evolution model. If the
effective mass of the dark matter haloes in which the sources reside
does not depend on cosmic time, as found for optically-selected
quasars \citep{2004MNRAS.355.1010P, 2004ASPC..311..457C}, the
predicted angular correlation function badly fails to reproduce the
observed one. This is because contributions to $w(\theta)$ on a given
angular scale come from both local, relatively close pairs of sources
and from high-redshift, more distant ones. But for $z\gsim 1$
angular scales $\theta\gsim 2^\circ$ correspond to linear scales where
the correlation function is negative. Since the contribution of
distant objects is overwhelming, we expect negative values of
$w(\theta)$, while observations give us positive values.

\begin{figure}
\includegraphics[width=0.75\textwidth]{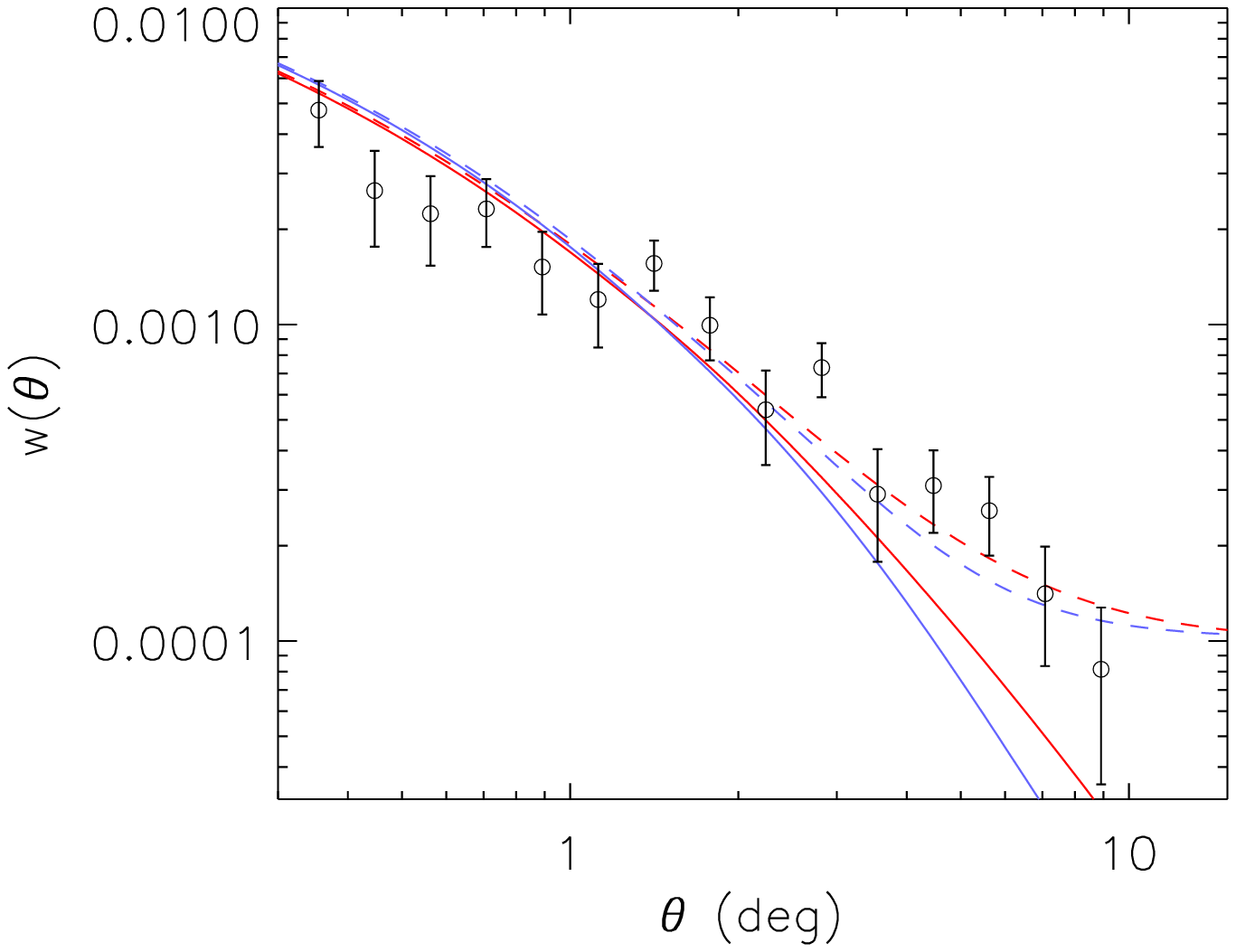}
\caption{Two-point angular correlation function of NVSS sources with
$S_{1.4\rm GHz}\ge 10\,$mJy as measured by \citet{2002MNRAS.337..993B}
compared with the model by \citet[][red curves]{2006MNRAS.368..935N}
and with the updated model (blue curves) fitting the redshift distribution by
\citet{2008MNRAS.385.1297B}. The dashed curves include the
contribution of a constant offset $\epsilon=0.0001$ to $w(\theta)$ in
order to account for the effect of possible spurious density gradients
in the survey.}
\label{fig:wth}
\end{figure}

The only way out appears to be a damping down of the contribution to
$w(\theta)$ of high-$z$ sources, and this can only be achieved through
$b(z)$. \citet{2006MNRAS.368..935N} found that the $w(\theta)$ data
can be reproduced by assuming an epoch-dependent effective mass
proportional to the mass scale at which the matter-density
fluctuations collapse to form bound structures. Such mass decreases
with increasing redshift, thus abating the negative high-$z$
contributions to $w(\theta)$. This assumption may be justified --
locally, AGN-powered radio galaxies are found mainly in very dense
environments such as groups or clusters of galaxies, and the
characteristic mass of virialized systems indeed decreases with
increasing redshift. The best fit to the data was obtained for a high
value of the local effective mass, $M_{\rm eff}(z=0) \simeq
10^{15}\,M_{\odot}$/h. However the CENSORS data
\citep{2008MNRAS.385.1297B} have shown that the redshift distribution
peaks at lower redshifts than predicted by \citet{1990MNRAS.247...19D}
PLE model (Fig.~\ref{fig:nz}). Using a smooth description of the
CENSORS redshift distribution
\begin{equation} {\mathcal N}(z)= 1.29 + 32.37z -32.89z^2 + 11.13z^3
-1.25z^4,
\label{eq:nzB}
\end{equation}
the best fit is obtained with a somewhat lower value for the local
effective mass, $M_{\rm eff}(z=0) \simeq 10^{14.5}\,M_{\odot}$/h
(Fig.~\ref{fig:wth}).

\subsection{Integrated Sachs--Wolfe (ISW) effect}

The ISW effect describes the influence of the evolution of the
gravitational potential in time-variable, linear, metric perturbations
on CMB photons that traverse them. When the CMB photons enter an
overdensity they are gravitationally blue-shifted, and they are
red-shifted when they emerge. In an Einstein-de Sitter universe the
density contrast grows as the linear scale, so that the gravitational
potential associated with the mass fluctuation is independent of
time. Hence the red- and the blue-shift exactly compensate each other
and the net effect is zero. However, a non-zero effect arises if the
gravitational potential decays, as in the case of an open universe
when the effect of the space curvature is important, or when the
dynamics of the universe are dominated by dark energy.

As first pointed out by \citet{1996PhRvL..76..575C}, a promising way
of probing the ISW effect is through correlating fluctuations in the
Cosmic Microwave Background (CMB) with large-scale structure. Since
the timescale for the decay of the potential is of the order of the
present-day Hubble time, the effect is largely canceled on small
scales, because photons travel through multiple density peaks and
troughs. This is why surveys covering large areas of the sky and
probing the large scale distribution up to $z\sim 1$ are necessary.

A high quality all-sky CMB temperature map has been provided by the
WMAP satellite \citep{2003ApJS..148....1B, 2007ApJS..170..288H,
2008S&T...115e..18H}. A particularly well-suited probe of the
large-scale structure is the NVSS survey, and indeed this has been
extensively exploited to look for the ISW signal
\citep{2004Natur.427...45B, 2005NewAR..49...75B, 2006PhRvD..74d3524P,
2007MNRAS.376.1211M, 2008MNRAS.384.1289M, 2008PhRvD..78d3519H,
2008PhRvD..77l3520G, 2008MNRAS.386.2161R}.

The comparison of the correlations inferred from the data with model
predictions requires once again the redshift distribution and the
bias parameter as a function of redshift. All analyses carried out
so far have used redshift distributions inconsistent with the
CENSORS results. The product of the latter redshift distribution
with the redshift-dependent bias factor best fitting the observed
$w(\theta)$ (see the previous sub-section), whose integral
determines the amplitude of the ISW effect, peaks at redshifts where
the contribution to the ISW signal in a $\Lambda$CDM cosmology also
peaks, namely $z\simeq 0.4$. This means that the NVSS sample is very
well suited to test the effects of dark energy on the growth of
structure. The predicted cross-correlation power spectrum between
the surface density fluctuations of NVSS sources and the CMB
fluctuations expected for the ``concordance'' $\Lambda$CDM cosmology
turns out to be in good agreement with the empirical determination
using the CMB map obtained from WMAP data. This conclusion is at
odds with that of \citet{2008PhRvD..78d3519H} who found that the
WMAP 3-year model predicts an ISW amplitude about $2\sigma$ below
their estimate. Hence we suggest that the amplitude of the ISW
cross-correlation does not support the case for new gravitational
physics on cosmological scales \citep{2008arXiv0812.2244A} or for a
large local primordial non-Gaussianity \citep{2008PhRvD..78l3507A}.

\section{The future}
\label{sec:future}

There are prospects for dramatic steps forward in radio and
millimeter-wave astronomy within the decade, thanks to a new
generation of large to gigantic interferometers as well as
refurbishment of old interferometers. Interferometric observations
gain over single-dish observations not just through resolution but
through improved sensitivity, because correlation of the signals from
the antennas can distinguish signal from noise and background. Long
integrations become possible without the limitation of systematic
errors. However, observing with interferometers requires careful
set-up of the antenna array (and its parameters in software) to image
sources correctly by measuring their flux densities on the appropriate
angular scales. There is also the dreaded problem of `missing flux'
from (lack of) low-order harmonics in the spatial transform,
corresponding to structure on the larger scales
(\S~\ref{sec:counts}). In addition, the amount of post-processing
required is large in comparison to single-dish measurements, to
correct for the many instrumental and atmospheric issues, to convert
the Fourier components into brightness images of radio sources, and
(because of incomplete sampling in the Fourier plane) to apply
algorithms to maximize image fidelity and dynamic
range. Interferometers offer an additional advantage: on the
increasingly noise-polluted surface of our planet, the processing can
excise radio-frequency interference (RFI), which, even at the remote
sites of future large arrays, would otherwise seriously compromise
observations.

The small beam size at millimeter wavelengths makes large-area deep
surveys extremely difficult because of the time penalty. New
scanning techniques will need to be developed to perform such
surveys. The next generation of interferometers, thanks to larger
collecting areas, broader bandwidths and faster scanning
capabilities are expected to produce deeper surveys of large sky
regions, both in total intensity and in polarization. Receiver
advances have resulted in huge bandwidths ($\geq 10\,$per cent) and
very low equivalent noise temperatures over these bandwidths. To
realize these gains in sensitivity with interferometers requires
development in correlator speed and in processing power. Moreover
instead of single-pixel feeds, the development of focal plane arrays
(FPAs, or phased-array feeds) looks to realize the long-standing
dream of making near-full use of the information brought to the
focal plane \citep[e.g. APERTIF,][]{2008AIPC.1035..265V}. There are
further implications for correlator- and processor-power
requirements.

The focus of current effort in the radio-astronomy community is
towards the realization of the Square Kilometer Array (SKA), the
largest and most sensitive radio telescope ever. The SKA stands to be
one of the iconic scientific instruments of the 21st century. It will
consist of an array of thousands of dishes, each 10--15 m in diameter,
as well as a complementary aperture array -- a large number of small,
fixed antenna elements plus receiver chains arranged in a regular or
random pattern on the ground. Between these two technologies, a
frequency range of 100 MHz to 25 GHz will be covered. The collecting
area will add up to approximately one million square meters, with
baselines ranging from $\sim$15 m to more than 3000 km. The SKA will
require super-fast data transport networks (of order the current total
internet capacity) and computing power far beyond current
capabilities. Indeed the concept is only feasible if Moore's Law (the
packing density of processing elements approximately doubling every
two years) continues to hold; and this in itself requires revolutions
in processor technologies. Site-testing has narrowed the choice to
remote regions either in Western Australia or in South Africa. The
telescope is expected to be fully operational after 2020, but a 10\%
SKA may be operating as early as 2015. Many different technological
solutions will be selected and integrated into the final instrument:
they will represent the results of developing the so-called SKA
pathfinders (ASKAP, MeerKAT, ATA, LOFAR, e-MERLIN, EVLA; www-page
descriptions are readily available for each).

These pathfinders carry the shorter-term excitement; they themselves
represent leaps forward in observational capability. Most of them will
be operational by 2015. They will realize dramatic improvements in
survey speed and sensitivity; ASKAP for example, is expected to
produce a survey similar to the NVSS \citep{1998AJ....115.1693C} in
sky coverage, but bettering it by a factor of 50 in sensitivity and of
5 in angular resolution \citep{2008ExA....22..151J}. These new tools
will allow us to distinguish the star-forming population from the AGN
population at low flux-density levels and to investigate source
populations at extreme redshifts.

The SKA itself will impact every area of astronomy and cosmology,
from detection and mapping of planetary systems, study of individual
stars, star clusters, pulsars, the structure of our Galaxy in both
baryons and in magnetic field, through to normal galaxies, AGNs,
proto-galaxies, and large-scale structure of the Universe. The
compelling science \citep[e.g.][and updates on the SKA
website]{2004NewAR..48..1C} to be realized makes irresistible
reading.

LOFAR \citep{2006astro.ph.10596R}, will open up the frequency window
at the low end of the radio spectrum, below 240 MHz. LOFAR will
survey the sky to unprecedented depths at low frequencies and will
therefore be sensitive to the relatively rare radio sources that
have very steep spectra, extreme luminosities and redshifts
(\S~\ref{intro}). A unique area of investigation will be the search
for redshifted 21cm line emission from the epoch of reionisation.

At the other end of the radio band, the Atacama Large Millimeter Array
(ALMA)
is eagerly anticipated by the mm continuum and molecular-line
community. Being built on a high (5000m), dry plain in the Atacama
desert, northern Chile, it is an international project, a giant
array of 50 12-m submillimetre quality antennas, with baselines of
several kilometres. An additional compact array (ACA) of 12 7-m and
4 12-m antennas is also foreseen. ALMA will be equipped with mm and
sub-mm receivers covering ultimately all the atmospheric windows at
5000m altitude in ten spectral bands, from 31 to 950 GHz. The array
will be operational by 2012 with a subset of the high-priority
receivers. The steep rise of the dust emission spectrum at mm and
sub-mm wavelengths implies that the K-correction compensates, at
$z\gsim 0.1$, for the dimming due to increasing distance
\citep{1993MNRAS.264..509B}, making the observed mm flux of dusty
galaxies of given bolometric luminosity only weakly dependent on
redshift up to $z \simeq 10$. This makes ALMA the ideal instrument
for investigating the origins of galaxies in the early universe,
with confusion made negligible by the high spatial resolution. Using
far-IR emission lines and CO rotational emission, ALMA will reveal
the astrophysics of early phases of galaxy formation and provide the
redshift of large numbers of obscured star-forming galaxies up to
very large distances. This will enable us to establish the
star-forming history of the universe, without the uncertainties
caused by dust extinction in optical studies.

Technological advances have resulted in upgrades of existing
telescopes that revolutionize performance. For examples:
\begin{enumerate}
\item The Australia Telescope Compact Array (ATCA), a 6 22m-dish
array, has recently completed the upgrade of 7 mm receivers
(working in the frequency range 30-50 GHz), and the increase of
the bandwidth from $2\times128\, \mathrm{MHz}$ to 4 GHz (thanks
to the new CABB system). These new capabilities together with
its fast scan rate ($15^\circ$/min at the meridian) will allow
the extension of the Australia Telescope 20 GHz (AT20G) Survey
to higher frequencies
or to lower flux densities. 

\item The Expanded Very Large Array (EVLA) is an upgrade of the
sensitivity and frequency coverage of the VLA. When completed, it will
use the 27 25m dishes of the VLA with 8-GHz bandwidth per polarization
in the frequency bands 18-26.5, 26.5-40, and 40-50 GHz. This is a 10
to 100-fold increase in sensitivity over the standard VLA. First
observations will be in 2010; after full commissioning (2013) the
(E)VLA is destined to remain at the forefront of radio astronomy for
at least a decade.

\end{enumerate}

Finally, several new survey instruments for the Sunyaev-Zeldovich
effect have either started operations or will shortly do so (see \S\,\ref{sect:sz}).

\section{Conclusions}
\label{concl}

For decades radio surveys have been a leading agent for extragalactic
research, as testified by the breakthroughs they triggered, from the
discovery of cosmic evolution, to quasars and the first high-$z$
galaxies. They continue at the forefront of astrophysics and
cosmology; e.g., via large-scale structure studies they pose
challenges for buildup of the cosmic web; and via downsizing and AGN
feedback deemed to produce this downsizing, they have come to the fore
in modelling galaxy formation and evolution.

But our physical understanding of the origin and evolution of the
AGN-powered radio emission is still poor. While physical models for
the cosmological evolution of galaxies and radio-quiet quasars have
been progressing rapidly in recent years, the main progress on the
radio side has been towards a phenomenological description of
evolution of various radio AGN types.

Even on the phenomenological side, there are aspects that are not fully understood. The epoch-dependent luminosity
functions of galaxies and radio-quiet quasars are now quite accurately determined up to high redshifts and there are
attempts to provide physical explanations for the evidence for earlier formation of the more massive objects.
Direct evidence for a substantial decline of the space densities of radio AGNs at $z
\gsim 2$ remains somewhat controversial, although modern evolutionary models accounting for the observed counts and
redshift distributions do include such a decline.

The origin of these uncertainties remains as it has been for the last
30 years -- a lack of concerted effort to obtain complete redshift
sets for radio AGN samples. The new redshift surveys (2dF, SDSS) have
helped greatly in defining local space densities. But beyond two or
three of the brightest samples such as 3CRR, there are {\it no}
samples with complete redshift data. Even samples of 100 to 200
objects would suffice for most purposes, and could be easily obtained
with 8- to 10m-class telescopes. Our deficiency in this regard is
highlighted by the reliance of most analyses needing complete redshift
information on the \citet{1990MNRAS.247...19D} model distributions --
even 20 years on. Doubtless the planned new deep optical and infrared
wide-field surveys such as PAN-STARRS \citep{2004SPIE.5489..667H} and
those with the VST and VISTA \citep{2007Msngr.127...28A} will
help. However there will remain a need for individual pursuit of the
faintest members in samples via deep imaging in different optical and
IR bands and via fast spectrographs to complete the redshift
information in samples of limited size.

Sixty years since their discovery, radio AGNs remain at the forefront of astrophysics and cosmology. Our continued
attempts to solve the mysteries which still surround them will doubtless lead to fresh discoveries of impact as great as
those which have distinguished the first 60 years.

\begin{acknowledgements}
We are grateful to the referee, Jim Condon, for a very careful
reading of the manuscript and for many very useful comments and
suggestions. Thanks are due to Benedetta Ciardi for having provided
the 1.4 GHz counts of GRB afterglows in tabular form. GDZ and MM
acknowledge partial financial support from ASI contracts I/016/07/0
``COFIS'' and ``Planck LFI Activity of Phase E2''. JVW acknowledges
support via Canadian NSERC Discovery Grants.
\end{acknowledgements}

\appendix

\centerline{\bf Appendix}

\medskip\noindent We present tabulations of the observed source
counts, including when relevant, an estimate of the clustering
contribution to the errors. The columns `norm. cts.' and `+,-' give,
respectively, $S^{2.5} \hbox{dN/dS} [\hbox{Jy}^{1.5}/\hbox{sr}]$, and
its positive and negative errors. The column `sampl. err.' gives, for
surveys covering less than $25\,\hbox{deg}^2$ the contribution to such
errors due to the sampling variance [eq.~(\ref{eq:sigmav})]; such
contributions are negligible for larger areas.

\begin{center}
\begin{longtable}{cccc}
\caption{Euclidean normalized differential source counts at 150 MHz.}\label{tab:150MHz} \\
\hline
$\log(S [\hbox{Jy}])$ & norm. cts. & + , - & ref. \\
\hline \endfirsthead \caption{\emph{continued}} \\ \hline
$\log(S [\hbox{Jy}])$ & norm. cts. & + , - & ref. \\
\hline \endhead \hline \multicolumn{4}{r}{\emph{continued on the
    next page}}
\endfoot \hline\endlastfoot
-0.978 &  765.720&  47.160,   55.080 &   mc90  \\
-0.908 &  957.960&  59.400,   68.760 &   mc90  \\
-0.859 & 1079.640&  66.600,   77.760 &   mc90  \\
-0.810 & 1104.120&  68.040,   79.560 &   mc90  \\
-0.772 & 1301.400&  80.280,   75.600 &   mc90  \\
-0.728 & 1330.560&  82.080,   86.400 &   mc90  \\
-0.685 & 1510.920&  93.240,   98.280 &   mc90  \\
-0.677 & 1305.360&  90.720,   88.920 &   ha88  \\
-0.636 & 1580.040&  97.561,   91.800 &   mc90  \\
-0.634 & 1459.440&  89.640,   68.040 &   ha88  \\
-0.596 & 1584.000&  84.960,   50.760 &   ha88  \\
-0.587 & 1652.400& 128.161,  107.639 &   mc90  \\
-0.552 & 1572.120&  84.240,   50.401 &   ha88  \\
-0.533 & 1947.600& 135.360,  126.720 &   mc90  \\
-0.515 & 1879.920&  85.680,   59.400 &   ha88  \\
-0.471 & 1922.040& 118.080,   32.400 &   ha88  \\
-0.467 & 1918.080& 133.560,   83.879 &   mc90  \\
-0.428 & 2070.720&  94.680,   79.560 &   ha88  \\
-0.391 & 2548.080& 197.641,  129.959 &   mc90  \\
-0.385 & 2230.560& 101.880,   85.320 &   ha88  \\
-0.347 & 2569.679& 117.360,  115.560 &   ha88  \\
-0.309 & 2768.040& 126.720,  123.479 &   ha88  \\
-0.304 & 2472.479& 172.081,  160.919 &   mc90  \\
-0.266 & 2872.800& 131.400,  127.800 &   ha88  \\
-0.228 & 2937.599& 157.321,  109.799 &   ha88  \\
-0.196 & 3187.080& 171.001,  207.360 &   mc90  \\
-0.184 & 3117.600& 191.519,   93.240 &   ha88  \\
-0.141 & 2981.160& 206.639,  153.720 &   ha88  \\
-0.109 & 3538.079& 216.721,  180.720 &   ha88  \\
-0.071 & 3618.000& 251.999,  184.679 &   mc90  \\
-0.060 & 4046.400& 248.400,  176.401 &   ha88  \\
-0.016 & 3726.000& 230.401,  189.000 &   ha88  \\
0.027 & 3866.400& 208.800,  194.399 &   ha88  \\
0.076 & 4136.401& 284.399,  237.602 &   ha88  \\
0.130 & 3726.000& 259.199,  242.640 &   ha88  \\
0.190 & 4388.399& 338.401,  251.998 &   ha88  \\
0.190 & 3837.600& 237.600,  223.201 &   mc90  \\
0.255 & 3895.199& 331.201,  280.800 &   ha88  \\
0.321 & 3808.800& 356.400,  274.681 &   ha88  \\
0.391 & 4978.800& 464.401,  388.799 &   ha88  \\
0.467 & 3614.399& 424.801,  336.238 &   ha88  \\
0.565 & 4068.000& 450.001,  457.201 &   ha88  \\
0.668 & 3805.201& 482.400,  454.321 &   ha88  \\
0.783 & 2823.120& 502.919,  500.760 &   ha88  \\
0.919 & 3349.800& 595.799,  613.080 &   ha88  \\
1.147 & 2927.520& 626.041,  554.399 &   ha88  \\
\hline \multicolumn{4}{l}{\begin{minipage}{8cm} {\vskip 5pt Reference
      codes. ha88: \citet{1988MNRAS.234..919H}; mc90:
      \citet{1990MNRAS.246..110M}.}
  \end{minipage} }
\end{longtable}
\end{center}

\begin{center}
\begin{longtable}{ccccc}
\caption{Euclidean normalized differential source counts at 325 MHz.}\label{tab:325MHz} \\
\hline
$\log(S [\hbox{Jy}])$ & norm. cts. & + , - & sampl. err. & ref. \\
\hline \endfirsthead \caption{\emph{continued}} \\ \hline
$\log(S [\hbox{Jy}])$ & norm. cts. & + , - & sampl. err. & ref. \\
\hline \endhead \hline \multicolumn{5}{r}{\emph{continued on the
    next page}}
\endfoot \hline\endlastfoot
  -3.372 &   25.400&   3.700,    3.700 &   0.044 &ow09   \\
  -3.270 &   25.900&   3.700,    3.700 &   0.044 &ow09   \\
  -3.125 &   16.900&   1.800,    1.800 &   0.029 &ow09   \\
  -2.949 &   17.200&   2.400,    2.400 &   0.030 &ow09   \\
  -2.776 &   19.300&   2.000,    2.000 &   0.033 &ow09   \\
  -2.602 &   27.300&   2.900,    2.900 &   0.047 &ow09   \\
  -2.426 &   30.800&   4.200,    4.200 &   0.053 &ow09   \\
  -2.250 &   58.000&   7.901,    7.901 &   0.100 &ow09   \\
  -2.175 &   33.935&   7.911,    7.911 &   0.241 &oo88   \\
  -2.077 &   48.700&   9.400,    9.400 &   0.084 &ow09   \\
  -1.873 &   75.660&   9.784,    9.784 &   0.538 &oo88   \\
  -1.824 &   89.000&  14.701,   14.701 &   0.153 &ow09   \\
  -1.573 &  166.577&  19.838,   19.838 &   1.185 &oo88   \\
  -1.523 &  243.600&  43.702,   43.702 &   0.418 &ow09   \\
  -1.272 &  450.850&  51.078,   51.078 &   3.207 &oo88   \\
  -1.222 &  495.600& 105.103,  105.104 &   0.851 &ow09   \\
  -0.971 &  468.133&  86.201,   86.201 &   3.330 &oo88   \\
  -0.921 &  398.100& 162.501,  162.502 &   0.684 &ow09   \\
  -0.670 & 1005.126& 213.841,  213.841 &   7.150 &oo88   \\
  -0.218 & 1535.170& 396.048,  396.048 &  10.921 &oo88   \\
\hline \multicolumn{5}{l}{\begin{minipage}{8cm} {\vskip 5pt
      Reference codes.  ow09: \citet{2009AJ....137.4846O}; oo88:
      \citet{1988A&AS...73..103O}.  \vskip 5pt}
  \end{minipage} }
\end{longtable}
\end{center}

\begin{center}
\begin{longtable}{ccccc}
\caption{Euclidean normalized differential source counts at 408 MHz.}\label{tab:408MHz} \\
\hline
$\log(S [\hbox{Jy}])$ & norm. cts. & + , - & sampl. err. & ref. \\
\hline \endfirsthead \caption{\emph{continued}} \\ \hline
$\log(S [\hbox{Jy}])$ & norm. cts. & + , - & sampl. err. & ref. \\
\hline \endhead \hline \multicolumn{5}{r}{\emph{continued on the
    next page}}
\endfoot \hline\endlastfoot
-1.857 &   86.625&  22.647,   22.647 &   2.577 &be82   \\
-1.673 &  124.875&  11.847,   11.847 &   3.715 &be82   \\
-1.488 &  200.250&  23.275,   23.275 &   5.957 &be82   \\
-1.304 &  304.875&  34.947,   34.947 &   9.069 &be82   \\
-1.120 &  430.875&  46.790,   46.790 &  12.817 &be82   \\
-1.111 &  437.400&  36.010,   36.010 &   0.835 &gr88   \\
-1.034 &  441.000&  27.013,   27.013 &   0.842 &gr88   \\
-1.021 &  491.908&  53.754,   59.034 &  -&ro77   \\
-0.936 &  478.125&  58.020,   58.020 &  14.223 &be82   \\
-0.910 &  619.861&  59.305,   47.119 &  -&ro77   \\
-0.903 &  540.000&  19.827,   19.827 &   1.031 &gr88   \\
-0.757 &  732.600&  23.442,   23.442 &   1.398 &gr88   \\
-0.752 &  835.875& 115.215,  115.215 &  24.865 &be82   \\
-0.746 &  762.302&  88.416,   66.481 &  -&ro77   \\
-0.604 & 1039.737& 149.211,  102.227 &  -&ro77   \\
-0.602 &  892.800&  27.054,   27.054 &   1.704 &gr88   \\
-0.568 &  789.750& 148.125,  148.125 &  23.493 &be82   \\
-0.456 & 1014.417& 152.595,  132.557 &  -&ro77   \\
-0.430 &  925.865&  70.137,   49.421 &  -&ro77   \\
-0.398 & 1116.000&  36.063,   36.063 &   2.130 &gr88   \\
-0.383 &  833.625& 215.184,  215.183 &  24.798 &be82   \\
-0.298 & 1020.239& 204.589,  185.769 &  -&ro77   \\
-0.271 & 1117.862&  77.449,   85.302 &  -&ro77   \\
-0.222 & 1143.000&  61.238,   61.239 &   2.182 &gr88   \\
-0.144 & 1711.456& 381.828,  319.545 &  -&ro77   \\
-0.123 & 1515.126& 155.710,  115.090 &  -&ro77   \\
-0.071 & 1278.000&  81.036,   81.037 &   2.440 &gr88   \\
0.024 & 1201.945& 172.364,  118.204 &  -&ro77   \\
0.051 & 1449.000& 118.832,  118.833 &   2.766 &gr88   \\
0.138 & 1342.800& 147.622,  147.623 &   2.563 &gr88   \\
0.167 & 1515.126& 227.471,  190.644 &  -&ro77   \\
0.243 & 1234.800& 135.021,  135.021 &   2.357 &gr88   \\
0.326 & 1850.918& 189.560,  181.435 &  -&ro77   \\
0.398 & 1008.000& 165.611,  165.612 &   1.924 &gr88   \\
0.474 & 1513.772& 227.473,  213.932 &  -&ro77   \\
0.602 & 1008.000& 165.611,  165.612 &   1.924 &gr88   \\
0.632 & 1171.481& 278.653,  264.436 &  -&ro77   \\
0.769 & 1291.174& 345.812,  339.041 &  -&ro77   \\
0.875 &  873.000& 205.206,  205.207 &   1.666 &gr88   \\
0.923 & 1493.462& 471.192,  494.210 &  -&ro77   \\
1.081 &  969.058& 118.882,   90.041 &  -&ro77   \\
1.223 &  543.089& 101.008,   85.167 &  -&ro77   \\
1.377 &  647.889& 153.950,  111.570 &  -&ro77   \\
1.525 &  768.124& 205.943,  169.792 &  -&ro77   \\
\hline \multicolumn{5}{l}{\begin{minipage}{8cm} {\vskip 5pt
      Reference codes.  be82: \citet{1982MNRAS.200..747B}; gr88:
      \citet{1988A&A...193...40G}; ro77:
      \citet{1977AuJPh..30..241R}.  \vskip 5pt}
  \end{minipage} }
\end{longtable}
\end{center}

\begin{center}
\begin{longtable}{ccccc}
\caption{Euclidean normalized differential source counts at 610 MHz.}\label{tab:610MHz} \\
\hline
$\log(S [\hbox{Jy}])$ & norm. cts. & + , - & sampl. err. & ref. \\
\hline \endfirsthead \caption{\emph{continued}} \\ \hline
$\log(S [\hbox{Jy}])$ & norm. cts. & + , - & sampl. err. & ref. \\
\hline \endhead \hline \multicolumn{5}{r}{\emph{continued on the
    next page}}
\endfoot \hline\endlastfoot
  -4.252 &    4.540&   1.034,    1.034 &   0.805 &ib09  \\
  -4.056 &    6.920&   0.780,    0.780 &   0.637 &ib09  \\
  -3.857 &    8.790&   0.745,    0.745 &   0.586 &ib09  \\
  -3.662 &   11.350&   0.875,    0.875 &   0.646 &ib09  \\
  -3.480 &    7.600&   0.559,    0.559 &   0.251 &ga08  \\
  -3.467 &   10.330&   0.938,    0.938 &   0.564 &ib09  \\
  -3.432 &   11.060&   1.103,    1.103 &   0.593 &bo07  \\
  -3.334 &    8.800&   0.578,    0.578 &   0.290 &ga08  \\
  -3.310 &   15.100&   2.735,    2.735 &   0.849 &mo07  \\
  -3.271 &   10.520&   1.212,    1.212 &   0.569 &ib09  \\
  -3.260 &    9.720&   1.137,    1.137 &   0.521 &bo07  \\
  -3.175 &   10.100&   0.686,    0.686 &   0.333 &ga08  \\
  -3.081 &   12.230&   1.665,    1.665 &   0.656 &bo07  \\
  -3.076 &   11.170&   1.710,    1.710 &   0.604 &ib09  \\
  -3.019 &   10.300&   0.869,    0.869 &   0.340 &ga08  \\
  -3.013 &   15.600&   2.093,    2.093 &   0.877 &mo07  \\
  -2.907 &   11.590&   2.113,    2.113 &   0.621 &bo07  \\
  -2.880 &   12.720&   2.583,    2.583 &   0.688 &ib09  \\
  -2.865 &   12.600&   1.176,    1.176 &   0.415 &ga08  \\
  -2.730 &   20.640&   3.814,    3.814 &   1.107 &bo07  \\
  -2.714 &   20.200&   3.396,    3.396 &   1.136 &mo07  \\
  -2.710 &   14.500&   1.670,    1.670 &   0.478 &ga08  \\
  -2.684 &   16.020&   4.161,    4.161 &   0.866 &ib09  \\
  -2.554 &   22.520&   5.309,    5.309 &   1.207 &bo07  \\
  -2.554 &   21.500&   2.599,    2.599 &   0.709 &ga08  \\
  -2.489 &   24.600&   7.371,    7.371 &   1.330 &ib09  \\
  -2.413 &   34.400&   7.166,    7.166 &   1.935 &mo07  \\
  -2.391 &   27.700&   3.908,    3.908 &   0.913 &ga08  \\
  -2.378 &   34.840&   8.908,    8.908 &   1.868 &bo07  \\
  -2.293 &   40.230&  13.467,   13.467 &   2.176 &ib09  \\
  -2.243 &   28.200&   4.987,    4.987 &   0.930 &ga08  \\
  -2.202 &   44.000&  13.478,   13.478 &   2.359 &bo07  \\
  -2.114 &   63.600&  15.421,   15.421 &   3.577 &mo07  \\
  -2.098 &   57.920&  23.470,   23.470 &   3.132 &ib09  \\
  -2.075 &   43.900&   8.425,    8.425 &   1.447 &ga08  \\
  -2.026 &   29.390&  14.784,   14.784 &   1.576 &bo07  \\
  -1.937 &   78.600&  10.802,   10.802 &   2.537 &ga08  \\
  -1.850 &   80.990&  33.354,   33.354 &   4.343 &bo07  \\
  -1.802 &   86.100&  14.273,   14.273 &   2.779 &ga08  \\
  -1.678 &  168.000&  24.703,   24.703 &   5.423 &ga08  \\
  -1.674 &  148.800&  61.272,   61.272 &   7.978 &bo07  \\
  -1.664 &   86.500&  26.451,   26.451 &   4.865 &mo07  \\
  -1.552 &  180.000&  36.000,   36.000 &   0.000 &ka79  \\
  -1.544 &  198.000&  33.613,   33.613 &   6.392 &ga08  \\
  -1.498 &   91.120&  64.615,   64.615 &   4.886 &bo07  \\
  -1.418 &  202.000&  41.713,   41.713 &   6.521 &ga08  \\
  -1.362 &  306.000&  45.000,   45.000 &   0.000 &ka79  \\
  -1.322 &  334.790& 168.360,  168.360 &  17.951 &bo07  \\
  -1.263 &  137.000&  45.913,   45.914 &   4.423 &ga08  \\
  -1.171 &  360.000&  54.000,   54.000 &   0.000 &ka79  \\
  -1.166 &  217.000&  65.874,   65.873 &   7.005 &ga08  \\
  -1.063 &  562.000& 189.653,  189.653 &  31.609 &mo07  \\
  -1.023 &  133.000&  66.838,   66.838 &   4.293 &ga08  \\
  -0.981 &  360.000&  63.000,   63.000 &   0.000 &ka79  \\
  -0.915 &  276.000& 113.351,  113.351 &   8.910 &ga08  \\
  -0.790 &  495.000&  99.000,   99.000 &   0.000 &ka79  \\
  -0.757 &  339.000& 170.352,  170.352 &  10.943 &ga08  \\
  -0.600 &  675.000& 153.000,  153.000 &   0.000 &ka79  \\
  -0.410 &  684.000& 216.000,  216.000 &   0.000 &ka79  \\
  -0.220 & 1314.000& 414.000,  414.000 &   0.000 &ka79  \\
   0.096 &  837.000& 369.000,  369.000 &   0.000 &ka79  \\
\hline \multicolumn{5}{l}{\begin{minipage}{8cm} {\vskip 5pt
      Reference codes.  bo07: \citet{2007A&A...463..519B};
      ga08: \citet{2008MNRAS.387.1037G};
      ib09: \citet{2009MNRAS.397..281I};
      ka79: \citet{1979A&A....73..107K};
      mo07: \citet{2007MNRAS.378..995M}. \vskip 10pt}
  \end{minipage} }
\end{longtable}
\end{center}

\begin{center}
\begin{longtable}{ccccc}
\caption{Euclidean normalized differential source counts at 1.4 GHz.}\label{tab:1d4GHz} \\
\hline
$\log(S [\hbox{Jy}])$ & norm. cts. & + , - & sampl. err. & ref. \\
\hline \endfirsthead \caption{\emph{continued}} \\ \hline
$\log(S [\hbox{Jy}])$ & norm. cts. & + , - & sampl. err. & ref.\\
\hline \endhead \hline \multicolumn{5}{r}{\emph{continued on the
    next page}}
\endfoot \hline\endlastfoot
  -4.770 &    6.480&   1.510,    1.510 &   0.011 & ow08 \\
  -4.668 &    6.500&   0.890,    0.890 &   0.011 & ow08 \\
  -4.602 &    3.210&   0.752,    0.752 &   0.676 & ib09 \\
  -4.523 &    6.110&   0.580,    0.580 &   0.010 & ow08 \\
  -4.467 &    2.600&   1.624,    1.624 &   0.278 & fo06 \\
  -4.398 &    4.140&   0.529,    0.529 &   0.471 & ib09 \\
  -4.373 &    4.100&   2.048,    2.048 &   0.439 & fo06 \\
  -4.347 &    5.750&   0.690,    0.690 &   0.010 & ow08 \\
  -4.340 &    0.980&   0.278,    0.278 &   0.171 &se08A \\
  -4.284 &    2.300&   0.688,    0.688 &   0.314 & ri00 \\
  -4.272 &    5.400&   2.372,    2.372 &   0.578 & fo06 \\
  -4.244 &    2.490&   0.509,    0.509 &   0.136 & ho03 \\
  -4.208 &    5.120&   0.511,    0.511 &   0.446 & ib09 \\
  -4.206 &    2.580&   0.528,    0.528 &   0.276 & ke08 \\
  -4.180 &    3.287&   0.245,    0.245 &   0.176 & bo08 \\
  -4.173 &    5.700&   2.476,    2.476 &   0.610 & fo06 \\
  -4.155 &    2.900&   0.803,    0.803 &   0.396 & ri00 \\
  -4.151 &    1.110&   0.285,    0.285 &   0.193 &se08A \\
  -4.143 &    3.270&   0.447,    0.447 &   0.179 & ho03 \\
  -4.114 &    7.010&   0.740,    0.740 &   0.012 & ow08 \\
  -4.092 &    4.454&   0.318,    0.318 &   0.239 & bo08 \\
  -4.077 &    7.000&   2.706,    2.706 &   0.750 & fo06 \\
  -4.066 &    3.660&   0.614,    0.614 &   0.201 & ho03 \\
  -4.060 &    3.100&   0.763,    0.763 &   0.423 & ri00 \\
  -4.021 &    2.520&   0.466,    0.466 &   0.270 & ke08 \\
  -4.009 &    5.060&   0.473,    0.473 &   0.374 & ib09 \\
  -4.000 &    4.434&   0.331,    0.331 &   0.238 & bo08 \\
  -4.000 &    3.910&   1.255,    1.255 &   0.110 & ho03 \\
  -3.971 &    2.100&   0.839,    0.839 &   0.254 & mi85 \\
  -3.960 &    5.200&   2.366,    2.366 &   0.557 & fo06 \\
  -3.955 &    2.700&   0.727,    0.727 &   0.368 & ri00 \\
  -3.927 &    1.610&   0.369,    0.369 &   0.280 &se08A \\
  -3.914 &    4.250&   0.908,    0.908 &   0.120 & ho03 \\
  -3.914 &    4.944&   0.371,    0.371 &   0.265 & bo08 \\
  -3.903 &    6.240&   1.110,    1.110 &   0.011 & ow08 \\
  -3.836 &    4.440&   0.429,    0.429 &   0.125 & ho03 \\
  -3.827 &    2.880&   0.546,    0.546 &   0.308 & ke08 \\
  -3.824 &    5.147&   0.438,    0.438 &   0.276 & bo08 \\
  -3.810 &    4.300&   0.460,    0.460 &   0.298 & ib09 \\
  -3.807 &    2.400&   0.698,    0.698 &   0.327 & ri00 \\
  -3.784 &    5.100&   2.364,    2.364 &   0.546 & fo06 \\
  -3.783 &    3.500&   0.905,    0.905 &   0.423 & mi85 \\
  -3.770 &    3.600&   0.404,    0.404 &   0.105 & ci99 \\
  -3.764 &    4.960&   0.500,    0.500 &   0.140 & ho03 \\
  -3.738 &    5.014&   0.499,    0.499 &   0.269 & bo08 \\
  -3.686 &    4.730&   0.412,    0.412 &   0.133 & ho03 \\
  -3.669 &    1.820&   0.487,    0.487 &   0.317 &se08A \\
  -3.650 &    4.020&   0.517,    0.517 &   0.216 & bo08 \\
  -3.648 &    4.460&   0.940,    0.940 &   0.008 & ow08 \\
  -3.629 &    4.400&   1.313,    1.313 &   0.532 & mi85 \\
  -3.627 &    2.900&   0.745,    0.745 &   0.396 & ri00 \\
  -3.611 &    3.880&   0.549,    0.549 &   0.266 & ib09 \\
  -3.597 &    4.670&   0.412,    0.412 &   0.131 & ho03 \\
  -3.569 &    5.300&   1.012,    1.012 &   0.158 & gr99 \\
  -3.561 &    5.300&   0.691,    0.691 &   0.284 & bo08 \\
  -3.543 &    4.230&   0.817,    0.817 &   0.453 & ke08 \\
  -3.509 &    3.800&   1.285,    1.285 &   0.460 & mi85 \\
  -3.495 &    4.610&   0.365,    0.365 &   0.134 & ci99 \\
  -3.487 &    4.370&   0.380,    0.380 &   0.123 & ho03 \\
  -3.472 &    4.490&   0.712,    0.712 &   0.241 & bo08 \\
  -3.413 &    5.480&   0.902,    0.902 &   0.376 & ib09 \\
  -3.384 &    4.470&   0.816,    0.816 &   0.240 & bo08 \\
  -3.378 &    6.200&   2.587,    2.587 &   0.664 & fo06 \\
  -3.377 &    2.530&   0.777,    0.777 &   0.441 &se08A \\
  -3.350 &    4.640&   0.411,    0.411 &   0.131 & ho03 \\
  -3.347 &    3.500&   0.853,    0.853 &   0.477 & ri00 \\
  -3.323 &    6.500&   1.519,    1.519 &   0.786 & mi85 \\
  -3.319 &    7.700&   0.642,    0.642 &   0.229 & gr99 \\
  -3.297 &    5.140&   1.008,    1.008 &   0.276 & bo08 \\
  -3.252 &    4.780&   0.414,    0.414 &   0.139 & ci99 \\
  -3.215 &    4.700&   1.185,    1.185 &   0.322 & ib09 \\
  -3.208 &    5.970&   1.319,    1.319 &   0.320 & bo08 \\
  -3.197 &    6.150&   0.548,    0.548 &   0.173 & ho03 \\
  -3.120 &    4.050&   1.190,    1.190 &   0.217 & bo08 \\
  -3.111 &    3.930&   1.275,    1.081 &   0.211 & bo08 \\
  -3.097 &    5.400&   1.821,    1.821 &   0.653 & mi85 \\
  -3.070 &    6.580&   1.666,    1.666 &   0.705 & ke08 \\
  -3.060 &    5.700&   0.528,    0.528 &   0.170 & gr99 \\
  -3.046 &    4.230&   1.260,    1.260 &   0.007 & ow08 \\
  -3.032 &    5.490&   1.607,    1.607 &   0.294 & bo08 \\
  -3.031 &    5.483&   1.646,    1.508 &   0.294 & bo08 \\
  -3.025 &    7.980&   0.697,    0.697 &   0.225 & ho03 \\
  -3.017 &    7.720&   2.175,    2.175 &   0.530 & ib09 \\
  -3.000 &    3.550&   0.040,    0.040 &   0.000 & wi97 \\
  -2.991 &    6.560&   0.611,    0.611 &   0.191 & ci99 \\
  -2.951 &    3.230&   0.050,    0.050 &   0.000 & wi97 \\
  -2.900 &    5.190&   0.060,    0.060 &   0.000 & wi97 \\
  -2.889 &   11.490&   1.705,    2.035 &   0.616 & bo08 \\
  -2.851 &    7.230&   0.080,    0.080 &   0.000 & wi97 \\
  -2.824 &    9.600&   3.217,    3.217 &   1.161 & mi85 \\
  -2.820 &   12.330&   3.922,    3.922 &   0.846 & ib09 \\
  -2.804 &    7.200&   0.828,    0.828 &   0.214 & gr99 \\
  -2.801 &    7.640&   0.090,    0.090 &   0.000 & wi97 \\
  -2.783 &   11.500&   0.994,    0.994 &   0.324 & ho03 \\
  -2.750 &    9.120&   0.110,    0.110 &   0.000 & wi97 \\
  -2.738 &   11.680&   1.151,    1.151 &   0.340 & ci99 \\
  -2.699 &   10.040&   0.120,    0.120 &   0.000 & wi97 \\
  -2.684 &    8.869&   2.457,    2.321 &   0.476 & bo08 \\
  -2.650 &   12.270&   0.150,    0.150 &   0.000 & wi97 \\
  -2.622 &   10.630&   5.677,    5.677 &   0.729 & ib09 \\
  -2.600 &   12.680&   0.170,    0.170 &   0.000 & wi97 \\
  -2.550 &   14.160&   0.190,    0.190 &   0.000 & wi97 \\
  -2.550 &   10.100&   1.432,    1.432 &   0.300 & gr99 \\
  -2.500 &   15.410&   0.220,    0.220 &   0.000 & wi97 \\
  -2.481 &   15.860&   2.033,    2.033 &   0.461 & ci99 \\
  -2.479 &   14.080&   4.228,    4.061 &   0.755 & bo08 \\
  -2.450 &   17.100&   0.250,    0.250 &   0.000 & wi97 \\
  -2.400 &   18.990&   0.290,    0.290 &   0.000 & wi97 \\
  -2.350 &   20.540&   0.330,    0.330 &   0.000 & wi97 \\
  -2.309 &   23.980&   2.073,    2.073 &   0.675 & ho03 \\
  -2.300 &   22.890&   0.370,    0.370 &   0.000 & wi97 \\
  -2.295 &   20.900&   3.260,    3.260 &   0.622 & gr99 \\
  -2.284 &   55.000&  14.272,   14.272 &   5.890 & ke08 \\
  -2.283 &   31.190&   8.643,    8.987 &   1.672 & bo08 \\
  -2.250 &   25.260&   0.430,    0.430 &   -- & wi97 \\
  -2.227 &   31.610&   4.436,    4.436 &   0.919 & ci99 \\
  -2.200 &   27.280&   0.490,    0.490 &   -- & wi97 \\
  -2.194 &   14.300&   3.820,    3.820 &   1.531 & fo06 \\
  -2.150 &   31.380&   0.570,    0.570 &   -- & wi97 \\
  -2.100 &   34.540&   0.650,    0.650 &   -- & wi97 \\
  -2.078 &   32.370&  12.085,   13.899 &   1.736 & bo08 \\
  -2.050 &   37.240&   0.740,    0.740 &   -- & wi97 \\
  -2.040 &   30.000&   6.066,    6.066 &   0.892 & gr99 \\
  -2.000 &   42.040&   0.850,    0.850 &   -- & wi97 \\
  -1.971 &   61.870&   9.600,    9.600 &   1.799 & ci99 \\
  -1.950 &   46.010&   0.970,    0.970 &   -- & wi97 \\
  -1.900 &   50.670&   1.110,    1.110 &   -- & wi97 \\
  -1.873 &   84.690&  31.638,   29.689 &   4.541 & bo08 \\
  -1.850 &   54.590&   1.260,    1.260 &   -- & wi97 \\
  -1.800 &   59.780&   1.430,    1.430 &   -- & wi97 \\
  -1.783 &   52.200&  12.398,   12.398 &   1.553 & gr99 \\
  -1.750 &   66.180&   1.650,    1.650 &   -- & wi97 \\
  -1.716 &   59.070&  14.433,   14.433 &   1.718 & ci99 \\
  -1.700 &   77.710&   1.940,    1.940 &   -- & wi97 \\
  -1.668 &   86.270&  41.191,   43.000 &   4.626 & bo08 \\
  -1.650 &   85.570&   2.220,    2.220 &   -- & wi97 \\
  -1.600 &   80.120&   2.350,    2.350 &   -- & wi97 \\
  -1.550 &  100.880&   2.870,    2.870 &   -- & wi97 \\
  -1.529 &  105.000&  27.279,   27.279 &   3.123 & gr99 \\
  -1.500 &  114.910&   3.340,    3.340 &   -- & wi97 \\
  -1.478 &   84.390&  11.429,   11.429 &   2.374 & ho03 \\
  -1.461 &  109.100&  30.416,   30.416 &   3.172 & ci99 \\
  -1.450 &  102.910&   3.450,    3.450 &   -- & wi97 \\
  -1.400 &  134.730&   4.300,    4.300 &   -- & wi97 \\
  -1.350 &  119.970&   4.420,    4.420 &   -- & wi97 \\
  -1.300 &  138.130&   5.170,    5.170 &   -- & wi97 \\
  -1.274 &   84.500&  37.883,   37.883 &   2.514 & gr99 \\
  -1.250 &  160.170&   6.070,    6.070 &   -- & wi97 \\
  -1.206 &  162.100&  57.503,   57.504 &   4.713 & ci99 \\
  -1.200 &  179.240&   7.000,    7.000 &   -- & wi97 \\
  -1.150 &  172.490&   7.490,    7.490 &   -- & wi97 \\
  -1.100 &  208.600&   8.980,    8.980 &   -- & wi97 \\
  -1.050 &  230.730&  10.290,   10.290 &   -- & wi97 \\
  -1.019 &  122.400&  70.794,   70.794 &   3.641 & gr99 \\
  -1.000 &  223.840&  11.050,   11.050 &   -- & wi97 \\
  -0.950 &  244.700& 109.631,  109.631 &   7.115 & ci99 \\
  -0.950 &  231.560&  12.260,   12.260 &   -- & wi97 \\
  -0.900 &  258.500&  14.120,   14.120 &   -- & wi97 \\
  -0.850 &  246.720&  15.030,   15.030 &   -- & wi97 \\
  -0.800 &  328.630&  18.920,   18.920 &   -- & wi97 \\
  -0.763 &  197.100& 139.523,  139.523 &   5.863 & gr99 \\
  -0.750 &  260.160&  18.350,   18.350 &   -- & wi97 \\
  -0.700 &  308.060&  21.770,   21.770 &   -- & wi97 \\
  -0.695 &  236.400& 167.241,  167.241 &   6.874 & ci99 \\
  -0.650 &  321.100&  24.230,   24.230 &   -- & wi97 \\
  -0.600 &  371.500&  28.410,   28.410 &   -- & wi97 \\
  -0.550 &  375.950&  31.160,   31.160 &   -- & wi97 \\
  -0.500 &  305.910&  30.640,   30.640 &   -- & wi97 \\
  -0.450 &  296.630&  32.890,   32.890 &   -- & wi97 \\
  -0.400 &  340.950&  38.440,   38.440 &   -- & wi97 \\
  -0.350 &  351.930&  42.580,   42.580 &   -- & wi97 \\
  -0.300 &  364.210&  47.220,   47.220 &   -- & wi97 \\
  -0.250 &  328.270&  48.880,   48.880 &   -- & wi97 \\
  -0.200 &  632.160&  73.940,   73.940 &   -- & wi97 \\
  -0.150 &  482.540&  70.430,   70.430 &   -- & wi97 \\
  -0.100 &  273.390&  57.790,   57.790 &   -- & wi97 \\
  -0.050 &  490.070&  84.350,   84.350 &   -- & wi97 \\
   0.002 &  286.500&  65.800,   80.300 &   -- & wi97 \\
   0.050 &  339.200&  81.900,   92.700 &   -- & wi97 \\
   0.100 &  286.400&  75.900,   97.000 &   -- & wi97 \\
   0.149 &  512.800& 111.800,  136.700 &   -- & wi97 \\
   0.198 &  317.500& 115.400,  107.500 &   -- & wi97 \\
   0.246 &  291.800& 133.000,  107.600 &   -- & wi97 \\
   0.296 &  436.900& 181.700,  139.400 &   -- & wi97 \\
   0.350 &  318.600&  29.375,   31.090 &   9.191 & br72 \\
   0.351 &  283.600& 114.100,  161.800 &   -- & wi97 \\
   0.449 &  851.100& 276.900,  287.400 &   -- & wi97 \\
   0.452 &  254.925&  23.457,   18.821 &   7.354 & br72 \\
   0.498 &  568.200& 213.800,  289.700 &   -- & wi97 \\
   0.547 &  488.800& 300.500,  246.900 &   -- & wi97 \\
   0.548 &  212.512&  32.643,   30.965 &   6.130 & br72 \\
   0.596 &  311.300& 295.300,  214.080 &   -- & wi97 \\
   0.700 &  332.400& 268.400,  299.410 &   -- & wi97 \\
   0.704 &  206.865&  31.751,   32.569 &   5.967 & br72 \\
   0.798 &  557.200& 646.800,  362.300 &   -- & wi97 \\
   0.891 &  141.885&  23.999,   22.338 &   4.093 & br72 \\
   1.359 &  150.570&  27.903,   25.461 &   4.343 & br72 \\
\hline \multicolumn{5}{l}{\begin{minipage}{8cm} {\vskip 5pt
      Reference codes.  bo08: \citet{2008ApJ...681.1129B};
      br72: \citet{1972AJ.....77..405B};
      ci99: \citet{1999MNRAS.302..222C};
      fo06: \citet{2006ApJS..167..103F};
      gr99: \citet{1999MNRAS.305..297G};
      ho03: \citet{2003AJ....125..465H};
      ib09: \citet{2009MNRAS.397..281I};
      ke08: \citet{2008ApJS..179...71K};
      mi85: \citet{1985AJ.....90.1957M};
      ow08: \citet{2008AJ....136.1889O};
      ri00: \citet{2000ApJ...533..611R};
      se08A: \citet{2008MNRAS.386.1695S};
      wi97: \citet{1997ApJ...475..479W}. \vskip 5pt}
  \end{minipage} }
\end{longtable}
\end{center}

\begin{center}
\begin{longtable}{ccccc}
\caption{Euclidean normalized differential source counts at 4.8 GHz.}\label{tab:4d8GHz} \\
\hline
$\log(S [\hbox{Jy}])$ & norm. cts. & + , - & sampl. err. & ref. \\
\hline \endfirsthead \caption{\emph{continued}} \\ \hline
$\log(S [\hbox{Jy}])$ & norm. cts. & + , - & sampl. err. & ref. \\
\hline \endhead \hline \multicolumn{5}{r}{\emph{continued on the
    next page}}
\endfoot \hline\endlastfoot
-4.750 &    0.810&   0.562,    0.562 &   0.156 &fo91  \\
-4.587 &    1.080&   0.341,    0.341 &   0.208 &fo91  \\
-4.347 &    1.890&   0.453,    0.453 &   0.364 &fo91  \\
-3.928 &    1.161&   0.455,    0.455 &   0.149 &do87  \\
-3.870 &    0.990&   0.331,    0.331 &   0.191 &fo91  \\
-3.862 &    1.975&   1.192,    0.977 &   0.381 &fo84  \\
-3.682 &    1.503&   0.416,    0.416 &   0.193 &do87  \\
-3.483 &    2.136&   0.967,    0.971 &   0.412 &fo84  \\
-3.319 &    3.807&   1.116,    1.116 &   0.489 &do87  \\
-2.919 &    4.383&   2.161,    2.134 &   0.845 &fo84  \\
-2.886 &    9.027&   2.823,    2.823 &   1.158 &do87  \\
-2.857 &    6.750&   3.160,    3.160 &   1.301 &fo91  \\
-2.721 &    6.210&   1.726,    1.726 &   0.235 &wr90  \\
-2.538 &    8.370&   2.987,    2.987 &   0.317 &wr90  \\
-2.352 &    9.630&   3.618,    3.618 &   0.364 &wr90  \\
-2.137 &   14.850&   6.235,    6.235 &   0.562 &wr90  \\
-1.961 &   29.700&  11.037,   11.037 &   1.124 &wr90  \\
-1.921 &   20.340&   6.982,    6.982 &   0.285 &pa80  \\
-1.796 &   28.080&   8.675,    8.675 &   0.394 &pa80  \\
-1.785 &   34.151&   2.586,    2.586 &   0.243 &al86  \\
-1.668 &   40.000&   2.003,    2.003 &   0.104 &gr96  \\
-1.668 &   44.165&   4.130,    4.130 &   0.314 &al86  \\
-1.648 &   46.350&   7.216,    7.216 &   0.650 &pa80  \\
-1.536 &   40.520&   4.083,    4.083 &   0.288 &al86  \\
-1.523 &   45.000&   2.001,    2.001 &   0.046 &gr96  \\
-1.389 &   60.322&   6.204,    6.204 &   0.429 &al86  \\
-1.387 &   54.270&   8.814,    8.814 &   0.761 &pa80  \\
-1.372 &   55.000&   3.001,    3.001 &   0.057 &gr96  \\
-1.222 &   64.000&   3.000,    3.001 &   0.056 &gr96  \\
-1.206 &   63.551&   7.664,    7.664 &   0.452 &al86  \\
-1.071 &   70.000&   4.000,    4.000 &   0.061 &gr96  \\
-1.020 &   73.650&  12.119,   12.119 &   0.524 &al86  \\
-0.903 &   78.000&   4.001,    4.001 &   0.068 &gr96  \\
-0.894 &   75.960&  15.542,   15.542 &   1.066 &pa80  \\
-0.757 &   81.000&   5.000,    5.001 &   0.070 &gr96  \\
-0.602 &   84.000&   5.001,    5.001 &   0.073 &gr96  \\
-0.398 &   90.000&   5.001,    5.001 &   0.078 &gr96  \\
-0.222 &   84.000&   6.000,    6.000 &   0.073 &gr96  \\
-0.071 &  107.000&   9.001,    9.000 &   0.093 &gr96  \\
0.032 &  103.410&  10.710,    8.100 & --&ku81  \\
0.097 &   96.000&   9.000,    9.000 &   0.083 &gr96  \\
0.102 &   92.250&   6.210,    7.263 & --&ku81  \\
0.173 &   84.987&   5.733,    5.400 & --&ku81  \\
0.243 &   83.000&  12.000,   12.000 &   0.072 &gr96  \\
0.278 &   79.587&   9.675,    8.622 & --&ku81  \\
0.398 &   62.000&  11.000,   11.000 &   0.054 &gr96  \\
0.412 &   87.813&   9.117,   12.042 & --&ku81  \\
0.539 &   79.587&  11.133,   16.317 & --&ku81  \\
0.764 &   62.244&  11.088,   12.762 & --&ku81  \\
\hline \multicolumn{5}{l}{\begin{minipage}{8cm}{\vskip 5pt
      Reference codes.  al86: \citet{1986A&AS...65..267A}; do87:
      \citet{1987ApJ...321...94D}; fo84:
      \citet{1984Sci...225...23F}; fo91:
      \citet{1991AJ....102.1258F}; gr96:
      \citet{1996ApJS..103..427G}; ku81:
      \citet{1981A&AS...45..367K}; pa80:
      \citet{1980A&A....85..329P}; wr90:
      \citet{1990ApJ...363...11W}.\vskip 5pt}
  \end{minipage} }
\end{longtable}
\end{center}

\begin{center}
\begin{longtable}{ccccc}
\caption{Euclidean normalized differential source counts at 8.4 GHz.}\label{tab:8d4GHz} \\
\hline
$\log(S [\hbox{Jy}])$ & norm. cts. & + , - & sampl. err. & ref. \\
\hline \endfirsthead \caption{\emph{continued}} \\ \hline
$\log(S [\hbox{Jy}])$ & norm. cts. & + , - & sampl. err. & ref. \\
\hline \endhead \hline \multicolumn{5}{r}{\emph{continued on the
    next page}}
\endfoot \hline\endlastfoot
-4.863 &    0.231&   0.050,    0.050 & --&fo02   \\
-4.717 &    0.520&   0.385,    0.385 &   0.106 &wi93   \\
-4.611 &    0.192&   0.043,    0.043 & --&fo02   \\
-4.480 &    0.530&   0.263,    0.263 &   0.108 &wi93   \\
-4.348 &    0.288&   0.064,    0.064 & --&fo02   \\
-4.192 &    1.320&   0.577,    0.577 &   0.270 &wi93   \\
-3.710 &    0.441&   0.096,    0.096 & --&fo02   \\
-3.678 &    0.834&   0.352,    0.352 &   0.089 &he05   \\
-3.481 &    1.190&   0.547,    0.547 &   0.244 &wi93   \\
-2.745 &    2.787&   1.125,    1.125 &   0.296 &he05   \\
-2.699 &    8.716&   3.107,    3.107 &   0.393 &he05   \\
-2.474 &   20.374&  11.961,   11.961 &   2.165 &he05   \\
-2.360 &   10.123&   4.550,    4.550 &   0.457 &he05   \\
-2.160 &   10.809&   5.426,    5.426 &   0.488 &he05   \\
-2.034 &   20.953&   9.918,    9.918 &   2.227 &he05   \\
-1.901 &   18.039&   8.108,    8.108 &   0.814 &he05   \\
-1.654 &   19.820&   3.769,    4.360 & --&wi93   \\
-1.618 &   41.222&  20.695,   20.695 &   1.860 &he05   \\
-1.457 &   40.823&   8.986,    6.514 & --&wi93   \\
-1.348 &   92.022&  65.202,   65.202 &   4.153 &he05   \\
-1.250 &   55.132&  15.549,   13.185 & --&wi93   \\
-1.052 &   31.982&   6.081,    5.758 & --&wi93   \\
-0.855 &   43.182&   8.211,    9.492 & --&wi93   \\
-0.658 &   62.820&  11.945,   11.321 & --&wi93   \\
-0.461 &   62.951&  19.796,   17.368 & --&wi93   \\
-0.263 &   82.912&  23.387,   19.831 & --&wi93   \\
-0.056 &   61.674&   9.908,    2.992 & --&wi93   \\
0.122 &   63.343&   6.625,    7.380 & --&wi93   \\
0.320 &   58.911&   6.161,    5.566 & --&wi93   \\
0.527 &   43.823&   8.309,    7.898 & --&wi93   \\
0.726 &   24.797&  10.319,    9.330 & --&wi93   \\
\hline \multicolumn{5}{l}{\begin{minipage}{8cm}{\vskip 5pt
      Reference codes.  fo02: \citet{2002AJ....123.2402F}; he05:
      \citet{2005ApJ...635..950H}; wi93:
      \citet{1993ApJ...405..498W}.\vskip 5pt}
  \end{minipage} }
\end{longtable}
\end{center}

\begin{table*}[h]
\begin{center}
\caption{Euclidean normalized differential 15.2 GHz source counts from the 9C survey
  \citep{2003MNRAS.342..915W,2009arXiv0908.0066W}.}\label{tab:15GHz}
\begin{tabular}{ccc}
  \hline
  $\log(S [Jy])$ & norm. cts. & + , -    \\
  \hline
  -2.229 &    8.680&   1.810,    1.810 \\
  -2.152 &    6.597&   1.440,    1.440 \\
  -2.051 &    8.054&   1.758,    1.758 \\
  -1.979 &   12.287&   1.993,    1.993 \\
  -1.923 &   10.112&   1.542,    1.542 \\
  -1.852 &   11.647&   1.842,    1.842 \\
  -1.757 &   10.083&   1.594,    1.594 \\
  -1.650 &   16.120&   2.581,    2.581 \\
  -1.561 &   12.667&   1.416,    1.416 \\
  -1.477 &   14.355&   4.053,    3.640 \\
  -1.314 &   17.579&   2.558,    1.136 \\
  -1.105 &   26.424&   3.156,    2.269 \\
  -0.826 &   26.424&   4.550,    2.819 \\
  -0.465 &   40.551&   9.108,    6.666 \\
  -0.140 &   48.529&  21.134,   13.935 \\
\hline
\end{tabular}
\end{center}
\end{table*}

\begin{table*}[h]
\begin{center}
\caption{Euclidean normalized differential 20 GHz source counts for the AT20G Bright Source
  Sample \citep{2008MNRAS.384..775M}.}\label{tab:20GHz}
\begin{tabular}{ccc}
  \hline
  $\log(S [Jy])$ & norm. cts. & + , -  \\
  \hline
  -0.108 &   37.150&   2.493,    2.494 \\
  0.275 &   50.314&   5.626,    5.625 \\
  0.658 &   35.456&   9.155,    9.155 \\
  1.042 &   17.768&  12.564,   12.564 \\
  1.808 &  125.487& 125.487,  125.487 \\
  \hline
\end{tabular}
\end{center}
\end{table*}

\begin{center}
\begin{longtable}{cccc}
\caption{Euclidean normalized differential source counts from WMAP maps
  \citep{2009MNRAS.392..733M}.}\label{tab:WMAP} \\
\hline
$\log(S [\hbox{Jy}])$ & norm. cts. & + , - &  $\nu$ (GHz)   \\
\hline \endfirsthead \caption{\emph{continued}} \\ \hline
$\log(S [\hbox{Jy}])$ & norm. cts. & + , - &  $\nu$ (GHz)  \\
\hline \endhead \hline \multicolumn{4}{r}{\emph{continued on the
    next page}}
\endfoot \hline\endlastfoot
  0.111 &   48.993&   3.720,    3.721&23 \\
  0.333 &   37.111&   4.639,    4.639&23 \\
  0.556 &   32.480&   6.371,    6.370&23 \\
  0.778 &   32.297&   9.324,    9.323&23 \\
  1.000 &   11.597&   7.538,   10.594&23 \\ \hline
  0.111 &   75.708&   8.316,    8.317&33 \\
  0.333 &   38.878&   4.646,    4.647&33 \\
  0.556 &   28.717&   5.863,    5.862&33 \\
  0.778 &   28.357&   8.550,    8.550&33 \\
  1.000 &   16.662&   9.097,   12.730&33 \\
  1.222 &   23.931&  15.556,   21.861&33 \\ \hline
  0.111 &   58.115&   5.670,    5.670&41 \\
  0.333 &   34.633&   4.364,    4.363&41 \\
  0.556 &   28.425&   5.802,    5.802&41 \\
  0.778 &   20.413&   7.068,    9.138&41 \\
  1.000 &   21.989&  10.544,   14.474&41 \\
  1.222 &   23.688&  15.397,   21.639&41 \\
  1.444 &   25.516&  21.459,   25.516&41 \\ \hline
  0.333 &   26.078&   3.805,    3.804&61 \\
  0.556 &   21.307&   5.022,    5.022&61 \\
  0.778 &   10.201&   4.892,    6.715&61 \\
  1.000 &   16.483&   9.000,   12.593&61 \\
  1.222 &   23.674&  15.388,   21.627&61 \\
  1.444 &   25.502&  21.444,   25.502&61 \\
  \hline
\end{longtable}
\end{center}

\bibliographystyle{aa} \bibliography{references}

\end{document}